\documentclass[fleqn,usenatbib]{mnras}

\usepackage{newtxtext,newtxmath}

\usepackage[T1]{fontenc}

\DeclareRobustCommand{\VAN}[3]{#2}
\let\VANthebibliography\thebibliography
\def\thebibliography{\DeclareRobustCommand{\VAN}[3]{##3}\VANthebibliography}

\usepackage{graphicx}	
\usepackage{amsmath}	
\usepackage{bm}
\usepackage{subcaption}
\usepackage{framed}
\graphicspath{{figures/}} 

\title[What makes a cosmic filament?]{What makes a cosmic filament? 
\\  the dynamical origin and identity of filaments I. fundamentals in 2D}

\author[J. Feldbrugge and R. van de Weygaert]{
Job Feldbrugge,$^{1}$\thanks{E-mail: job.feldbrugge@ed.ac.uk}
Rien van de Weygaert$^{2}$
\\
$^{1}$University of Edinburgh, Higgs Centre for Theoretical Physics, James Clerk Maxwell Building, Edinburgh EH9 3FD, UK\\
$^{2}$Kapteyn Astronomical Institute, University of Groningen, PO Box 800, 9700 AV Groningen, The Netherlands}

\date{Accepted XXX. Received YYY; in original form ZZZ}

\pubyear{2024}

\begin{document}
\label{firstpage}
\pagerange{\pageref{firstpage}--\pageref{lastpage}}
\maketitle

\begin{abstract}

Cosmic filaments are the main transport channels of matter in the Megaparsec universe, and represent the most prominent structural feature in the matter and 
galaxy distribution. 

Here we describe and define the physical and dynamical nature of cosmic filaments. It is based on the realization that the complex spatial pattern and connectivity of the cosmic web are already visible in the primordial random Gaussian density field, in the spatial pattern of the primordial tidal and deformation eigenvalue field. The filaments and other structural features in the cosmic web emerging from this are multistream features and structural singularities in phase-space. The caustic skeleton formalism allows a fully analytical classification, identification, and treatment of the nonlinear cosmic web. The caustic conditions yield the mathematical specification of weblike structures in terms of the primordial deformation tensor eigenvalue and eigenvector fields, in which filaments are identified -- in 2D -- with the so-called cusp caustics. These are centered around points that are maximally stretched as a result of the tidal force field.

The resulting mathematical conditions represent a complete characterization of filaments in terms of their formation history, dynamics, and orientation. We illustrate the workings of the formalism on the basis of a set of constrained $N$-body simulations of protofilament realizations. These realizations are analyzed in terms of spatial structure, density profiles, and multistream structure and compared to simpler density or potential field saddle point specifications. 

The presented formalism, and its 3D generalization, will facilitate the mining of the rich cosmological information contained in the observed weblike galaxy distribution, and be of key significance for the analysis of cosmological surveys such as SDSS, DESI, and Euclid.

\end{abstract}

\begin{keywords}
    large-scale structure of Universe -- cosmology: theory -- dark matter
\end{keywords}


\section{Introduction}
\label{sec:introduction}
The present study aims to infer and present a fundamental physical definition for the identity of cosmic filaments. On the basis of the analytical fully non-linear Caustic Skeleton model of the cosmic web \citep{Feldbrugge:2018}, the outline of its filamentary spine is inferred from the eigenvalues of the primordial tidal and deformation field. It yields an unequivocal, solid physical definition of the principal constituent of the cosmic web, the pervasive web-like network that constitutes the system of cosmic arteries along which matter is transported from low-density voids toward massive compact nodes. 

On scales of a few to hundreds of Megaparsec dark matter, gas, and galaxies have organized into a complex spatial network, defining an intricate spatial pattern of elongated filaments and flattened walls surrounding near-empty void regions and connecting at cluster nodes \citep{Bond:1996,Weygaert:2008}. By far the most prominent feature in the \textit{cosmic web} are the filaments \citep{Haynes:1986, Colberg:2005, Cautun:2014, Galarraga-Espinosa:2024}. The pervasive interconnecting skeleton of filamentary ridges defines the spine of the cosmic web \citep{Aragon:2010a,Aragon:2010b}. More than 50\% of matter, gas, and galaxies have aggregated in the filamentary arteries of the cosmic web \citep{Cautun:2014, Ganeshaiah:2020}.

The prominence of filaments is particularly striking in large-scale galaxy surveys. Their existence and importance was already noticed by early galaxy redshift surveys \citep{Joeveer:1978,Haynes:1986,Lapparent:1986,Shectman:1996}. Subsequent generations of wide redshift surveys have  increasingly enforced the realization that cosmic filaments form a fundamental building block of the universe's infrastructure. The 2dFGRS survey \citep{Colless:2003}, the Sloan Digital Sky Survey SDSS \citep{Tegmark:2004} and the 2MASS redshift survey \citep{Huchra:2012} all emphasized this observation, culminating in the recently published astounding vista offered by the DESI survey, 

In recent years, numerous studies have addressed the influence of the filamentary environment on the properties, formation, and evolution of galaxies. This ranges from the direct impact on the generation of rotation and spin orientation of halos and galaxies \citep{Porciani:2002a,Aragon:2007a,Hahn:2007,Aragon:2007c,Schaefer:2009,Hahn:2009,Hahn:2010,Tempel:2013,Codis:2012,Codis:2015,Codis:2018a,Ganeshaiah:2018,Ganeshaiah:2019,Lopez:2021,Lopez:2024} to the question in how far the mass, stellar content and star formation activity are influenced \citep{Cautun:2014,Kraljic:2018, Kuutma:2017,Laigle:2018,Hellwing:2020,Parente:2024}. Also, while representing the dominant morphological feature of the spatial organization of matter in the Universe, they function as the main transport channels on Megaparsec scales \citep{Courtois:2013, Pomarede:2017,Tully:2023}. As mass streams out of lower-density regions to higher-density areas, filaments funnel matter, gas, and galaxies toward the compact high-density nodes of the cosmic web, the location of massive clusters. Within this context, it is of considerable importance that filaments also dominate the gravitational and tidal force field on Megaparsec scales \citep{Kugul:2020}, the same large-scale force field that even induces the rotation of filaments \citep{Wang:2021,Xia:2021}. Moreover, various subtle gas dynamical processes involved in the anisotropic flow into and along filaments may have a decisive influence on the star formation activity and history of galaxies residing in or migrating towards them \citep{Aragon:2019,Lu:2024}.

The filamentary network covers a wide spectrum of scales. In terms of mass density, we find filaments over at least five orders of magnitude \citep{Cautun:2014}. Most outstanding are the heavy and thick filamentary extensions of the cluster nodes, the outer edges of the most prominent arteries in the cosmic web. In the local Universe, the Pisces-Perseus chain is a beautiful specimen of such features \citep{Haynes:1986}. A representative sample of filaments, identified in the SDSS survey, is the filament catalogue of \cite{Tempel:2014}. Most filaments are more moderate-density arms, often branching out of the major filaments. Towards low-density voidlike regions, this culminates in a branching forest of tenuous tendrils \citep{Cautun:2014,Alpaslan:2014,Shivashankar:2016,Jaber:2024,Aragon:2024}. This multiscale pattern is a direct manifestation of the hierarchical fashion in which structure has assembled itself over the evolution of the Universe, in which structure gradually builds up through the merging of smaller-scale features that emerged earlier. By implication, at earlier cosmic eons we find a similar web-like arrangement on smaller scales as we find currently on Megaparsec scales. In the context of the formation of galaxies, we see that galaxies and galaxy halos formed at the node of an intricate network of filamentary extensions, with substantial repercussions for the outcome and process of galaxy formation \citep[see \textit{e.g.}][]{Aquarius:2008,Bi:2024}. On a more general note, we may observe that the complex multiscale cosmic network resembles that of other intricate ridgelike networks in nature, not unlike that of nerve and artery systems \citep{Neyrinck:2018,Vazza:2020}. 

\subsection{Filament identification and classification}
While the cosmic web represents the largest structure that emerged in the Universe, its complex multiscale, morphological, and topological properties have formed a challenge for its classification and analysis. Furthering our insight into the role of filaments in the formation of structure \citep{Icke:1973,Eisenstein:1997,Cautun:2014,Kugul:2020,Wang:2024}, their influence on the formation and evolution of galaxies, and in the sensitivity of their structure and dynamics to global cosmological factors \citep[see \textit{e.g.}][]{Bonnaire:2022} makes it necessary to establish an objective physical definition of what constitutes a filament. Hence, understanding out of which primordial configurations in the initial Gaussian random field these structures emerged, and how different global cosmological conditions -- including that of the nature of dark matter, dark energy, and the mass of neutrinos -- impact their evolution and structure is of key importance. Yet, a physically motivated, unequivocal understanding and agreement on what defines a filament, or any of the other structural components of the cosmic web, has as yet still to be established.

To this adds the complication that we do not yet have a complete picture of how different galaxy populations may yield a 
biased view of the spatial structures and connections in the cosmic matter distribution. It may imply that filaments and voids
detected in galaxy surveys may not always be representative for those in the underlying dark matter distribution. A few recent
studies do indicate subtle yet significant structural and topological differences \citep{Zakharova:2023,Bermejo:2024}. 

Driven by these considerations, over the past decades, a range of methods and formalisms have been forwarded for the detection and classification of filaments. \cite{Libeskind:2018} provide a review and comparison of more than a dozen formalisms. One may distinguish at least five classes of filament finders. Geometric filament finders are usually based on the Hessian of the density or gravitational potential at each location. It includes the Twerb and Vweb formalisms \citep{Hahn:2007,Foreroromero:2009,Hoffman:2012}. The most sophisticated ones explicitly take into account the multiscale nature of the mass distribution, of which MMF/Nexus is a particular example \citep{Aragon:2007,Cautun:2013}. Topological methods address the connections between structural singularities in the mass distribution. They are amongst the most widely used formalisms in the studies of the cosmic web, in particular, that of Disperse \citep{Sousbie:2011a,Sousbie:2011b}. Other representatives are Spineweb \citep{Aragon:2010b} and Felix \citep{Shivashankar:2016}. Possible biases that may affect the various means of filament detection have been investigated in a recent study by \cite{Dhawalikar:2024}. 

In addition to the geometric and topological formalisms, several alternative methods have played a substantial role in the study of the cosmic web. Bisous is a well-known stochastic formalism, involving Bayesian exploration based on stochastic geometric modeling of filaments \citep{Tempel:2014}. It forms the basis for the widely used filament catalogue extracted from the SDSS survey \citep{Tempel:2014}. More recent developments often incorporate machine learning codes \citep[see \textit{e.g.}][]{Awad:2022}. Perhaps the oldest representatives for a systematic analysis of filamentary patterns are graph-based methods. The Minimal Spanning Tree (MST) is a prime example and is figuring prominently in the cosmic web analysis of the GAMA survey \citep{Alpaslan:2014}. There is even a class of cosmic web identifiers that exploit the resemblance of the cosmic web to biological branching networks. The Monte Carlo Physarum Machine, inspired by the growth of Physarum polycephalum 'slime mold', has been successfully applied to the structural analysis of the cosmic web in both simulations and observations \citep{Elek:2021a,Elek:2021b,Wilde:2023}. 

Possibly the most profound techniques for classification of the cosmic web are those emanating from the analysis of the 6D phase-space structure of the cosmic mass distribution \citep{Shandarin:2011,Shandarin:2012,Neyrinck:2012,Abel:2012}. Restricted to situations in which the initial conditions are known, they yield the identification of the matter streams constituting the migration of mass in the buildup of structure. It allows the definition of an objective physical criterion for what constitutes the various structural elements of the cosmic web. The present study pursues the ideal of understanding the cosmic web in terms of the gravitational physics underlying its formation and development by elaborating on the phase-space evolution of the cosmic matter distribution and inferring the implied emergence, branching, and connectivity of the cosmic web and its structural components.

\subsection{Tides, Deformation and the Caustic Skeleton}
The cosmic web represents a key phase in the dynamical buildup of structure in the Universe. It emerges when the original long phase of linear evolution of the primordial density and velocity field turns into a more advanced non-linear stage involving contraction and collapse of mass inhomogeneities. As such, it marks the transition from the primordial (Gaussian) random field to highly non-linear structures that have fully collapsed into halos and galaxies. The formation and evolution of the characteristic anisotropic structures, \textit{i.e.}, the filaments and walls, are the result of anisotropic deformations. It is the anisotropy of the force field and the resulting deformation of the matter distribution which are at the heart of the emergence of the web-like structure of the mildly non-linear mass distribution, which has been recognized and accurately described in the mildly non-linear stage by the Zel'dovich formalism \citep{Zeldovich:1970}. Hence, by implication, the gravitational tidal force field, induced by the inhomogeneous mass distribution, is the key to modeling and understanding the structure of the cosmic web.

The seminal role of gravitational tidal force fields in shaping the anisotropic wall-like and filamentary structures in the cosmic web has been recognized for many years \citep{Zeldovich:1970,Weygaert:1996,Bond:1996,Hahn:2007,Weygaert:2008,Paranjape:2018,Paranjape:2021}. In recent years the dominant position of the tidal force field in laying out the spatial structure of the cosmic web has led to the formulation of an analytical model for the fully non-linear evolution of the cosmic web, the \textit{caustic skeleton} \citep{Arnold:1982b,Hidding:2014,Feldbrugge:2018,Feldbrugge:2023a,Feldbrugge:2023b}. It is based on the realization that the evolution of the cosmic web can be understood in detail in terms of the singularities and caustics that are arising in the matter distribution as a result of the structure of the corresponding multi-stream flow field. The geometric pattern of the large-scale cosmic matter distribution is to be understood in terms of the folding of the dark matter sheet in phase-space \citep{Shandarin:2011,Shandarin:2012,Neyrinck:2012,Abel:2012}. A hierarchy of non-linear structures emerges after shell crossing in the corresponding multi-stream regions. The mathematical framework of catastrophe theory \citep{Thom:1972,Zeeman:1977,Poston:1978,Arnold:1984} defines and classifies the stable folding configurations, whose geometrically assemble into the connecting framework of the caustic skeleton \citep{Arnold:1982b,Hidding:2014,Feldbrugge:2018}.  The resulting phase-space based \textit{caustic skeleton} description of the evolving web-like pattern in the cosmic matter distribution is codified in terms of a complete set of caustic conditions. The study by \cite{Feldbrugge:2018} demonstrated that a full understanding of the cosmic web structure is obtained through the spatial characteristics of the \textit{eigenvalue} and \textit{eigenvector} fields of the cosmic tidal force field. Underlining this is the realization that the embryonic outline of the cosmic web, in particular its filamentary network, can already be seen in the primordial tidal eigenvalue field \citep{Feldbrugge:2023b, Wilding:2022} (see figures \ref{fig:Caustic_Skeleton} and \ref{fig:eigenvalues}).

In the current study, we focus on the nature of filaments within the caustic skeleton framework. Of key importance is the observation that it implies that the proper dynamical identity of filaments is dictated by the distribution and connectivity of the relevant singularities in the tidal eigenvalue field. Instrumental is the realization that this concerns the complex phase-space singularities that are inferred from the Lagrangian description of the evolving mass distribution \citep{Feldbrugge:2018,Feldbrugge:2023a}.

As such it differs fundamentally from studies -- and identification methods -- that base themselves on the visual impression of filaments as bridges between maxima in the mass distribution \citep[see e.g][]{Colberg:2005}, identifying filaments with the saddle points and corresponding integral lines in the corresponding cosmic density field \citep{Pogosyan:2009a,Pogosyan:2009b,Sousbie:2011a,Sousbie:2011b,Codis:2018b}. However, the implicit identification of morphological and structural features on the basis of density contour levels is probably based on an unwarranted assumption \citep[see \textit{e.g.}][figure 13]{Cautun:2014}: filaments span no less than five orders of magnitude in density, ranging from the massive prominent filamentary extensions of clusters \citep{Kuchner:2020,Kuchner:2021} to the tenuous and small intravoid tendrils \citep{Park:2009,Shivashankar:2016,Jaber:2024}. This renders any density threshold-based identification unrepresentative. This may also concern more sophisticated topological identifications on the basis of density field filtrations \citep{Sousbie:2011b}, which was demonstrated by the study of \cite{Shivashankar:2016} and which may substantially affect the detection of filaments in the galaxy distribution \citep[see \textit{e.g.}][]{Kuchner:2020,Kuchner:2021}.

Of even more fundamental importance is the issue of the nature of the filamentary network, and the question of whether filaments are exclusively \textit{Bridges} or whether there is a substantial population of \textit{Branches} that fork off from the main arteries, gradually fading into lower density areas without revealing any connection to a higher density tip. In particular, in low-density void regions we tend to find a system of tenuous tendrils, emanating from major filamentary arteries at the boundary of voids, which bear a striking resemblance to a network of branches \citep[see \textit{e.g.}][]{Rieder:2013,Cautun:2014,Alpaslan:2014,Shivashankar:2016,Jaber:2024,Aragon:2024}. This may actually be a direct manifestation of the multiscale nature of the filamentary web and its hierarchical evolution. The migration of mass towards high-density regions along the small-scale filaments culminated in the aggregation of mass in surrounding large-scale filaments and the evacuation of mass from the low-density tips of the small-scale filaments, turning these into tendrils. The direct implication is that we do need to take care that any identification of filaments takes into account the fact that filaments are not necessarily \textit{Bridges}, but may also be \textit{Branches}. 

Based on the caustic skeleton formalism, we seek to identify the objective and dynamical identity of filaments in the cosmic matter distribution. In previous studies, we found that filaments can be identified -- in the three-dimensional context -- with two distinct singular features, the $A_4$ \textit{swallowtail} singularity, and the $D_4$ \textit{elliptic/hyperbolic} umbilic. Visual inspection of $N$-body simulations reveals that the latter are the high-density filamentary extensions from the cluster nodes in the cosmic matter distribution. In the present study, we elaborate on this by addressing the question of the precise nature of the identity and outline of the filamentary network of the cosmic web. The first aspect is to establish the formal criteria of the filamentary spine of the cosmic web, on the basis of the related caustic singularities and their connections. Subsequently, we study the gravitationally contracted matter distribution around the filamentary spine, in Eulerian space, on the basis of the recently developed non-linear constrained formalism \citep{Feldbrugge:2023b}. 

\subsection{Filamentary explorations}
The ultimate intention of the presented formalism is to establish a physically based definition of filaments that allows us
(1) to explore in a targeted laboratory approach the properties of filaments as a function of different
primordial conditions, different spatial scales and at different cosmic epochs, (2) to identify the rich cosmological
information contained in the intricate and complex filamentary network of the cosmic web and (3) to investigate the formation, evolution and properties of dark halos and galaxies as a function of filamentary environment.

With respect to the latter aspect, the present study provides a physically based template for systematic studies of halo and galaxy formation as a function of large scale environment. In this sense, it forms an an elaboration and extension of earlier systematic studies, ranging from the study of disk galaxy orientation in and around a zoom-in simulation of a cosmic filament \citep{Hahn:2010}, of (galaxy) halos forming in a large cosmic void \citep{Rieder:2013} to explicit and systematic techniques involving the splicing and tuning of large-scale and small-scale fluctuations in the primordial Gaussian density field. Following up on a similar ideas employed in the peak-patch formalism of
\cite{BondMyers:1996}, the principal example of this is the MIP ensemble simulation formalism forwarded by \cite{Aragon:2016} and
its elaborations \citep[e.g.][]{Cadiou:2021}. 

In the present study, we limit ourselves to the two-dimensional situation, in order to establish the key characteristics of the phase-space-based filamentary network in the cosmic web. The considerably more complex three-dimensional situation, involving a larger number of caustic singularities, is the subject of the sequel to this study. 

\subsection{Outline}
Towards this end, the present paper on the nature of filaments is organized as follows. In section~\ref{sec:caustic_skeleton} we summarize the essential elements of the theory of the caustic skeleton and its definition in terms of the deformation tensor eigenvalue and eigenvector field. In section~\ref{sec:scalespace} we discuss the initial conditions for structure formation and consider the caustic skeleton in scale space to focus attention on structures at a specific length scale. In section~\ref{sec:filament} we identify the location of the ridge and center of filaments. Subsequently, in section~\ref{sec:filamentstat}, we assess the statistical properties of the filament population focusing on the filament formation time and number density. This leads to the treatment in section~\ref{sec:cosmicweb} of the relation and connectivity of filaments within the context of the cosmic web. In section~\ref{sec:alternatives}, we compare the caustic definition of cosmic filaments with different suggestions for the identity of filaments using $N$-body simulations. Finally, in section~\ref{sec:3d} we outline the generalization of the current two-dimensional analysis to the full three-dimensional situation.

\section{The caustic skeleton}
\label{sec:caustic_skeleton}
The Caustic Skeleton is an analytical model that describes the emergence and fully non-linear buildup of the cosmic web. The formalism is based on the assessment of the evolving cosmic mass distribution in 6D phase-space. It focuses on the folding of the dark matter sheet in phase-space and the emergence and formation of multi-stream regions as the mass distribution evolves from its pristine Gaussian configuration towards an ever more complex non-linear arrangement of mass. For a detailed exposition of caustic skeleton theory and its mathematical nuances see \cite{Arnold:1982a, Arnold:1982b, Arnold:1984, Hidding:2014, Feldbrugge:2018, Feldbrugge:2023a, Feldbrugge:2023b}.

In the current study we consider a heuristic two-dimensional model of large-scale structure formation in the context of the Caustic Skeleton model, in order to elucidate the mathematical structure of filaments in the cosmic web. We consider an Einstein-deSitter (EdS) cosmology. In an EdS Universe, the growing mode of (linear) structure growth is proportional to the scale factor,
\begin{equation}
b_+(t)\,\propto\,a(t)\,,
\end{equation}
so that it will suffice to parametrize structure formation in terms of the growing mode, ranging from $b_+(0)=0$ to $b_+(t_0)=1$ at the current epoch. 

We start our investigation by discussing the theoretical concepts that underpin the caustic skeleton model. Fundamental is the description of the cosmic mass distribution in a Lagrangian context, formulated in terms of Lagrangian fluid dynamics. The folding of phase-space sheets leads to families of Lagrangian singularities, whose identity is fully determined by the eigenvalues and eigenvectors of the deformation tensor \citep{Feldbrugge:2018,Feldbrugge:2023b}. The expressions for the specific two-dimensional context of the present study are specified in a separate subsection.  Within the context of cosmic structure formation, the first-order Lagrangian scheme is of fundamental importance, the Zel'dovich approximation \citep{Zeldovich:1970}. It plays a central role in the work described in this study and allows a fully analytical treatment of the Lagrangian singularities. The description of these also provides the notation used in the present study.

\subsection{Lagrangian fluid dynamics}
\label{sec:lagrangian}
The caustic skeleton describes the cosmic web in terms of the stable singularities of the dark matter sheet in Lagrangian fluid dynamics. Lagrangian fluid dynamics describes the evolution of the cold dark matter fluid in terms of the Lagrangian map $\bm{x}_t:L\to E$, mapping initial positions in Lagrangian space $L$ to final positions in Eulerian space $E$ at time $t$, \textit{i.e.}, a mass element starting at the initial position $\bm{q}\in L$ moves in time $t$ to the position
\begin{align}
    \bm{x}_t(\bm{q}) = \bm{q} + \bm{s}_t(\bm{q})\,,
\end{align}
with he displacement map $\bm{s}_t$ describing its relative motion. The associated density field is a derived property of the fluid following from the conservation of mass of the mass elements,
\begin{align}
    \rho(\bm{x}) 
    &= \sum_{\bm{q} \in \bm{x}_t^{-1}(\bm{x})} \frac{\bar{\rho}}{|\det (\nabla \bm{x}_t(\bm{q}))|}\nonumber\\
    &= \sum_{\bm{q} \in \bm{x}_t^{-1}(\bm{x})} \frac{\bar{\rho}}{|\det(I + \nabla \bm{s}_t(\bm{q}))|}\,,
    \label{eq:jacbdens}
\end{align}
where $\det (\nabla \bm{x}_t(\bm{q}))$ is the relative signed volume of a mass element at time $t$, with the mean density $\bar{\rho}$, and the preimage 
\begin{align}
    \bm{x}_t^{-1}(\bm{x}) =\{ \bm{q} \in L\,|\, \bm{x}_t(\bm{q}) = \bm{x}\}\,,
\end{align}
consisting of the Lagrangian points reaching $\bm{x}$ in time $t$. The number of streams $n$ (defined as the number of elements in the preimage $\bm{x}_t^{-1}(\bm{x})$) partitions Eulerian space into $n$-stream regions\footnote{The number of streams is an odd natural number by the topology of folding.}. At the boundaries of a multi-stream region, the deformation tensor $\nabla \bm{x}_t$ is singular and the density spikes to infinity. This spike marks the onset of the non-linear gravitational evolution of the mass element and is known as the phenomenon of shell-crossing. More formally, this is known as the \textit{fold caustic}. Subsequently, we see the emergence of a family of ever more complex and intricate \textit{Lagrangian singularities}. Their definition, identification with structural features of the cosmic web and mathematical expressions form the core of the Caustic Skeleton model \citep{Feldbrugge:2018}.   

\subsection{Lagrangian singularities}
\label{sec:lagrangian_singularities}
The formal definition and expressions for the Lagrangian singularities can be fully specified in terms of the eigenvalue and eigenvector fields of the deformation tensor $\nabla \bm{s}_t$. The eigenvalue and eigenvector fields of the deformation tensor $\nabla \bm{s}_t$ are defined by the eigen equation
\begin{align}
    [\nabla \bm{s}_t(\bm{q})] \bm{v}_{t,i}(\bm{q}) = \mu_{t,i}(\bm{q})\bm{v}_{t,i}(\bm{q})\,,
\end{align}
with the ordering $\mu_{t,1}(\bm{q}) \leq \mu_{t,2}(\bm{q})$. On the basis of the eigenvalues and eigenvectors we can identify two families of singularities, the $A$ singularities that depend on only one eigenvalue, and the umbilic D singularities that are defined by two eigenvalues. For the two-dimensional situation, we deal with two eigenvalues $\mu_1$ and $\mu_2$, and identify three A singularities -- the fold caustic $A_2$, the cusp caustic $A_3$ and the swallowtail caustic $A_4$ -- and one $D$ singularity, the umbilic $D_4$ singularity. 

\bigskip
The \textit{fold caustic} $A_2$ in Lagrangian space corresponding to the eigenvalue field $\mu_{t,i}$ (with $i=1,2$) consists of the manifold 
\begin{align}
    A_2^i = \{ \bm{q} \in L \,|\, 1+ \mu_{t_0,i}(\bm{q}) = 0\}\,,
\end{align}
with $t_0$ the current time. The density spikes at the fold manifold. Conversely, the fold manifold bounds the mass elements having undergone shell-crossing along the $\bm{v}_{t_0,i}$ direction. 

\medskip
Interestingly, the structural complexity and classification of caustics do not stop at the fold caustic. Caustic theory provides a refined description of the intricate folding of the dark matter sheet in phase-space \citep{Arnold:1982b,Zeeman:1977,Thom:1972}. The next-order caustic in this hierarchy of singularity is the \textit{cusp caustic} $A_3$,
\begin{align}
    A_3^i = \{ \bm{q} \in L \, |\,& 1+ \mu_{t,i}(\bm{q}) = 0\text{ and } \bm{v}_{t,i}(\bm{q})\cdot \nabla \mu_{t,i}(\bm{q})=0\nonumber\\
    & \text{ for some } t \leq t_0\}\,.
\end{align}
In addition to $\nabla \bm{x}_t$ being singular due to the condition $1+\mu_{t,i} = 0$, the derivative of the eigenvalue field $\mu_{t,i}$ in the direction of the corresponding eigenvector $\bm{v}_{i,t}$ vanishes. The cusp manifold marks the mass elements that form a kink in the boundary between the different $n$-stream regions. Within the context of the formation of the cosmic web, (in the 2D context) it corresponds to filaments.

\medskip
In addition to the cusp caustic, a two-dimensional Lagrangian fluid generically also contains the higher order $A_4$ \textit{swallowtail caustics}.  The swallowtail caustic consists of the points for which 
\begin{align}
    A_4^i = \{ \bm{q} \in L \, |\,& 1+ \mu_{t,i}(\bm{q}) = 0\,, \bm{v}_{t,i}(\bm{q})\cdot \nabla \mu_{t,i}(\bm{q})=0\,,\nonumber\\
    & \text{ and } 
    \bm{v}_{t,i}(\bm{q}) \cdot \nabla (\bm{v}_{t,i}(\bm{q})\cdot \nabla \mu_{t,i}(\bm{q}))=0 \nonumber\\
    &\text{ for some } t \leq t_0\}\,.
\end{align}

\begin{figure*}
    \centering
    \begin{subfigure}[b]{0.49\textwidth}
        \includegraphics[width=\textwidth]{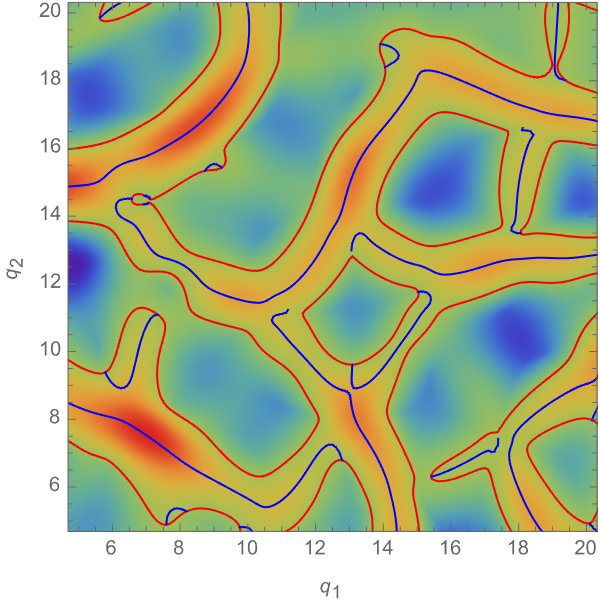}
        \caption{$\lambda_1$}
    \end{subfigure}
    ~ 
    \begin{subfigure}[b]{0.49\textwidth}
        \includegraphics[width=\textwidth]{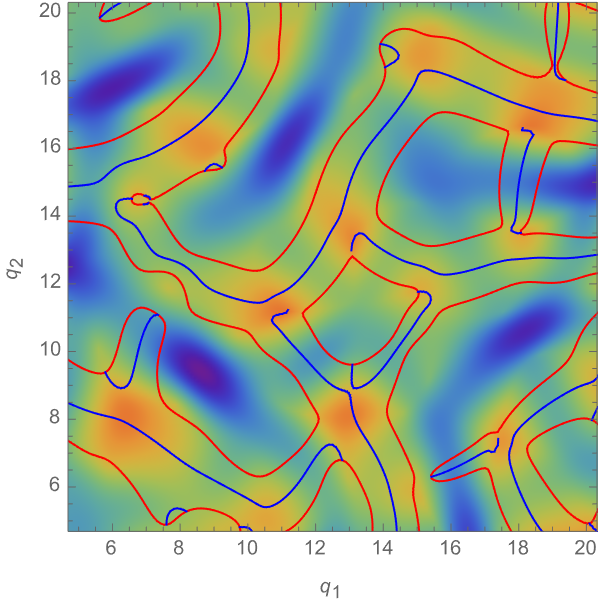}
        \caption{$\lambda_2$}
    \end{subfigure}\\
    \begin{subfigure}[b]{0.49\textwidth}
        \includegraphics[width=\textwidth]{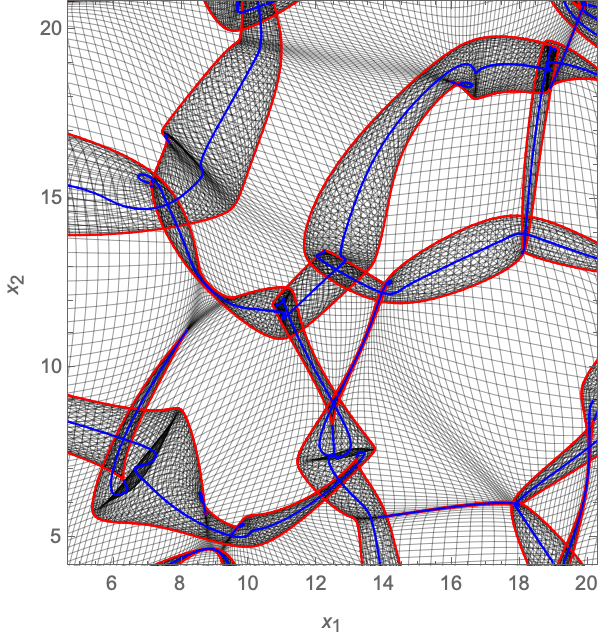}
        \caption{Zel'dovich approximation}
    \end{subfigure}
    ~
    \begin{subfigure}[b]{0.49\textwidth}
        \includegraphics[width=\textwidth]{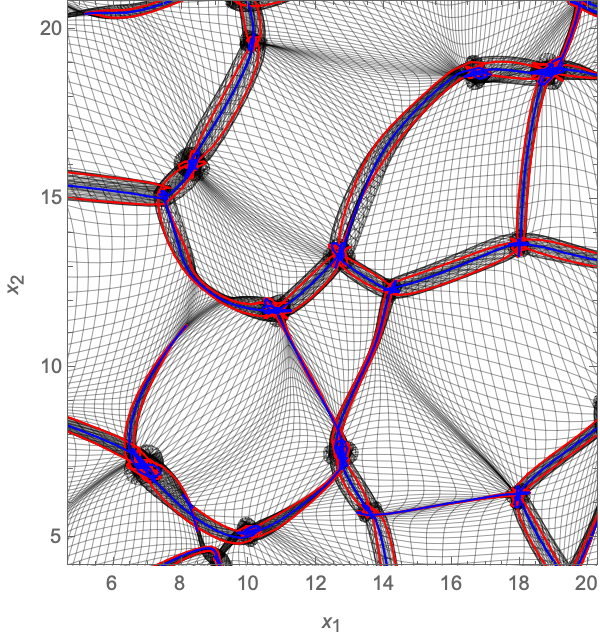}
        \caption{$N$-body simulation}
    \end{subfigure}
    \caption{The caustic skeleton in Lagrangian (upper panels) and Eulerian space (lower panels). The upper left and upper right panels show the Lagrangian skeleton in relation to the first ($\lambda_1$) and second eigenvalue field ($\lambda_2$). The lower left and right panel show the skeleton in Eulerian space, based on the Zel'dovich approximation and an $N$-body simulation.}\label{fig:Caustic_Skeleton}
\end{figure*}
\medskip
Finally, the \textit{umbilic caustic} $D_4$ is defined by the points for which the two eigenvalue fields coincide
\begin{align}
    D_4 = \{ \bm{q} \in L \, |\,& 1+ \mu_{t,1}(\bm{q}) = 0\text{ and }  1+ \mu_{t,2}(\bm{q}) = 0\,. \nonumber\\
    &\text{ for some } t \leq t_0\}\,,
\end{align}
Around the umbilic caustic, the density spikes more strongly. This follows from the fact that the matrix $\nabla \bm{x}_t$ is singular along two directions.

\medskip
\begin{table}
\center
  \begin{tabular}{lll}
    \hline
    \mbox{Singularity} & \mbox{Singularity} & \mbox{Feature in the } \\
  \mbox{class} & \mbox{name} & \mbox{2D cosmic web} \\ \hline 	
$A_2$		&\mbox{fold} & \mbox{collapsed region}  \\ 
$A_3$          &\mbox{cusp} & \mbox{filament} \\ 
$A_4$          & \mbox{swallowtail}& \mbox{cluster or knot}  \\ 
 &  \ \\
$D_4$          &\mbox{hyperbolic/elliptic}& \mbox{cluster or knot} \\ 
\hline
\end{tabular}
  \caption{The identification of caustic singularities with cosmic web components, for (for the $2$-dimensional situation)}\label{table:caustic_web}
\label{table:caustic_web}
\end{table}
\medskip

\subsection{Caustic skeleton \& Cosmic Web}
The embryonic caustic skeleton in Lagrangian space can be mapped onto Eulerian space by means of the Lagrangian map. The resulting varieties $\bm{x}_{t_0}(A_2^i),$ $\bm{x}_{t_0}(A_3^i),$ $\bm{x}_{t_0}(A_4^i),$ and $\bm{x}_{t_0}(D_4)$ delineate the various structural components of the cosmic web. Table~\ref{table:caustic_web} lists the identification of the caustic singularities with the well-known structural features of the cosmic web. In other words, the Lagrangian maps capture the geometry of the cosmic web, and their evolution their formation history. 

\subsection{The Zel'dovich approximation}
The focus of the present study is on the exploration and presentation of the mathematical structure of filaments in the cosmic web. To this end, we center the analysis on a fully analytical heuristic model of cosmic structure formation. To optimize transparency and insight into the implied structure and dynamics of filaments, we here explore the two-dimensional situation and relegate the full three-dimensional situation to the follow-up paper. 

The Zel'dovich approximation provides a fully analytical description of the displacement of mass elements and hence entails a fully analytical expression for the deformation tensor. Within the context of the caustic skeleton formalism, it therefore allows a fully analytical treatment. The Zel'dovich approximation describes the evolution of mass elements in terms of their ballistic motion, in which time evolution and spatial structure are separated. The mass elements follow linear trajectories determined by the primordial gravitational potential. This follows from the first-order approximation of the displacement $\bm{s}_t(\bm{q})$ of the mass element with Lagrangian coordinate $\bm{q}$, yielding  
\begin{align}
    \bm{s}_t(\bm{q}) = - b_+(t) \nabla \Psi(\bm{q})\,. 
\end{align}
The time evolution is represented by the growing mode $b_+(t)$, while the spatial structure is encapsulated by the displacement potential $\Psi$, expressed in terms of the current total energy density and the linearly extrapolated gravitational potential $\phi_0$, 
\begin{align}
  \Psi(\bm{q}) \,=\,\frac{2}{3\Omega_0 H_0^2} \phi_0(\bm{q})\,,
\end{align}
contains the geometric information of the evolution of the fluid. 

The Zel'dovich approximation accurately describes single-stream regions from early times up to surprisingly advanced phases of quasi-linear structure formation \citep{Zeldovich:1970,Shandarin:1989,Hidding:2014}. For our purpose, it represents an ideal context, given the focus on the first shell-crossing events in the caustic skeleton formalism. It fails at late times in multi-stream regions when the gravitational backreaction of the mass elements becomes important. This follows directly from the observation that the Zel'dovich approximation describes a ballistic motion, and hence unable to capture gravitational interactions at later stages of structure formation. In particular of relevance in the present context of phase-space evolution, is that the Zel'dovich approximation does not capture the turnaround typically occurring during phase mixing. 

\bigskip
For conciseness, we define the eigenvalues $\lambda_i(\bm{q})$ of the Hessian of the displacement potential $\Psi(\bm{q})$
\begin{align}
    [\mathcal{H} \Psi ] \bm{v}_i = \lambda_i \bm{v}_i\,,
\end{align}
with the ordering $\lambda_1(\bm{q}) \geq \lambda_2(\bm{q})$. The corresponding (non-linear) density field in the Zel'dovich approximation assumes the form (see eq.~\eqref{eq:jacbdens}),

\begin{align}
    \rho(\bm{x}) 
    = \sum_{\bm{q} \in \bm{x}_t^{-1}(\bm{x})} \frac{\bar{\rho}}{|1-b_+(t) \lambda_1(\bm{q})||1-b_+(t) \lambda_2(\bm{q})|}\,.
\end{align}

\bigskip
\noindent The \textit{caustic conditions}, presented in section~\ref{sec:lagrangian_singularities}, in the Zel'dovich approximation assume the following form for the caustic manifolds corresponding to the first eigenvalue field (in which we have dropped the eigenvalue subscript),

\bigskip
\noindent\rule[0.5ex]{\linewidth}{1pt}
\begin{align}
    A_2 = \{ \bm{q} \,|\,  \lambda_1(\bm{q}) = 1/b_+(t_0)\}\,,
\end{align} 
\begin{align}
    A_3 = \{ \bm{q} \, |\, \lambda_1(\bm{q}) \geq 1/b_+(t_0)\,, \bm{v}_{1}(\bm{q})\cdot \nabla \lambda_{1}(\bm{q})=0\}\,,
\end{align} 
\begin{align}
    A_4 &= \{ \bm{q} \,|\, \lambda_1(\bm{q}) \geq 1/b_+(t_0)\,, \bm{v}_{1}(\bm{q})\cdot \nabla \lambda_{1}(\bm{q})=0\,,\nonumber\\ 
    &\phantom{=\{\bm{q}\,|\,\,}\bm{v}_1(\bm{q}) \cdot \nabla(\bm{v}_{1}(\bm{q})\cdot \nabla \lambda_{1}(\bm{q}))=0\}\,,
\end{align} 
\begin{align}
    D_4 = \{ \bm{q}\, |\, \lambda_{1}(\bm{q}) = \lambda_{2}(\bm{q}) \geq 1/b_+(t_0)\}\,.
\end{align}
\noindent\rule[0.5ex]{\linewidth}{1pt}
\bigskip

An illustration of a realization of the caustic skeleton of the two-dimensional cosmic web is shown in figure~\ref{fig:Caustic_Skeleton}. The two upper panels show the outline of the caustic skeleton in Lagrangian space. The two lower panels show the skeleton in Eulerian space, both for the Zel'dovich approximation as well as for the fully non-linear outline inferred from the corresponding $N$-body simulation. Each of the various caustic features is indicated using different colours, red for the fold caustics $A_2$, blue for cusp caustics $A_3$ that delineate the filamentary spine of the cosmic web. 

\section{the Multiscale Skeleton}\label{sec:scalespace}
The cosmic web has evolved gravitationally from a field of primordial density, gravity, and potential perturbations. In the present study, we assume the primordial density fluctuations to be a realization of a statistically homogeneous and isotropic stationary Gaussian random field. 

As a result, the emerging cosmic web is an intricate multiscale structure, with structure and features over a vast range of scales. In order to focus attention on features around a specific length scale, we invoke the concept of the \textit{Scale Space} of the Caustic Skeleton. 

\subsection{Primordial Density and Potential field}
A Gaussian random field is fully specified by its second-order moment. The power spectrum fully quantifies the statistical properties of the Fourier modes $\hat{\delta}(\bm{k})$ of the density field,
\begin{align}
    \hat{\delta}(\bm{k}) = \int \delta(\bm{q})e^{i\bm{k}\cdot \bm{x}}\mathrm{d}\bm{q}\,,
\end{align}
in terms of their covariance, 
\begin{align}
    \langle \hat{\delta}(\bm{k}_1) \hat{\delta}^*(\bm{k}_2)\rangle = (2 \pi)^2 \delta_D^{(2)}(\bm{k}_1-\bm{k}_2) P_{\delta}(\|\bm{k}_1\|)\,,
\end{align}
in which $\delta_D^{(2)}$ is the two-dimensional Dirac delta function. 

For the purpose of the present study, we use a heuristic power spectrum that consists of a scale-free power spectrum in combination with a Gaussian cutoff at scale $R_s$ at high frequencies,
\begin{align}
    P_\delta(k) = \mathcal{N_\delta} k^{n} e^{-R_s^2 k^2}\,.
    \label{eq:powerpk}
\end{align}
The scale-free part of the spectrum is characterized by the spectral index $n$, the cutoff part by the Gaussian smoothing scale $R_s$, while $\mathcal{N_\delta}$ is the normalization factor. Notice that the the Gaussian spectral cutoff is meant to represent an intrinsic physical spectral aspect, not unlike that seen in warm dark matter or neutrino-dominated cosmologies and to some extent not unlike the more gradual high-frequency downturn seen in ($\Lambda$)CDM cosmologies. 

By implication, also the displacement potential $\Psi$ of the Zel'dovich approximation is a Gaussian random field. Given the Poisson equation,
\begin{equation}
  \nabla^2 \Psi \propto \delta\,,
\end{equation}
it follows that the power spectrum
\begin{align}
  P_\Psi(k) = \mathcal{N}_\Psi k^{n-4}e^{-R_s^2 k^2}\,.
  \label{eq:power}
\end{align}

In the present study, we take the primordial gravitational potential and displacement potential to follow the Harrison-Zel'dovich scale-invariant spectrum on large scales \citep{Harrison:1970, Zeldovich:1972, Mukhanov:1981, Guth:1982, Starobinsky:1982, Bardeen:1983}, for which the spectral index $n=2$. On small scales, we invoke a cutoff filter scale $R_s$ to smooth the fields. In our calculations, we normalize the power spectrum on the basis of the amplitude of the density perturbations, 
\begin{equation}
  \langle \delta^2(\bm{q})\rangle = \alpha^2\,,
\end{equation}
which sets the normalization $\mathcal{N_\delta}$ to
\begin{align}
    \mathcal{N_\delta} =  \frac{\alpha^2 4 \pi R_s^{n+2}}{\Gamma \left(1+\frac{n}{2}\right)}\,,
\end{align}
with the gamma function $\Gamma$. For our calculations, we set the density field amplitude to $\alpha^2=10$. While this sets the time scale of the simulations, the employed cutoff length scale $R_s=1$ sets the spatial scale of the simulations. 

\begin{figure*}
    \centering
    \begin{subfigure}[b]{0.49\textwidth}
        \includegraphics[width=\textwidth]{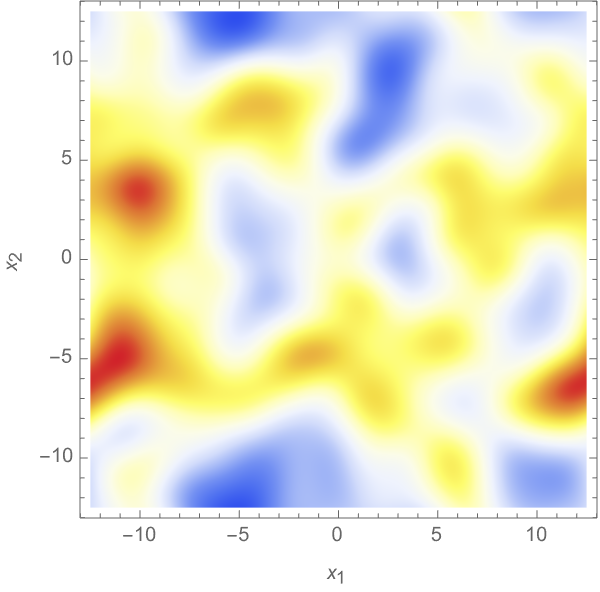}
        \caption{$\Psi$}
    \end{subfigure}
    \hfill
    \begin{subfigure}[b]{0.49\textwidth}
        \includegraphics[width=\textwidth]{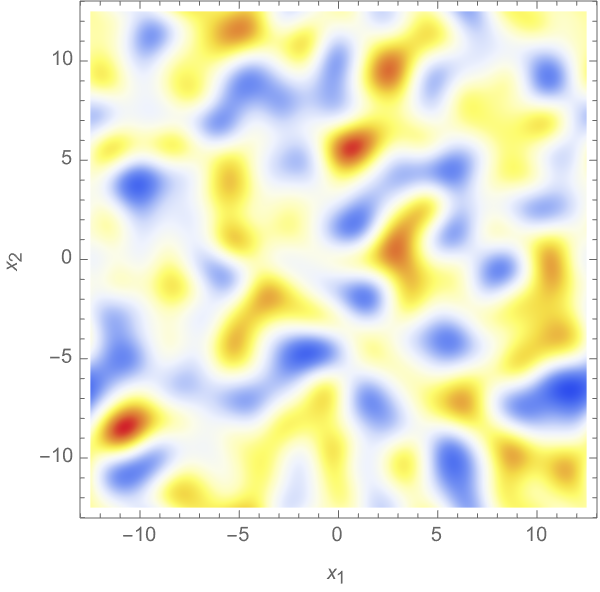}
        \caption{$\delta$}
    \end{subfigure}\\
    \begin{subfigure}[b]{0.49\textwidth}
        \includegraphics[width=\textwidth]{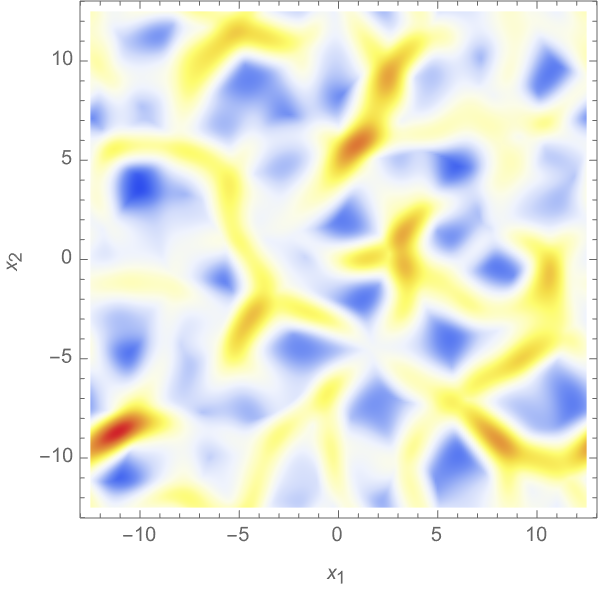}
        \caption{$\lambda_1$}
    \end{subfigure}
    \hfill
    \begin{subfigure}[b]{0.49\textwidth}
        \includegraphics[width=\textwidth]{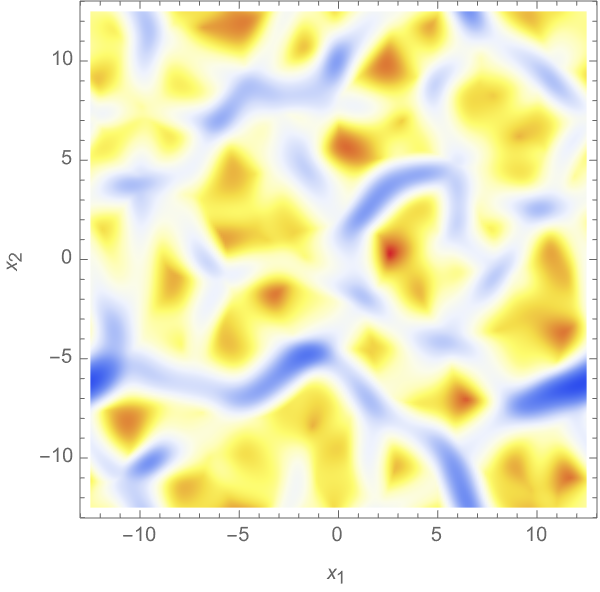}
        \caption{$\lambda_2$}
    \end{subfigure}
    \caption{The displacement potential (upper left), with the associated primordial density perturbation (upper right), and associated eigenvalue fields (lower left and lower right). The primordial eigenvalue fields display the progenitors of the late-time cosmic web.} \label{fig:eigenvalues}
\end{figure*}

\subsection{Scale Space Cosmic Web}
While the filamentary structure of the multiscale cosmic web is that of a complex and intricate aggregate of elongated bridges and branches, in the present study we chose to zoom in on features of a specific length scale $\sigma$, and the implied corresponding mass scale.

To direct attention to features on a scale $\sigma$, we chose to follow the strategy of studying the formation of the caustic skeleton from correspondingly smoothed primordial conditions. A similar strategy has been followed in various studies on the multiscale evolution and structure of the cosmic web by \citep[e.g.][]{Aragon:2010a,Aragon:2013,Jaber:2024,Aragon:2024}. While the non-linear gravitational evolution of cosmic structure generates its structure over a spectrum of scales, the cosmic web finds itself in the transition from linear to non-linear scales and as such retains a considerable level of memory of its primordial spatial structure. This is in particular so for the spatial pattern of the tidal/deformation field that outlines the cosmic web \citep[see][]{Feldbrugge:2023b}. 

\bigskip
To accomplish this \textit{Scale Space} approach, we smooth the displacement field on a Gaussian scale $\sigma$. In a sense, this linearly takes into account the nonlocal influences on the emergence of the caustic skeleton: the skeleton is not only sensitive to the displacement at one single point but emerges through the influence of the neighborhood on the mass element. 

To this end, we convolve the displacement potential $\Psi$, 
\begin{align}
    \Psi_\sigma(\bm{q}) =  \int \Psi(\bm{q}-\bm{q}')W_\sigma(\bm{q}')\, \mathrm{d}\bm{q}'\,,
\end{align}
with a Gaussian kernel\footnote{An alternative choice frequently used in late-time cosmology is to invoke a spherical (or rather circular in the two-dimensional context) tophat kernel 
\begin{align}
    W_\sigma(\bm{q}) = \frac{\Theta_H(\|\bm{q}\| - \sigma)}{\pi \sigma^2}\,,
\end{align}
with the Heaviside step function $\Theta_H$. While we do not expect the results to significantly alter for similar alternative smoothing kernels, here we only consider Gaussian smoothing.} on scale $\sigma$,
\begin{align}
    W_\sigma (\bm{q}) = \frac{1}{2 \pi \sigma^2} e^{- \frac{\bm{q}^2}{2\sigma^2}}\,.\label{eq:kernel}
\end{align}
The resulting scale-space power spectrum of the smoothed displacement potential $\Psi_\sigma$ assumes the form
\begin{align}
    P_{\Psi_\sigma}(k) = \mathcal{N} k^{n-4} e^{-(R_s^2+\sigma^2)k^2}\,.
\end{align}
We emphasize the difference between the smoothing scales $R_s$ and $\sigma$: the spectral smoothing scale $R_s$ is a physically intrinsic aspect of the fluctuation spectrum, while $\sigma$ is a user-imposed smoothing scale meant to study the multiscale structure of the cosmic web. 

\medskip
Within the context of the scale dependence of the filamentary web, and our corresponding sampling schemes, an important role is played by the generalized moments,
\begin{align}
    \sigma_i^2 &= \frac{1}{(2\pi)^2} \int \|\bm{k}\|^{2i} P_{\Psi_\sigma}(\bm{k}) \mathrm{d}\bm{k}\label{eq:moments}\\
    &= \frac{1}{2\pi} \int_0^\infty k^{2i +1} P_{\Psi_\sigma}(k) \mathrm{d}k\nonumber
\end{align} 
\begin{align}
    &=\frac{\alpha^2 R_s^{n+2}}{\left(R_s^2+\sigma ^2\right)^{i+\frac{n}{2}-1}}\frac{\Gamma \left(i+\frac{n}{2}-1\right) }{\Gamma \left(\frac{n}{2}+1\right)}\,.\nonumber
\end{align}
These represent convenient parameterizations of the statistical properties of Gaussian random fields, 

As described below, these generalized moments play an important role in the properties of filaments and our sampling schemes. For the particular case of a 2D Harrison-Zel'dovich primordial spectrum, with $n=2$, we have 
\begin{align}
    \sigma_2 = \frac{\alpha R_s^{2}}{R_s^2+\sigma ^2}\,.
\end{align}

\subsection{Scaling considerations: space vs. time}
A scaling in real space $\bm{x} \mapsto L \bm{x}$ for some $L>0$ leads to a scaling $\bm{k} \mapsto \bm{k}/L$ in Fourier space. Under this transformation, the power spectrum scales as
\begin{align}
    P_{\Psi_\sigma}(k/L) = L^{4-n} \mathcal{N} k^{n-4}e^{- (R_s/L)^2 k^2 }\,.
\end{align}
Rescaling in position space $\bm{x} \mapsto L \bm{x}$ thus corresponds to a Gaussian smoothing with the length scale
\begin{align}
    \sigma = \frac{\sqrt{1 - L^2}}{L}R_s\,,
\end{align}
and a scaling of the amplitude of the Gaussian random field
\begin{align}
    \alpha \mapsto L^{2-\frac{n}{2}}\alpha\,.
\end{align}
The latter can be interpreted as a rescaling of the growing mode, 
\begin{align}
    b_+(t) \mapsto L^{2-\frac{n}{2}}b_+(t)\,.
\end{align}
For these simple power law power spectra, it thus suffices to study the properties of the caustic skeleton at a fixed smoothing scale $\sigma$ as a function of time $t$, or conversely for a fixed time $t$ as a function of the scale $\sigma$. For more general power spectra this degeneracy is generically broken.

\section{Caustic Filaments}\label{sec:filament}
Within the context of the caustic skeleton formalism, the Zel'dovich approximation enables us to infer specific analytical expressions for filaments and relate their configuration directly to the Gaussian primordial initial conditions. Given the fact that the cosmic web is a non-linear complex geometric structure, this is a remarkable observation. It emanates from the realization that the spatial outline of the cosmic web may already be recognized in the primordial Universe when assessing the primordial tidal/deformation eigenvalue field (see fig.\ \ref{fig:eigenvalues} and \textit{e.g.} \cite{Feldbrugge:2023b}).

\begin{figure}
    \centering
    \includegraphics[width=\linewidth]{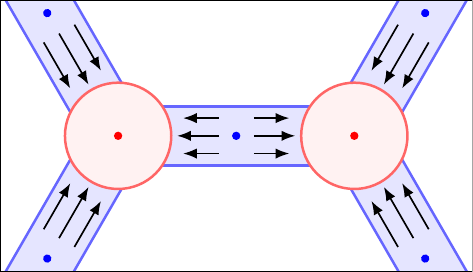}\\
    \vskip 0.2cm
    \includegraphics[width=\linewidth]{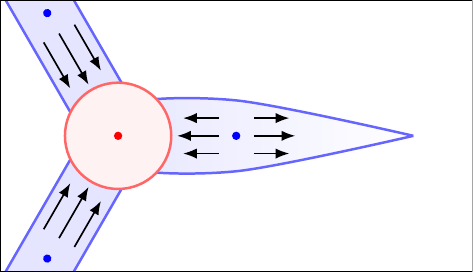}
    \vskip 0.1cm
    \mbox{\hskip -0.15cm\includegraphics[width=1.03\linewidth]{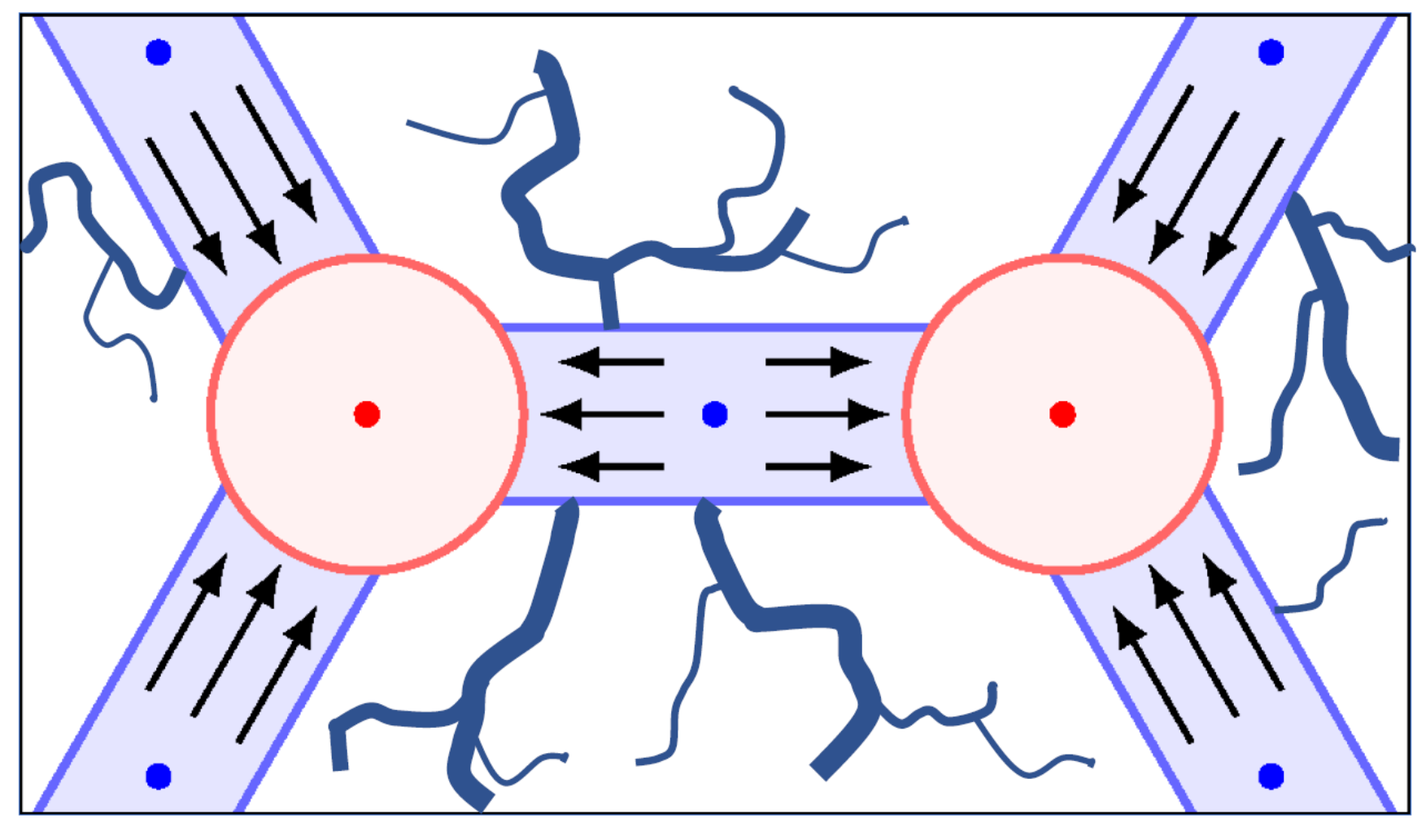}}
    \caption{Top: a sketch of a filament (blue) connecting two clusters (red). Matter flows into the clusters along the filaments (black arrows). The stretching centers of the filaments, where due to the shearing migration flow along the filament mass elements get maximally stretched, are indicated by the blue points. The left panel illustrates a filament between two clusters. Centre: The right filament branching off via cusp caustics. Bottom: multiscale filament network and filament branching.}\label{fig:sketch}
\end{figure}

As a result of gravitationally driven evolution of the primordial mass distribution, the embryonic pattern of the cosmic web is shaped into
a complex multiscale weblike network. It includes major arteries suspended between clusters at the nodes of the network  (top panel, figure~\ref{fig:sketch}), 
arteries tapering off while stretching into low density void regions (central panel, figure~\ref{fig:sketch} and configurations of tendrils that branch out 
from more massive peers (lower panel, figure~\ref{fig:sketch}). 

\subsection{Filament Caustic}
Filaments of the two-dimensional cosmic web correspond to the $A_3$ cusp manifold of the first eigenvalue field $\lambda_1$. In the Zel'dovich approximation, in Lagrangian space, these assume the form 

\bigskip
\noindent\rule[0.5ex]{\linewidth}{1pt}
\begin{align}
  A_3 = \{ \bm{q} \,|\,& \lambda_1(\bm{q}) \geq 1/b_+(t_0)\,, \bm{v}_1(\bm{q}) \cdot \nabla \lambda_1(\bm{q}) = 0\}\,,
  \label{eq:filcaustic}
\end{align} 
with an orientation in Lagrangian space given by the normal 
\begin{align}
    \bm{n} = \nabla (\bm{v}_1(\bm{q})\cdot \nabla \lambda(\bm{q}))\,.
\end{align}
\noindent\rule[0.5ex]{\linewidth}{1pt}
\bigskip

The fact that the caustic skeleton model allows the determination of $A_3$ filaments solely based on the initial eigenvalue and eigenvector field represents a remarkable simplification concerning a fully non-linear gravitational assessment. Amongst others, it facilitates a statistical assessment of filament properties, by its direct relation to the Gaussianity of the primordial field (see section \ref{sec:filamentstat}). 

\bigskip
The caustic equation~\eqref{eq:filcaustic} identifies all Lagrangian points ${\bm{q}}$ that belong to the $A_3$ singularity. Its spatial outline in the evolving cosmic matter distribution is obtained through the Lagrangian map towards Eulerian space, which translates each mass element $\bm{q}_c$ of the Lagrangian singularity to the Eulerian position $\bm{x}_t(\bm{q}_c)$ to which it has migrated at time $t$. A few complications arise in the pure translation of the complete $A_3$ caustic in Lagrangian space to Eulerian space. This makes it necessary to distinguish between the main ridge of the resulting filament, the \textit{trunk} of the filament, and the parts of the caustic that form its connection to the higher-density regions.

Firstly, as the trunks of the filaments stretch while the connecting regions compress during the evolution of the cosmic web, a general point on the cusp manifold in Lagrangian space is biased to end up in a cluster-like region. In other words, a general point on the cusp manifold is not the most representative point on this manifold of mass elements ending up in the filamentary structures at the current epoch.

Secondly, in the practical implementation within the context of the Zel'dovich approximation which we use for the analytical description in the present study, we are not able to follow the development of phase-space and the resulting caustic structure in the high-density contracting region\footnote{Note that more accurate higher order descriptions or the use of $N$-body simulation inferred deformation values would allow a more accurate caustic description in the evolving phase-space in and around high-density regions.}. In these regions, mass elements are part of an ever more complex wrapping phase-space structure \citep[see \textit{e.g.}][]{Colombi:2015}, which the straight ballistic motion of mass elements implied by the Zel'dovich approximation entirely fails to capture.

The combination of both aspects outlined above makes it necessary to augment the straightforward projection of the mass elements along the $A_3$ caustic from Lagrangian to Eulerian space with an additional criterion on the stretching of mass elements along the trunk of the filament. To this end, we introduce a local criterion for the central stretching region of a filament, which is outlined in detail below. 

\subsection{Filament Trunk \& Stretching Center}
\label{sec:stretchcenter}
To infer a complete criterion for the main ridge of filaments within the context of caustic skeleton theory we invoke a straightforward dynamical criterion concerning the effect of the large-scale tidal force field on the mass elements in a filament and the induced migration flows along the filament. To this end, we base ourselves on the realization that filaments are formed and shaped by the large-scale tidal force field.  These induce characteristic shearing migration flows towards and along the emerging filaments.

As a result of the tidally induced migration streams, mass elements in the main ridge get stretched. By contrast, those at the higher density regions experience contraction and find themselves in a region in which the phase-space structure quickly evolves towards an ever more complex configuration. It is the stretching part of the caustic as defining the cosmic filament which we would recognize visually as such: the trunk of the filament. Following the notion of the induced shearing flow along the filaments's ridge, we see a variation of flow velocity along the filament, increasing towards the high-density nodes towards which mass is migrating (see fig.~\ref{fig:sketch}).

On the basis of this, we employ the natural dynamical definition for filaments as the regions along which the matter distribution is being stretched along its spine (see fig.\ \ref{fig:sketch} for a sketch of this configuration). We may note that there are also regions along a filament that are not stretching, such as near, around, and in the cluster nodes at the tips of a filament. Instead, as mass elements approach such high-density nodes, subsequently, pass through them and accrete on them, they may get compressed. Hence, we are principally only interested in the main stretching part of the ridge a filament, the \textit{trunk} of a filament. 

Following this definition, we may identify a characteristic dynamical center of a filament, the \textit{stretching center of a filament} -- or, shortly, the \textit{center of a filament} -- as the \textit{location along the filament where the flow corresponds to a maximum stretching of the spatial mass distribution along the filament's ridge}. Within our study, these filament centers are a key handle for allowing a mathematical analysis of the distribution and statistical properties of filaments. This is particularly true when invoking the Zel'dovich approximation, which allows a fully analytical formalization in terms of implied elementary algebraic conditions. 

\subsection{the Filament Stretching Center: Conditions}
\label{sec:stretchcenter_conditions}
To infer the formal conditions to specify the location of the trunk and stretching center of a filament, we invoke and combine several constraints on the eigenvalue and eigenvector fields:

\subsubsection{Two Eigenvalues}
To develop a definition for the stretching center of a filament, we first note that clusters -- in 2D --  are typically identified with the regions in Lagrangian space that undergo shell-crossing along at least two directions. In the Zel'dovich approximation, this implies that in a cluster region, both eigenvalues $\lambda_1$ and $\lambda_2$ exceed $1/b_+(t_0)$. On the basis of this realization, we may already conclude that a full filament criterion should involve both the first and second eigenvalue fields. 

\subsubsection{Tangent Space and Stretching Condition}
The Taylor series of the Lagrangian map in the Lagrangian point $\bm{q}_0$, 
\begin{align}
    \bm{x}_t(\bm{q}_1) - \bm{x}_t(\bm{q}_0)= \nabla \bm{x}_t(\bm{q}_0)(\bm{q}_1-\bm{q}_0)+\dots
\end{align}
demonstrates that a vector in the tangent space of Lagrangian space $\bm{T} \in T_{\bm{q}}L$ in the point $\bm{q}\in L$ is mapped to the vector $\nabla \bm{x}_t(\bm{q}) \bm{T}$ in the tangent space in Eulerian space $T_{\bm{x}_t(\bm{q})}E$.

In other words, as $\bm{q}_1 \to \bm{q}_0$ the vector
\begin{equation}
  T=\bm{q}_1 - \bm{q}_0
\end{equation}
is mapped to the vector
  $\nabla \bm{x}_t \bm{T}$
in Eulerian space. Restricting the map to the points $\bm{q}_c$ located on the filamentary cusp caustic $A_3$, the space $\bm{T}$ is identified with the unit tangent vector of the cusp manifold. This is identified by the condition
\begin{align}
  \bm{T} \cdot \bm{n} = 0\,,
\end{align}
with the normal of the cusp manifold $\bm{n}$. We will assume a unit tangent vector, \textit{i.e.}, $\|\bm{T}\|=1.$ The key observation is that when the norm $\|[\nabla\bm{x}_t(\bm{q}_c)]\bm{T}\|$ exceeds unity, the filament stretches along the tangent direction, \textit{i.e.}, the filament stretches along its ridge. On the other hand, when the norm is smaller than one, the filamentary region contracts at the time $t$.

On the basis of this relation we are therefore led to the formal definition of the stretching center of a filament: \textit{the stretching center of the filament is the location on the cusp manifold of maximum and enduring dilation.} Analytically, this point of maximum stretch can be inferred analytically within the framework of the Zel'dovich approximation.

\subsubsection{Tangent -- Second Eigenvalue alignment}
\begin{figure}
    \centering
    \includegraphics[width=\linewidth]{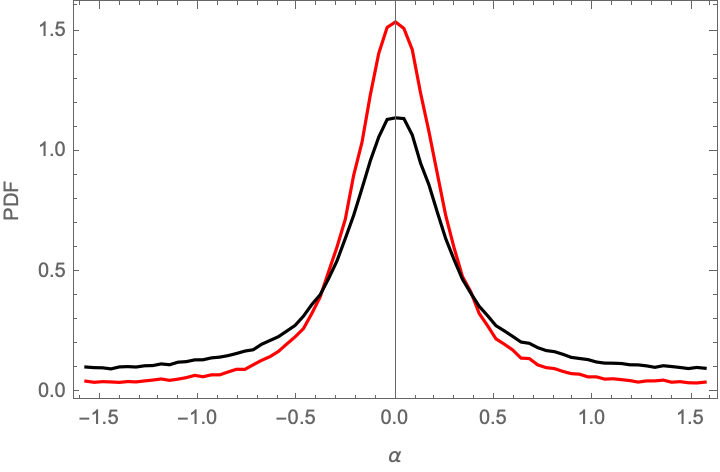}
    \caption{A comparison of the distribution of the angle between $\bm{T}$ and $\bm{v}_2$ (in degrees) for general points on the cups line (black) and points on the cusp line for which $\lambda_2<0$ (red) sampled in Lagrangian space for $b_+(t_c)=0.5$.}
    \label{fig:angle}
\end{figure}

A complicating factor for the practical implementation of the stretching conditions on the tangent space $\bm{T}$ is that they are algebraically and computationally very hard to evaluate. Though geometrically elegant, the maximum stretch condition on the tangent vector $\bm{T}$ yields a complicated algebraic identity that cannot be straightforwardly implemented using linear constrained theory.

Instead, we follow a simplifying approximation based on the observation of the alignment between the second eigenvector field $\bm{v}_2(\bm{q}_c)$ and the tangent vector $\bm{T}$. As demonstrated in our earlier study \citep{Feldbrugge:2023a}, on average these are aligned. This strong correlation between the orientation of the tangent vector and the second eigenvector is shown in the diagram in figure~\ref{fig:angle}, for general points on the cusp curve $A_3$ (black) and those with negative second eigenvalue field $\lambda_2$ (red). It suggests us to follow the simplifying approximation of a perfect alignment,
\begin{equation}
  \bm{T} = \bm{v}_2(\bm{q}_c)\,.
\end{equation}
This translates into the identity 
\begin{align}
    \|\nabla \bm{x}_t(\bm{q}_c) \bm{T}\| 
    &= \| (I - b_+(t) \mathcal{H} \Psi (\bm{q}_c))\bm{v}_2(\bm{q}_c)\| \\
    &=  |1 - b_+(t) \lambda_2(\bm{q}_c)|\,,
\end{align} 
with the unit two-by-two matrix $I$.

For positive eigenvalue $\lambda_2$, the line element contracts and undergoes shell crossing at 
\begin{equation}
  1-b_+(t)\lambda_2=0\,.
\end{equation}
Subsequently, with increasing growing mode the mass element expands. 

On the other hand, when the eigenvalue $\lambda_2$ is negative, following the Zel'dovich approximation the filament continues to stretch. We may immediately observe that the location -- or, rather, mass element -- where the filaments get maximally stretched corresponds to the mass element whose eigenvalue $\lambda_2$ along the cusp manifold $A_3$ represents a local minimum. 

\subsubsection{Minimum Second Eigenvalue}
Following the relations and approximations on the filament stretching conditions, we arrive at the set of conditions identifying the stretching center of filaments. They are identified with the points on cusp manifolds for which  the second eigenvalue obeys the conditions: 
\begin{itemize}
\item[1.] the second eigenvalue field is negative, 
  \begin{align} 
    \ \ \ \ \ \ \ \ \lambda_2(\bm{q}_c)<0\,,
\end{align} 
\item[2.] has a vanishing directional first derivative
\begin{align}
    \ \ \ \ \ \ \ \ \bm{v}_2(\bm{q}_c) \cdot \nabla \lambda_2(\bm{q}_c) &=0\,,
\end{align} 
\item[3.] and has a positive second directional derivative
\begin{align}
    \ \ \ \ \ \ \ \ \bm{v}_2(\bm{q}_c)^T [\mathcal{H} \lambda_2(\bm{q}_c)] \bm{v}_2(\bm{q}_c) &>0\,.
\end{align}
\end{itemize}
Note that the conditions $\lambda_1(\bm{q}_c)=1/b_+(t_c)$ and $\lambda_2(\bm{q}_c)<0$ indicate that -- in the Zel'dovich approximation -- at time $t_c$ the mass elements only undergone a single shell-crossing event. Non-linear gravitational collapse may include a set of further shell crossings.

\subsection{Filament Centers -- Umbilic and Swallowtail Nodes}
The stretching condition of a negative second eigenvalue, $\lambda_2<0$, ensures that the central region of the filament is far away from umbilic clusters. This may be directly inferred from the observation that umbilics are specified by the simultaneous requirement -- at some time $t_0$ -- on the first and second eigenvalue,
\begin{align}
1-b_+(t)\lambda_1=0\,,\nonumber
1-b_+(t) \lambda_2=0\,,
\end{align}
for some $t<t_0$. Evidently, these points are substantially removed from the minimum stretching criterion listed above. 

A similar observation can be made for swallowtail clusters. That these are well separated from the filament's stretching center can be inferred from the strong alignment between tangent $\bm{T}$ and second eigenvector $\bm{v}_2$ guarantees the stretching center to be well separated from the swallowtail clusters. In the swallowtail caustic the Lagrangian map satisfies the condition
\begin{align}
    \bm{v}_1(\bm{q}_c) \cdot \nabla(\bm{v}_1(\bm{q}_c) \cdot \nabla \lambda_1(\bm{q}_c)) = \bm{v}_1(\bm{q}_c) \cdot \bm{n}=0\,.
\end{align}
This condition implies the alignment of the first eigenvector field $\bm{v}_1$ with the tangent vector $\bm{T}$ field. In combination with an approximation of $\bm{T}$ being parallel to $\bm{v}_2$ leads to the conclusion that the stretching center of a filament is maximally distanced from the swallowtail caustics $A_4$. 

\begin{figure}
    \centering
    \includegraphics[width=\linewidth]{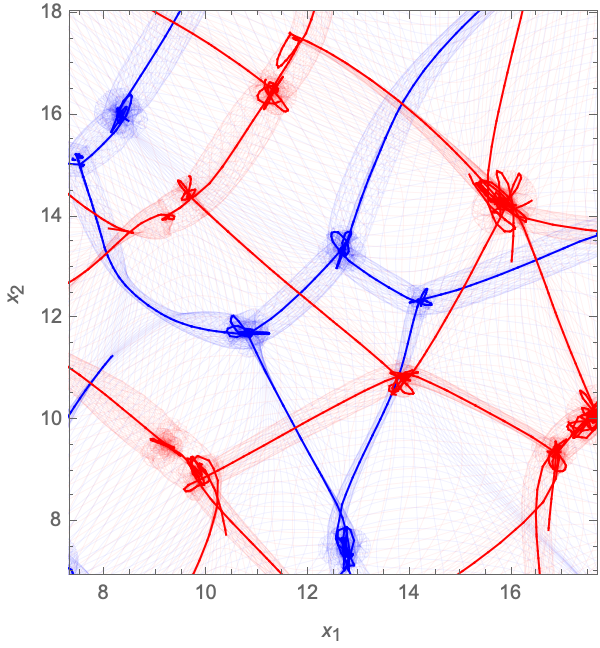}
    \caption{The caustic skeleton (blue) and the mirror skeleton (red) with the corresponding cusp lines in two $N$-body simulations in Eulerian space.} \label{fig:AntiSkeleton}
\end{figure}

\subsection{Stretching Filaments and Mirror Crossings}
In addition to the formal definition of the trunk and stretching center of a filament, there is also an alternative, interesting, insightful, and perhaps even surprising interpretation of the proposed condition for the \textit{stretching center of a filament}. 

To arrive at this interpretation, we notice that the conditions 
\begin{align}
    \bm{v}_1 \cdot \nabla \lambda_1 &=0\,,\\
    \bm{v}_2 \cdot \nabla \lambda_2 &=0\,,
\end{align}
display a neat symmetry between the first and the second eigenvalue fields $\lambda_1$ and $\lambda_2$. The key observation is that when we take the negative of the displacement field,
\begin{equation}
  \Psi \mapsto - \Psi\,,
  \end{equation}
also the eigenvalue fields flip sign,
\begin{align}
  \lambda_1 \mapsto - \lambda_1\,,\nonumber\\
  \lambda_2 \mapsto - \lambda_2\,.
  \end{align}
Given that this entails the initial Gaussian conditions that flipped sign, on the basis of these flipped eigenvalue fields we may infer the \textit{mirror caustic skeleton}. It defines the \textit{mirror cosmic web}. A telling illustration of a corresponding cosmic web realization and its mirror cosmic web is shown in figure~\ref{fig:AntiSkeleton}.

Inspecting the corresponding eigenvalue conditions (discussed above), we may directly see that the \textit{central stretching regions of filaments} correspond to the intersection of filaments in the \textit{caustic skeleton} and \textit{mirror caustic skeleton}. In a sense, the cosmic web and its mirror web may be seen as dual patterns, in which the nodes of the cosmic web correspond to voids in the mirror web, and the edges and web and the mirror web cross roughly at the midpoints. In this interpretation, clusters and filaments have the same role as that of vertices and edges in a dual (geometric) tessellation. This realization resembles the duality between the Voronoi tessellation and its dual Delaunay tessellation \citep{Okabe:2000,Icke:1987, Weygaert:1994}. Recently, \cite{Pontzen:2016, Shim:2021, Stopyra:2021, Shim:2023, Stopyra:2024} used a similar realization to identify voids in the cosmic web with clusters in the mirror web.

\section{Filament Statistics}\label{sec:filamentstat}

The caustic conditions enable a statistical analysis of the walls, filaments and clusters in the present-day cosmic web using a combination of stochastic geometry and Gaussian random field theory \cite{Rice:1944, Rice:1945, Longuet-Higgins:1957, Adler:1981, Bardeen:1986, Adler:2009}. This may ultimately shed light on several fundamental questions of structure formation and enable the exploitation of the substantial amount of cosmological information stored in the intricate and complex structure of the cosmic web. After all, it would allow us to obtain insight into the dependence of filament and cosmic web properties on the global cosmology.

Some of the most prominent key issues that are in need of a firm quantitative statistical assessment, in terms of distribution functions and statistical moments, are:
\begin{itemize}
\item[1)] at what time do filaments form? \\
\item[2)] what is the number density of filaments?\\
\item[3)] what is the mean curvature, areas, and lengths of filaments?\\
\item[4)] what is the topology of the cosmic web over cosmic time?\\
\item[5)] how are these filament properties influenced by the power spectrum of the primordial density perturbations?
\end{itemize}
In this paper, we study the statistical properties of filaments, extending the work presented in \cite{Feldbrugge:2016, Feldbrugge:2019, Feldbrugge:2023b}. We limit the analysis to the simplest statistics, exploring how the caustic conditions yield solid analytical predictions.

A full statistical analysis that extends these results to three-dimensional fluids is the subject of an upcoming paper. In the 3D study, the extended results include a more detailed analysis including the mean area and curvature of walls, the mean length and curvature of filaments, the merging history of filaments, and the implied number density of cluster nodes. 

\subsection{Potential derivatives: Definitions}
The caustic skeleton describes the geometry of the cosmic web in terms of the eigenvalues and eigenvectors of the Hessian $T_{ij}$ of the displacement potential $\Psi(\bm{q})$,
\begin{align}
T_{ij}(\bm{q}_c)\,=\,\frac{\partial}{\partial q_i}\frac{\partial}{\partial q_j} \Psi_\sigma(\bm{q}_c)\,,
\end{align}

The eigenvalue $\lambda_i$ fields and eigenvector $\bm{v}_i$ fields represent fields of non-linear functions of the second-order derivatives $T_{ij}$ of the displacement potential. Also, by Leibniz's rule, the gradients of the eigenvalue fields, $\nabla \lambda_i$, are non-linear functions of the second and third-order derivatives at a point. In addition, the normal of the cusp filament
\begin{equation}
  \bm{n} = \nabla (\bm{v}_i \cdot \nabla \lambda_i)\,,
\end{equation}
includes fourth-order derivatives of the displacement potential. In the following, we denote the higher order partial derivatives of the smoothed displacement potential $\Psi_\sigma$ at the point $\bm{q}_c$ as
\begin{align}
    T_{ij\dots k} =\frac{\partial}{\partial q_i}\frac{\partial}{\partial q_j} \dots \frac{\partial}{\partial q_k}  \Psi_\sigma(\bm{q}_c)\,.
\end{align}

\subsection{Potential Derivatives: Gaussian Statistics}
With the displacement potential representing a Gaussian random field, also linear functionals of the potential are Gaussian random fields. This includes derivatives of the potential $T_{ij\dots k}$ as well as convolutions of the potential field. Moreover, any linear combination of Gaussian variates has a Gaussian functional distribution. The probability distribution function $p(\bm{Y})$ for such a set of Gaussian variates $\bm{Y} \in \mathbb{R}^n$ follows the multi-dimensional Gaussian distribution 
\begin{align}
p(\bm{Y})\,=\,\frac{\exp\left[-\frac{1}{2}  \Delta \bm{Y}^T M^{-1} \Delta \bm{Y}\right]}{[(2\pi)^n \det M]^{1/2}}\,.
\end{align}
in which $\Delta \bm{Y}$ specifies the deviation from the mean,
\begin{equation}
\Delta \bm{Y} = \bm{Y} - \langle \bm{Y} \rangle\,,
\end{equation}
and $M$ is the covariance matrix
\begin{align}
M = \text{cov}(\bm{Y},\bm{Y}) = \langle \Delta \bm{Y}^T \Delta \bm{Y}\rangle\,.
\end{align}

\bigskip
For example, the second order derivatives $T_{ij}$,
\begin{equation}
  \bm{Y}_2=(T_{11}, T_{12}, T_{22})\,,
\end{equation}
are normally distributed random variables with zero mean, and covariance matrix $M_{2,2}$,
\begin{align}
    M_{2,2} = \left \langle \Delta\bm{Y}_2^T\Delta\bm{Y}_2\right \rangle
    = \frac{\sigma_2^2}{8}\begin{pmatrix}
        3 & 0 & 1\\
        0 & 1 & 0\\
        1 & 0 & 3
    \end{pmatrix}.\label{eq:M22}
\end{align}
where the $\sigma_2$'s is the second order spectral moment of the primordial displacement potential $\Psi_\sigma(\bm{q})$ (see eq.~\eqref{eq:moments}). Likewise, the third order derivatives $\bm{T}_{ijk}$,
\begin{equation}
  \bm{Y}_3=(T_{111}, T_{112}, T_{122},T_{222})\,,
\end{equation}
comprise a set of Gaussian random variables with covariance matrix $M_{3,3}$, 
\begin{align}
    M_{3,3} = \left \langle \Delta\bm{Y}_3^T\Delta\bm{Y}_3\right \rangle
    = \frac{\sigma_3^2}{16}\begin{pmatrix}
         5 & 0 & 1 & 0 \\
         0 & 1 & 0 & 1 \\
         1 & 0 & 1 & 0 \\
         0 & 1 & 0 & 5 
    \end{pmatrix}.\label{eq:M33}
\end{align}
Following this further for the fourth order derivatives $ T_{ijkl}$ 
\begin{equation}
  \bm{Y}_4=(T_{1111}, T_{1112}, T_{1122},T_{1222},T_{2222})\,,
\end{equation}
the corresponding Gaussian distribution is specified by the covariance matrix $M_{4,4}$,
\begin{align}
    M_{4,4} = \left \langle \Delta\bm{Y}_4^T\Delta\bm{Y}_4\right \rangle
    = \frac{\sigma_4^2}{128}\begin{pmatrix}
         35 & 0 & 5 & 0 & 3 \\
         0 & 5 & 0 & 3 & 0 \\
         5 & 0 & 3 & 0 & 5 \\
         0 & 3 & 0 & 5 & 0 \\
         3 & 0 & 5 & 0 & 35 \\
    \end{pmatrix}.\label{eq:M44}
\end{align}

In these Gaussian distributions, the influence of the power spectrum of the primordial fluctuations enters via the spectral moments  $\sigma_2, \sigma_3,$ and $\sigma_4$ defined by \eqref{eq:moments}.

\medskip
In all, we may consider the Gaussian distribution for a full set of second-, third-, and fourth-order derivatives of the displacement potential
\begin{equation}
  Y\,=\,(T_{11},\dots,T_{2222})\,,
\end{equation}
which corresponds to a Gaussian distribution with a vanishing mean and covariance matrix
\begin{align}
    M = \begin{pmatrix}
        M_{2,2} & 0 & M_{2,4}\\
        0 & M_{3,3} & 0 \\
        M_{2,4}^T & 0 & M_{4,4}
    \end{pmatrix},\label{eq:cov234}
\end{align}
in which $M_{i,j}$ are the covariance matrices of the $i$-th and $j$-th order derivatives of the displacement potential, which includes $M_{2,2}$, $M_{3,3}$ an $M_{4,4}$ specified above in equations~\eqref{eq:M22}, \eqref{eq:M33}, \eqref{eq:M44} and with the additional matrix $M_{2,4}$,
\begin{align}
    M_{2,4} &=-\frac{\sigma_3^2}{16}
    \left(
        \begin{array}{ccccc}
         5 & 0 & 1 & 0 & 1 \\
         0 & 1 & 0 & 1 & 0 \\
         1 & 0 & 1 & 0 & 5 \\
        \end{array}
    \right),
\end{align}
With respect to the Gaussian distribution for the combined second, third, and fourth-order gradients we may note from the covariance matrix (eqn.~\eqref{eq:cov234}) that the third-order gradients and the second-order and fourth-order gradients are independently distributed. In general, two derivatives of a Gaussian random field $T_{i_1 j_1}$, $T_{i_2 j_2}$ at a given point $\bm{q}_c$ are independently distributed in case $i_1+i_2$ or $j_1 + j_2$ is odd are independently distributed. It manifests itself in the checkerboard pattern in the covariance matrix, such as for example seen in expression~\eqref{eq:cov234}.


\subsection{Filament number density}
One of the most prominent aspects of the filament population concerns the issue of the number of filaments and the evolution of the number of filaments. Here, we equate the number density of filaments to that of the number of centers of filament trunks (see section~\ref{sec:filament}). 

\begin{figure}
    \centering
    \includegraphics[width=\linewidth]{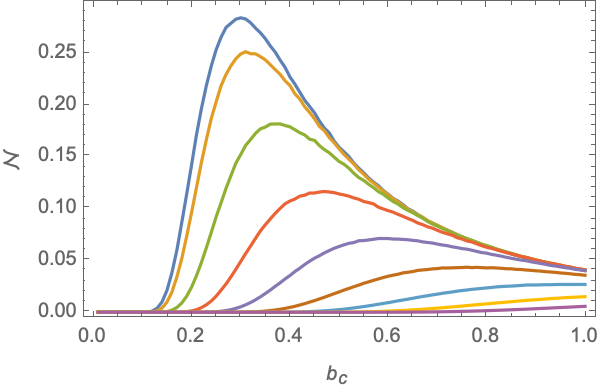}
    \caption{The number density of filament centers as a function of their formation times $b_c$ and the smoothing scale $\sigma=0,0.25,0.5,\dots,2$ from top to bottom for a scale-invariant power spectrum with a Gaussian cutoff at $R_s=1$.} \label{fig:numberdensity}
\end{figure}

Following Rice's formula \citep{Rice:1944, Rice:1945, Longuet-Higgins:1957, Adler:1981, Bardeen:1986, Adler:2009}, the number density of filament trunk centres, at length scale $\sigma$, forming at growing mode $b_c$  is given by the expectation value
\begin{align}
    &\mathcal{N}_\sigma (b_c) = b_c^{-2} \bigg\langle
        \left|\det \begin{pmatrix}
            \partial_1 (\bm{v}_1 \cdot \nabla \lambda_1) & \partial_2(\bm{v}_1 \cdot \nabla \lambda_1)\\
            \partial_1 (\bm{v}_2 \cdot \nabla \lambda_2) & \partial_2(\bm{v}_2 \cdot \nabla \lambda_2)\\
        \end{pmatrix}\right|\nonumber\\
        &\times \delta_D^{(1)}(\bm{v}_1 \cdot \nabla \lambda_1)
        \delta_D^{(1)}(\bm{v}_2 \cdot \nabla \lambda_2)
         \delta_D^{(1)}(\lambda_1-b_c^{-1})\nonumber\\
        &\times \Theta_H(-\lambda_2)
        \Theta_H\left(\bm{v}_2^T [\mathcal{H} \lambda_2] \bm{v}_2\right)
        \bigg\rangle\\
        &= b_c^{-2}\bigg\langle
            \left|\det \begin{pmatrix}
                T_{1111} + \frac{3 T_{112}^2}{\lambda_1-T_{22}} & T_{1112} + \frac{3 T_{112}T_{122}}{\lambda_1-T_{22}}\\
                T_{1222} - \frac{3T_{112}T_{122}}{\lambda_1-T_{22}} & T_{2222} - \frac{3T_{122}^2}{\lambda_1-T_{22}}\\
            \end{pmatrix}\right|\nonumber\\
        &\times \pi |\lambda_1-T_{22}|\delta_D^{(1)}(T_{12})\Theta_H(-T_{22})\delta_D^{(1)}(T_{111})
        \delta_D^{(1)}(T_{222})\nonumber\\
        & \times\Theta_H\left(T_{2222} - \frac{2T_{122}^2}{\lambda_1-T_{22}}\right)
        \delta_D^{(1)}(T_{11} - b_c^{-1})
            \bigg\rangle\,,
\end{align}
In the second equality, we change coordinates from the eigenvalue fields to the derivatives of the displacement field in the eigenframe. The resulting number density curves are plotted in fig.\ \ref{fig:numberdensity} for smoothing scales $\sigma =0,0.25,\dots,2$. The number density is the largest at the smallest smoothing scale around the formation time $b_c=0.3$. As the smoothing scale increases, the number density of filament trunks gradually decreases. Also, we see that they form at later times, corresponding to larger growing modes. Moreover, in the next section, we show that the large filaments emerge over a wider range of formation times. 

\subsection{Formation time of a filament}\label{sec:formation_time}
Considering the emergence of the centre of the filament's trunk as the formation time of a filament, we may assess the formation times of the entire filament population and explore its dependence on scale, smoothing scale, power spectrum, and other characteristics of the filaments. 

In our evaluation of the filament formation time properties, we therefore focus on the emergence of the centers of the filaments. The caustic skeleton framework provides a detailed quantitative definition for this. It is identified with the moment shell-crossing occurs. Specifying time $t$ in terms of the growing mode of structure formation, $b_+(t)$, for a location $\bm{q}_c$ caustic conditions specify the occurrence of shell-crossing to happen at growing mode factor $b_c=b_+(t_c)$,
\begin{equation}
\lambda_1(\bm{q}_c) = 1/b_c\,,
\end{equation}
with the additional condition that
\begin{equation}
  \bm{v}_1(\bm{q}_c) \cdot \nabla \lambda_1(\bm{q}_c) = 0\,.
\end{equation}
Restricting our attention to specific filament center locations, we have the additional requirement of a mass element at that location to expand maximally along the ridge of the filaments. This means that
\begin{itemize}
\item[1.] the second eigenvalue is negative
  \begin{equation}
    \hskip 0.75cm \lambda_2(\bm{q}_c)<0\,,
    \end{equation}
\item[2.] its first derivative in the $\bm{v}_2$ direction vanishes
\begin{equation}
  \hskip 0.75cm \bm{v}_2(\bm{q}_c) \cdot \nabla \lambda_2(\bm{q}_c) = 0
  \end{equation}
\item[3.] its second-order derivative is positive,
  \begin{equation}
    \hskip 0.75cm \bm{v}_2(\bm{q}_c)^T [\mathcal{H} \lambda_2(\bm{q}_c)] \bm{v}_2(\bm{q}_c) >0\,.
   \end{equation}
  \end{itemize}

\bigskip
\noindent Given the isotropy of the Gaussian random field, for the mathematical treatment of the problem it is convenient to rotate to the \textit{eigenframe of the Hessian} $\mathcal{H} \Psi_\sigma(\bm{q}_c)$. It allows us to infer the corresponding identities without loss of generality. For the details of the derivation of the corresponding identities, we refer to \cite{Feldbrugge:2023a}. In terms of the second-order derivatives $T_{ij}$ of the displacement potential the eigenvalues within this coordinate system are given by the identities
\begin{align}
  \lambda_1(\bm{q}_c) &= T_{11}\,,\nonumber\\
  T_{12}&=0,\\
  \lambda_2(\bm{q}_c) &= T_{22}\,.\nonumber
\end{align}
The directional derivatives of the eigenvalue fields coincide with the third-order derivatives of the displacement field
\begin{align}
    \bm{v}_1 \cdot \nabla \lambda_1 &= T_{111}\,,\\
    \bm{v}_2 \cdot \nabla \lambda_2 &= T_{222}\,,\nonumber
\end{align}
while the second-order derivative of the second eigenvalue field $\lambda_2$ in the $\bm{v}_2$ direction yields the identity
\begin{align}
    \bm{v}_2^T [\mathcal{H}\lambda_2] \bm{v}_2 = T_{2222} - \frac{2T_{122}^2}{\nu-T_{22}}\,.
\end{align}

\bigskip
Using these inferred identities, we find that the probability distribution function of the time at which the center of the filament emerges, in terms of the corresponding growing mode $b_c$, is equivalent to the conditional probability $p(\lambda_1\,|\, \dots)$ for the first eigenvalue $\lambda_1$,
\begin{align}
    &p(\text{center of filament emergest at growing mode } b_c)\\
    &=\lambda_1^2 p(\lambda_1 \,|\, \lambda_2<0<\lambda_1, \bm{v}_1 \cdot \nabla \lambda_1= \bm{v}_2 \cdot \nabla \lambda_2=0) \big|_{\lambda_1=1/b_c}\,,\nonumber
\end{align}
involving the conditions specified in the listing above. For simplicity we dropped the positivity condition $\bm{v}_2^T [\mathcal{H} \lambda_2] \bm{v}_2 >0$, as a critical point of $\lambda_2$ in the $\bm{v}_2$ direction is likely to be minimum when $\lambda_2<0$. In this expression, the pre-factor $\lambda_1^2$ results from the Jacobian of the transformation $b_c=1/\lambda_1$, preserving probability \textit{i.e.}, $|p(b_c)\mathrm{d} b_c| = |p(\lambda_1) \mathrm{d}\lambda_1|$.

\begin{figure}
    \centering
    \includegraphics[width=\linewidth]{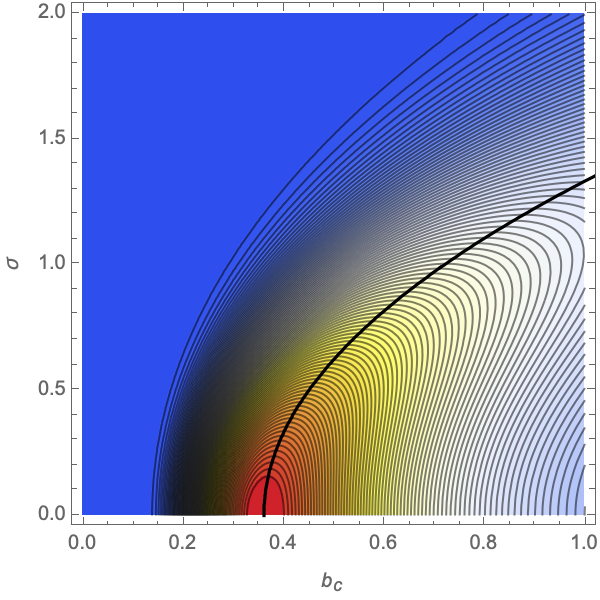}
    \caption{The PDF of the formation time $b_c$ of the centers of the filaments as a function of the smoothing scale $\sigma$ for a scale-invariant power spectrum with a Gaussian cutoff at $R_s=1$. The peak of the distribution is plotted by the black curve.} \label{fig:formationtimes}
\end{figure}

Since the eigenvalue field and its first order derivatives along the eigenvector directions are independent -- following from their relations to the derivatives of the primordial displacement potential in the eigenframe -- we find that 
\begin{align}
    &p(\text{center of filament forms a growing mode } b_c)\\
    &=\lambda_1^2  p(\lambda_1 | \lambda_2<0<\lambda_1)\big|_{\lambda_1=1/b_c} \\
    &=  \frac{\lambda_1^2 \int_{-\infty}^{0} p(\lambda_1, \lambda_2)\mathrm{d}\lambda_2 }{\int_{-\infty}^0\int_{0}^\infty p(\lambda_1,\lambda_2)\mathrm{d}\lambda_1 \mathrm{d}\lambda_2}\bigg|_{\lambda_1=1/b_c}\\
    &=  \frac{2 \lambda_1^2 \Theta_H(\lambda_1)}{3 \sigma_2^2} \bigg[2 \lambda_1 e^{-\frac{4 \lambda_1^2}{3 \sigma_2^2}} \text{erfc}\left(\frac{\lambda_1}{\sqrt{6} \sigma_2}\right)
    +\sqrt{\frac{6}{\pi }} \sigma_2 e^{-\frac{3 \lambda_1^2}{2 \sigma_2^2}}\bigg]\bigg|_{\lambda_1=1/b_c}\,,
\end{align}
using the Doroshkevich formula for the distribution of the eigenvalue fields \citep{Doroshkevich:1970,Feldbrugge:2023b}
\begin{align} 
    p(\lambda_1,\lambda_2) = &\sqrt{\frac{2}{\pi}} \frac{2}{\sigma_2^3}(\lambda_1-\lambda_2) e^{-\frac{1}{2 \sigma_2^2}\left(3(\lambda_1+\lambda_2)^2- 8 \lambda_1 \lambda_2\right)}\nonumber\\
    &\times \Theta_H(\lambda_1-\lambda_2)\,.
\end{align}

\bigskip
Figure~\ref{fig:formationtimes} visually summarizes the resulting filament formation time. It shows the probability value as a function of formation time $b_c$ and smoothing scale $\sigma$. From the inferred formation time distribution, we may conclude that in the case of no smoothing, \textit{i.e.}, $\sigma=0$, the filaments are likely to form at a growing mode of $b_c \approx 0.36$. Most of the filament centers have emerged at the current time $b_c =1$.

Another interesting observation is that filaments are more likely to form at later times as the smoothing length $\sigma$ increases. Also, at higher values of $\sigma$, we see a larger variation in formation time: large filaments have a larger diversity in formation time and history. Moreover, at smoothing length $\sigma=2$ a filament is unlikely to have formed yet in the cosmic web at the current epoch. 

\section{Filaments in the cosmic web:\\ \ \ \ \ \ Constrained Realizations}\label{sec:cosmicweb}
Based on the dynamical definition of a cosmic filament, in principle, we may use the corresponding conditions to generate constrained simulations of filaments within any cosmological setting. In practice, we may do so by investigating the primordial conditions and configurations out of which filaments emerged given the primordial density and gravity field is a Gaussian random field. Such a setup would enable a targeted "laboratory" study that explores the setting, connections, characteristics, development, and evolution of filaments in the cosmic web. The constrained filament simulations will allow a systematic study of these aspects as a function of the spatial scale $\sigma$ of the filaments and cosmic epoch (specified by growing mode $b_c$). Particularly important is that such a setup would allow a systematic exploration of the influence of background cosmology and the power spectrum of primordial fluctuations on emerging filaments. It will be the most efficient path towards gaining insight into the cosmological information content of the cosmic web, and recognizing the cosmological signatures in its observed structure and properties.

The formalism for constrained filament simulations has been described in detail in \cite{Feldbrugge:2023a}. It is an elaboration of the linear constraint Hoffman-Ribak formalism \citep{Hoffman:1991} for density and gravity fields towards non-linear constraints, following the definitions and relations outlined in \cite{Weygaert:1996}. The non-linear constraints are non-linear functions of Gaussian random field variables. Tidal or deformation eigenvalues, and their gradients, are telling examples of such non-linear functions of Gaussian quantities, in this case, the components of the Hessian of the gravitational potential. 

In the following, we outline the formalism for the constrained generation of cosmic filaments in the cosmic web, based on their physical definition. The latter includes the phase-space-based caustic conditions specifying their spine in terms of the cusp $A_3$ singularity \citep{Feldbrugge:2018}, in combination with the tidal dilation of the mass content along their ridge (section~\ref{sec:filament}). 

\begin{figure*}
    \centering
    \begin{subfigure}[b]{0.32\textwidth}
        \includegraphics[width=\textwidth]{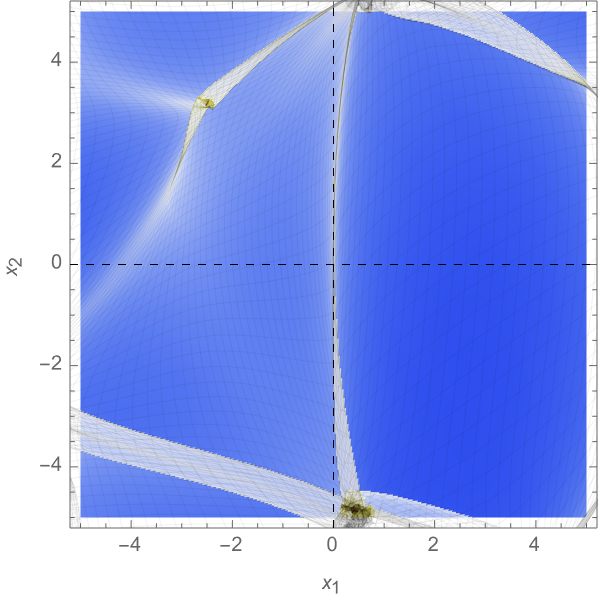}
    \end{subfigure}
    ~ 
    \begin{subfigure}[b]{0.32\textwidth}
        \includegraphics[width=\textwidth]{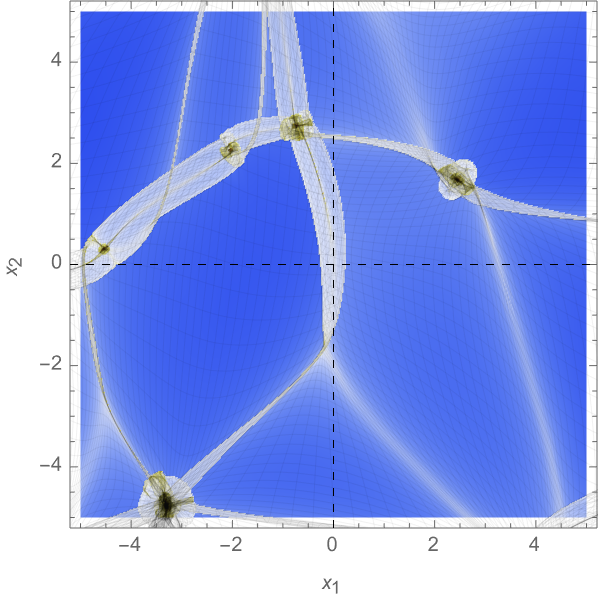}
    \end{subfigure}
    ~
    \begin{subfigure}[b]{0.32\textwidth}
        \includegraphics[width=\textwidth]{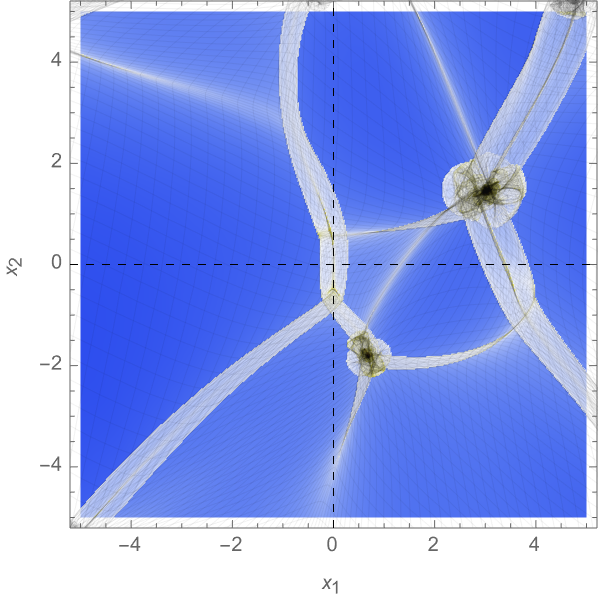}
    \end{subfigure}\\
    \begin{subfigure}[b]{0.32\textwidth}
        \includegraphics[width=\textwidth]{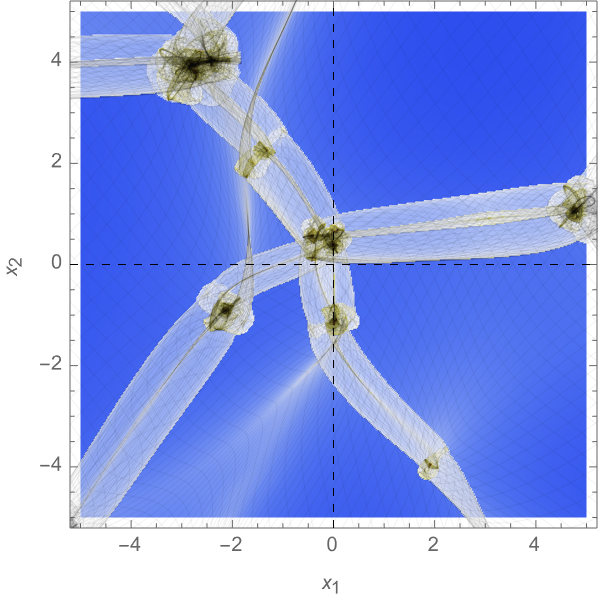}
    \end{subfigure}
    ~ 
    \begin{subfigure}[b]{0.32\textwidth}
        \includegraphics[width=\textwidth]{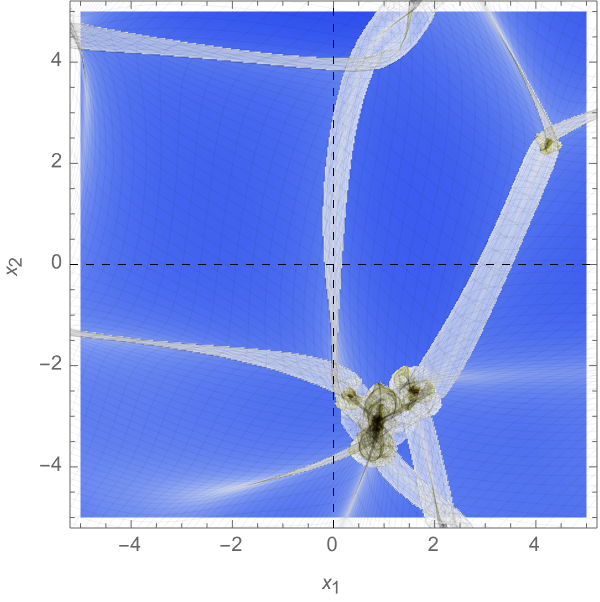}
    \end{subfigure}
    ~
    \begin{subfigure}[b]{0.32\textwidth}
        \includegraphics[width=\textwidth]{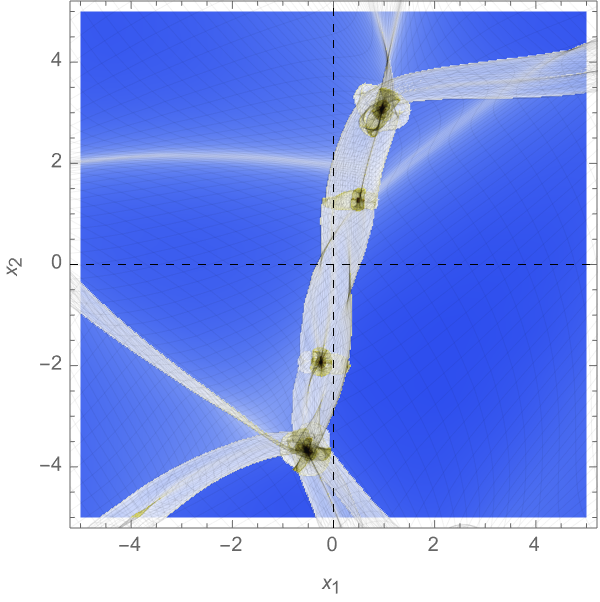}
    \end{subfigure}
    \caption{Six realizations of the cusp filaments with the formation time $b_c=0.5$ at the scale $\sigma=0.5$. We plot the $N$-body particles and the initial mesh on the corresponding density field $\log(\rho + 1)$.} \label{fig:realizations}
\end{figure*}

\begin{figure*}
    \centering
    \begin{subfigure}[b]{0.24\textwidth}
        \includegraphics[width=\textwidth]{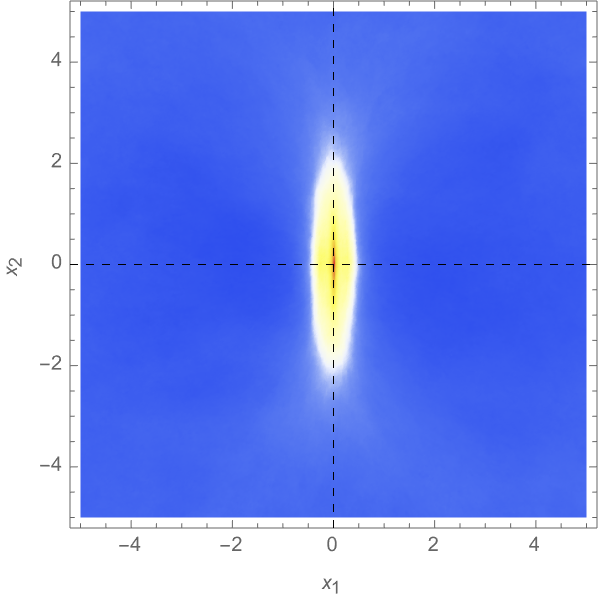}
    \end{subfigure}
    \begin{subfigure}[b]{0.24\textwidth}
        \includegraphics[width=\textwidth]{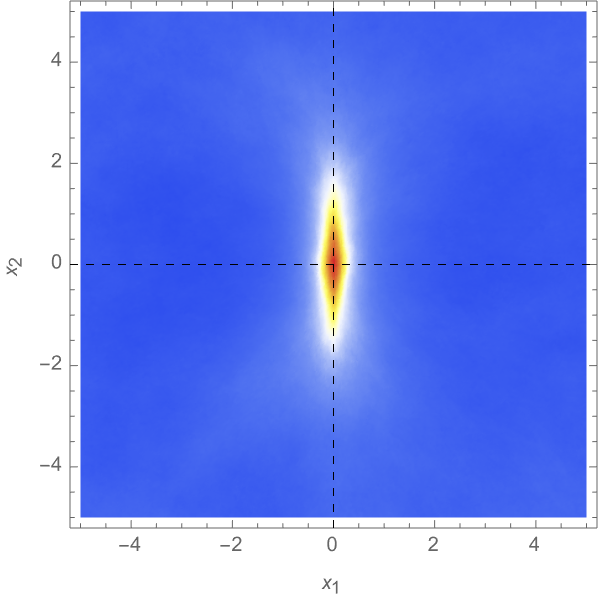}
    \end{subfigure}
    \begin{subfigure}[b]{0.24\textwidth}
        \includegraphics[width=\textwidth]{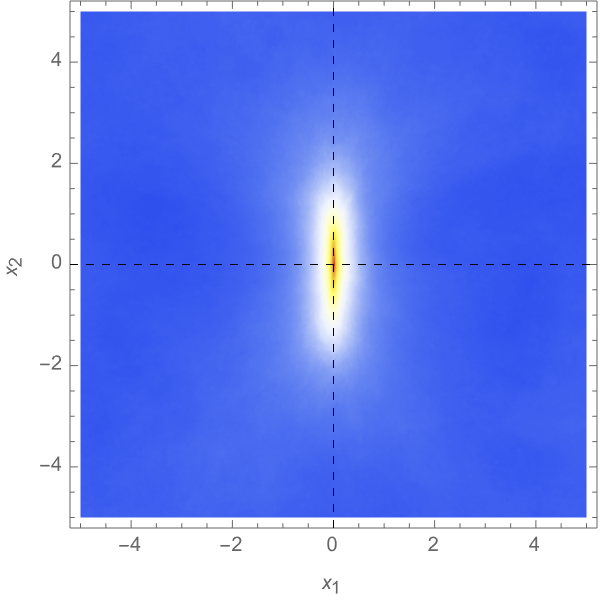}
    \end{subfigure}
    \begin{subfigure}[b]{0.24\textwidth}
        \includegraphics[width=\textwidth]{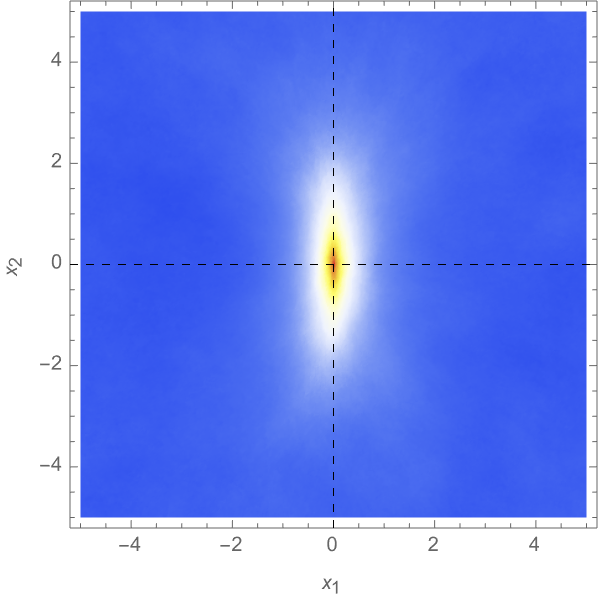}
    \end{subfigure}\\
    \begin{subfigure}[b]{0.24\textwidth}
        \includegraphics[width=\textwidth]{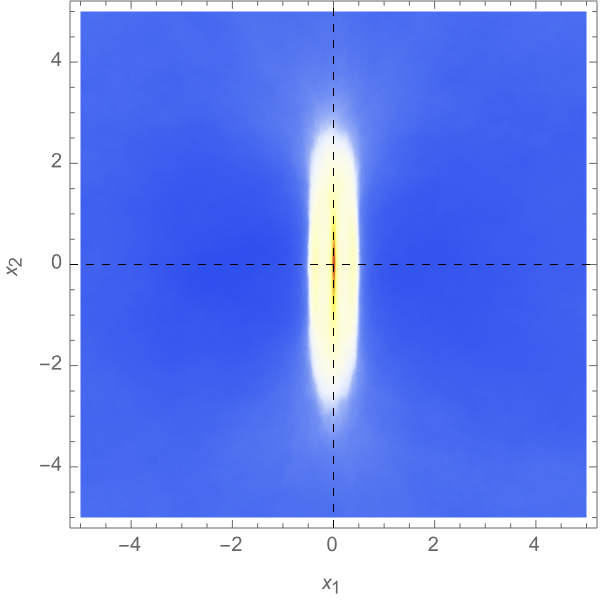}
    \end{subfigure}
    \begin{subfigure}[b]{0.24\textwidth}
        \includegraphics[width=\textwidth]{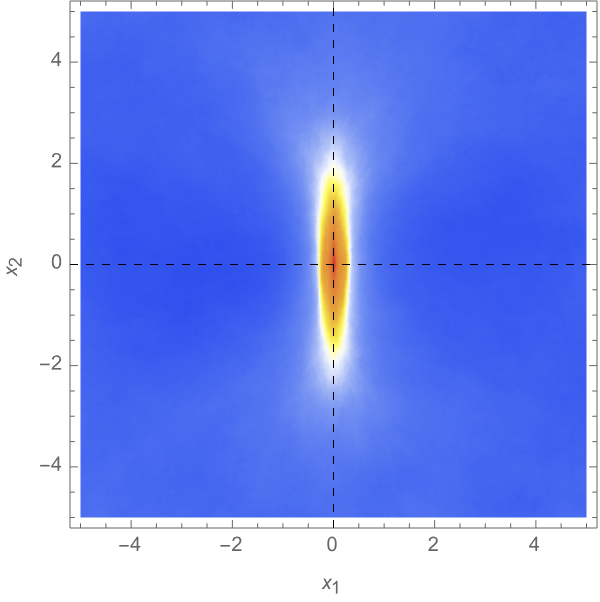}
    \end{subfigure}
    \begin{subfigure}[b]{0.24\textwidth}
        \includegraphics[width=\textwidth]{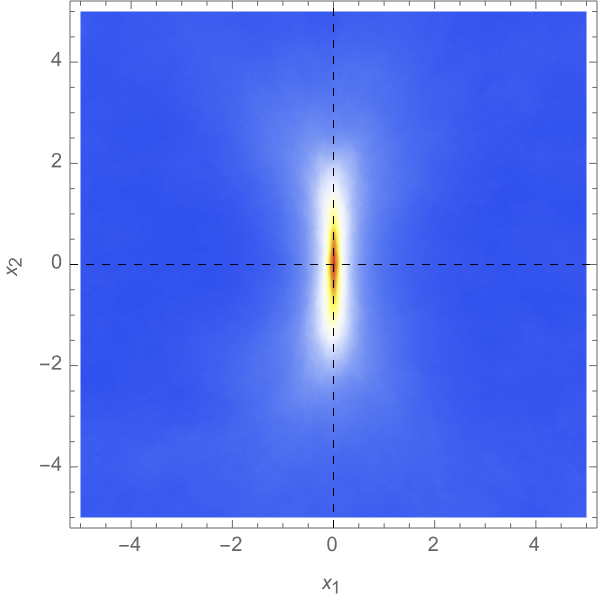}
    \end{subfigure}
    \begin{subfigure}[b]{0.24\textwidth}
        \includegraphics[width=\textwidth]{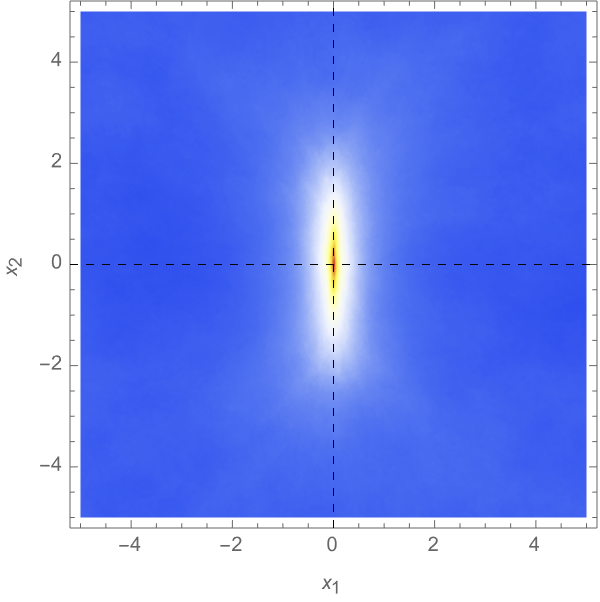}
    \end{subfigure}\\
    \begin{subfigure}[b]{0.24\textwidth}
        \includegraphics[width=\textwidth]{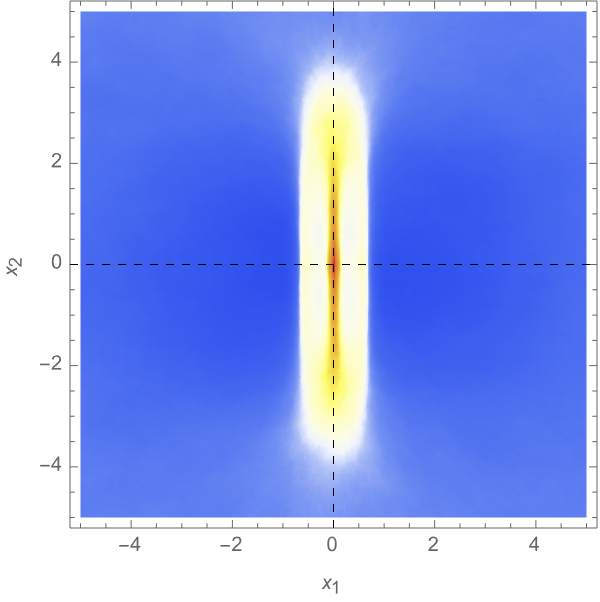}
    \end{subfigure}
    \begin{subfigure}[b]{0.24\textwidth}
        \includegraphics[width=\textwidth]{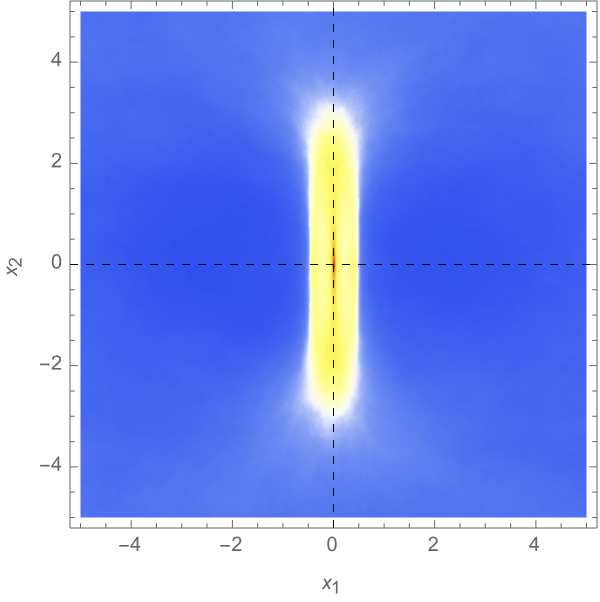}
    \end{subfigure}
    \begin{subfigure}[b]{0.24\textwidth}
        \includegraphics[width=\textwidth]{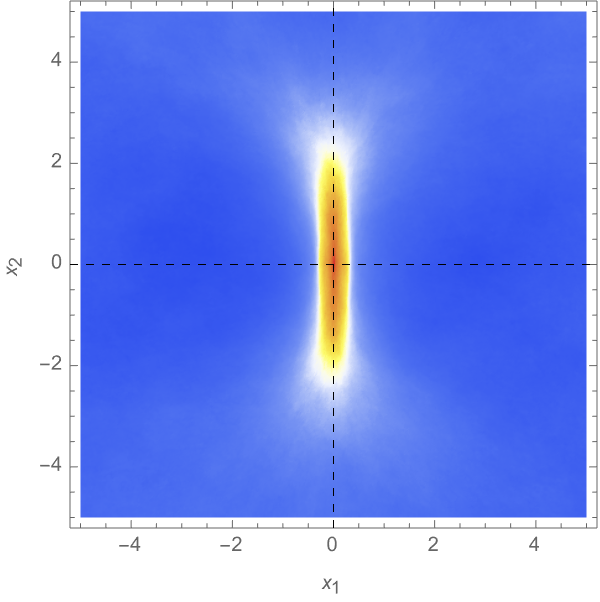}
    \end{subfigure}
    \begin{subfigure}[b]{0.24\textwidth}
        \includegraphics[width=\textwidth]{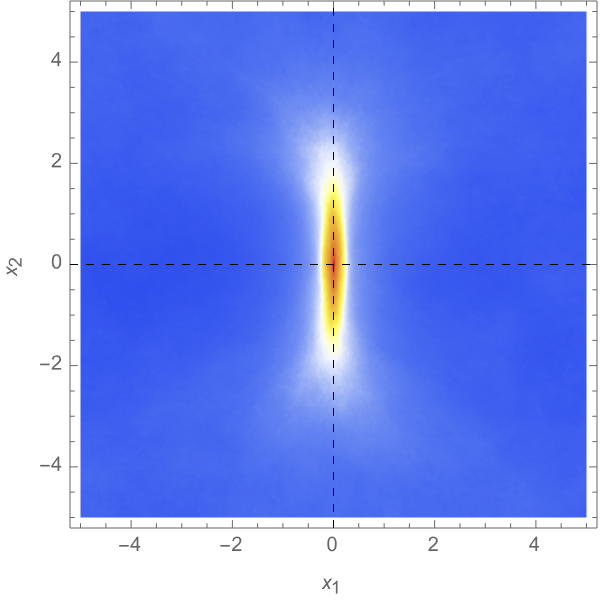}
    \end{subfigure}
    \caption{The median density field of the realizations of the centers of the filament as a function of the formation time $b_c=0.3,0.5,0.8,1.0$ from left to right and length scale $\sigma=0.1,0.5,1.0$ from top to bottom. We plot the median density field on a logarithmic scale $\log(\rho_{median} + 1)$.} \label{fig:Median_CoF}
    \centering
    \begin{subfigure}[b]{0.32\textwidth}
        \includegraphics[width=\textwidth]{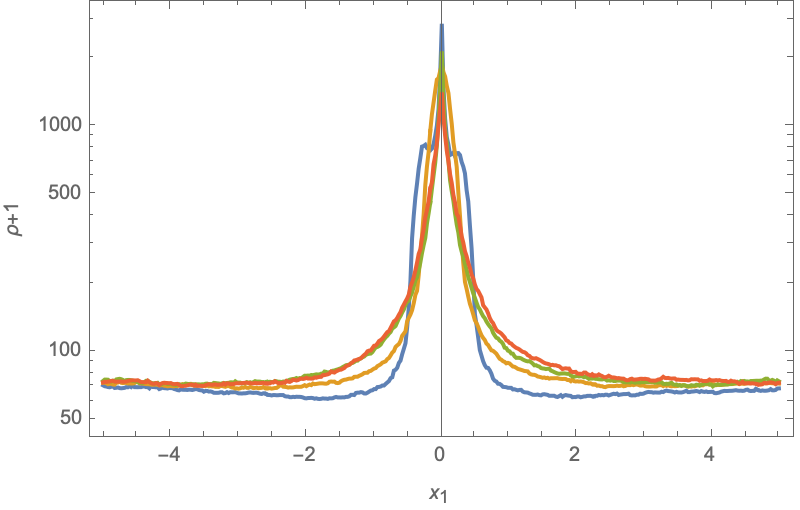}
    \end{subfigure}
    ~ 
    \begin{subfigure}[b]{0.32\textwidth}
        \includegraphics[width=\textwidth]{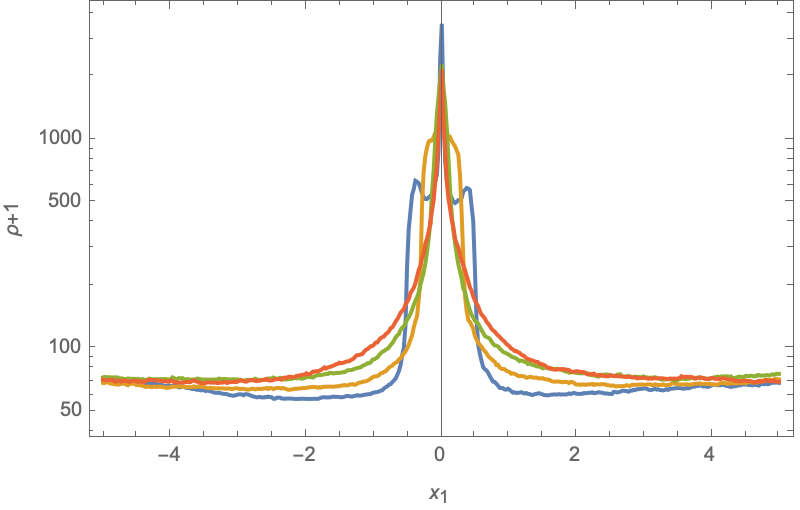}
    \end{subfigure}
    ~
    \begin{subfigure}[b]{0.32\textwidth}
        \includegraphics[width=\textwidth]{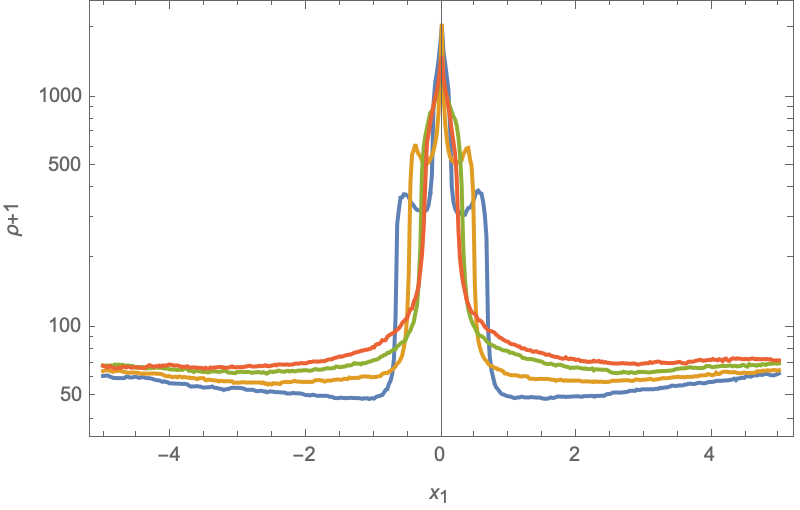}
    \end{subfigure}
    \caption{The density profile perpendicular to the filament as a function of formation time $b_c=0.3, 0.5, 0.8, 1$ (blue yellow green and red) for the length scale $\sigma=0.1, 0.5, 1.0$ (left to right).} \label{fig:prof}
    \centering
    \begin{subfigure}[b]{0.32\textwidth}
        \includegraphics[width=\textwidth]{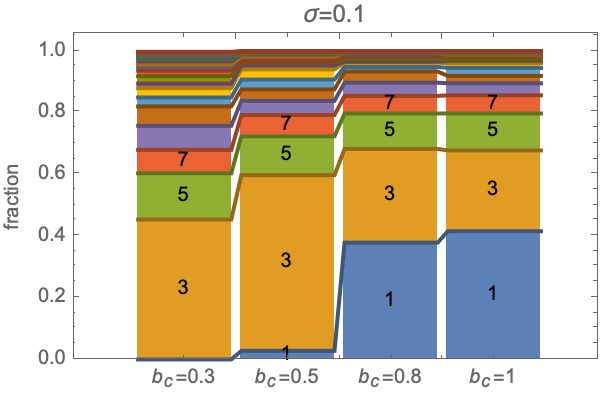}
    \end{subfigure}
    ~ 
    \begin{subfigure}[b]{0.32\textwidth}
        \includegraphics[width=\textwidth]{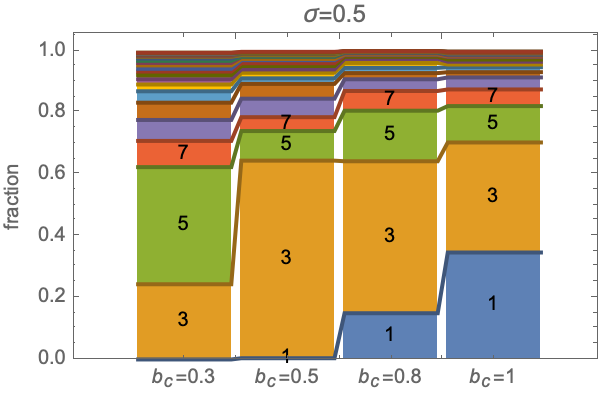}
    \end{subfigure}
    ~
    \begin{subfigure}[b]{0.32\textwidth}
        \includegraphics[width=\textwidth]{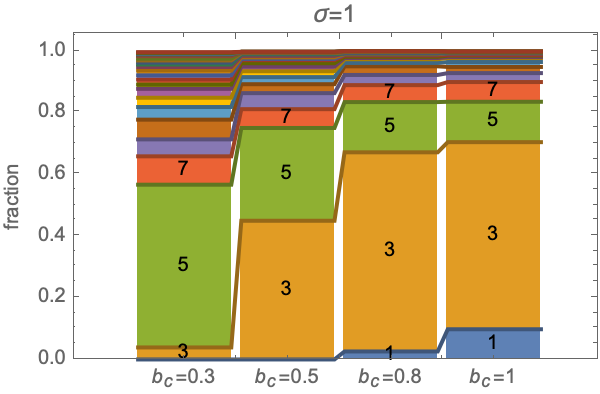}
    \end{subfigure}
    \caption{The distribution of the number of streams as a function of the formation time for the length scale $\sigma=0.1, 0.5, 1.0$ (left to right).} \label{fig:streams}
\end{figure*}

\subsection{Constrained filament initial conditions: \\ \ \ \ \ \ \ \ specification}
Initial conditions for filaments consist of two aspects: the cusp $A_3$ caustic constraints, in combination with supplementary constraints on the tidally induced stretching of its mass content. Following the previous section, we focus the filament constraint on the point representing the stretching center of the filament. 

\bigskip
In the eigenframe of the Hessian $\mathcal{H}\Psi_\sigma$ the cusp conditions $\lambda_1 = 1/b_c,$ and $\bm{v}_1 \cdot \nabla \lambda_1=0$ (see sect.~\ref{sec:formation_time}) assume the form
\begin{align}
    T_{11} &= 1/b_c\,,\,\nonumber\\ 
    T_{12} &= 0\,,\,\nonumber\\
    T_{22} &<T_{11}\,,\,\\
    T_{111} &= 0\,.\nonumber
\end{align}
The normal $\bm{n}$ of the cusp line in Lagrangian space is given by 
\begin{equation}
    \bm{n} = \left(T_{1111} + \frac{3 T_{112}^2}{T_{11}-T_{22}}, T_{1112} + \frac{3 T_{112}T_{122}}{T_{11}-T_{22}}\right)\,.
    \label{eq:normal}
\end{equation}

\bigskip
Along the cusp manifold, a mass element undergoes maximal stretch when it fulfills the additional conditions $\lambda_2 <0$, $\bm{v}_2 \cdot \nabla \lambda_2 =0$, $\bm{v}_2^T[\mathcal{H}\lambda_2 ]\bm{v}_2 >0$ which in the eigenframe assumes the form
\begin{align} 
    T_{22} &< 0\,,\,\nonumber\\
    T_{222} &= 0\,,\,\\
    0 &< T_{2222} - \frac{2 T_{122}^2}{T_{11}-T_{22}}\,.\nonumber
\end{align}

\subsection{Constrained filament initial conditions: \\ \ \ \ \ \ \ \ procedure}
We generate initial conditions for a set of $N$-body simulations by sampling Gaussian random fields subject to the constraints specified above. Outlined in detail in \cite{Feldbrugge:2023a} this is accomplished by the combination of a rejection sampling scheme for the non-linear function of second- third- and fourth-order derivatives of the smoothed displacement potential with that of the Hoffman-Ribak algorithm for sampling Gaussian random field  \cite{Bertschinger:1987,Hoffman:1991,Weygaert:1996}. It results in the following procedure:

\begin{enumerate}
\item Sample the statistic $Y=(T_{11},\dots,T_{2222})$. The probability of $Y$ is that of a multidimensional normal distribution with vanishing mean and covariance matrix \eqref{eq:cov234} subject to the linear constraints
\begin{equation}
  (T_{11},T_{12},T_{111},T_{222}) = (1/b_+(t_c),0,0,0)\,,
\end{equation}
multiplied with the Jacobian factor $|T_{11}-T_{22}|$. 
\item Reject samples that do not satisfy the conditions
\begin{align}
  T_{22}<0\,,\quad
  T_{2222}- \frac{2 T_{122}^2}{T_{11}-T_{22}} >0\,.
  \end{align}
\item For each remaining sample of $Y$ in the eigenframe, we evaluate the normal of the cusp manifold in the eigenframe using expression~\eqref{eq:normal},
\item Subsequently, the derivatives of the deformation potential are rotated to align the normal of the cusp manifold $\bm{n}$ with the vertical axis, following the transformations derived in appendix E of \cite{Feldbrugge:2023a}. 
\item For each sampled $Y$, we sample a corresponding Gaussian random field with power spectrum $P_\Psi(k)$ using the Hoffman-Ribak algorithm \citep{Bertschinger:1987, Hoffman:1991,Weygaert:1996}.
\end{enumerate}

\subsection{Constrained Filament Inventory}
Having specified filament constraints for the primordial Gaussian density and potential fields, we invoke these for an extensive and systematic study of the properties and large-scale environment of the specified filaments in their cosmological setting. 

Here we make an inventory of filaments for a range of configurations within a global cosmological setting of an Einstein-de Sitter Universe. Our interest concerns in particular the the morphology of the filament configurations and the multi-stream nature of the emerging filaments.

\subsubsection{Filament Configurations}
For each constrained filament configuration we set up $1000$ random realizations in a two-dimensional box of size $25 R_s \times 25 R_s$, with $R_s$ the spectral Gaussian cutoff scale (see eqn.~\eqref{eq:powerpk}), specifying the presence of the specific filaments at the centre of the box according to the constrained field formalism outlined above. We also arrange the filament orientation, such that all realizations contain a filament running through the center of the box, oriented in the vertical direction. This is arranged using a rather special procedure. For each sampled initial condition, two different $N$-body simulations are run:
\begin{enumerate}
\item The first simulation evolves the sampled initial conditions. The second simulation evolves a smoothed version of the initial conditions, convolving the entire field with a Gaussian kernel \eqref{eq:kernel} whose filter scale $\sigma$ is equal to that of the imposed filament constraint.
\item From the smoothed $N$-body simulation, we evaluate the position $\bm{x}_t(\bm{q}_c)$ and orientation of the cusp filament in Eulerian space using the equation $\nabla \bm{x}_t(\bm{q}_c) \bm{T}$. Subsequently, we center and orient the filament in the unsmoothed $N$-body simulation along the vertical direction using the position and orientation of the filament in the smoothed $N$-body simulation.
\end{enumerate}

The filament configurations of which we generate each $1000$ realizations have the following range of smoothing lengths $\sigma$: $\sigma=0.1,0.5,1.0$  In addition, for each of these we explore five different amplitudes, specified in terms of the implied formation (emergence) time $b_c=0.3,0.5,0.8,1.0$. 

\subsubsection{Filament Realizations}
A representative sample of six filament realizations is shown in figure~\ref{fig:realizations}. The density and particle map concern the matter distribution in and around the specified filament at the current epoch.

The filament realization in the four central and righthand panels show fully developed filaments, assembled around the stretching center at the center of the box, with their trunk neatly oriented along the vertical axis. While the mass distribution at the top lefthand panel stretches along a long vertical ridge, it has not yet reached the advanced stage in which it developed into a multi-stream region. In other words, the filament has not yet emerged as such. In this situation, the Zel'dovich approximation apparently did not properly predict the filament formation time. 

A more complex situation is seen in the bottom lefthand panel. Strong non-linear evolution of the surrounding mass distribution has entangled the specified central filament, oriented vertically, with a horizontally running filament. A massive and compact cluster has formed at their crossing point, at the center of the simulation box: the specified filament is one of the principal high-density filamentary extensions of the cluster. We should also note that the cusp filament of the Zel'dovich approximation is unable to fully capture the highly non-linear dynamical evolution of this region. In principle, one might opt for a post-processing step in which such configurations are removed from further consideration, by \textit{e.g.} restricting the filamentary regions in the smoothed $N$-body simulation to $3$- or $5$-stream regions. In the present analysis, we do not perform such a post-selection.

\subsubsection{Filament Density Profiles}
On the basis of the large, statistically representative, sample of constrained filament realizations, we assess the generic mass distribution in the vicinity of filaments. The density field around each filament realization is evaluated on a lattice around the stretching center of the filament. They are computed by means of the high-resolution self-adaptive phase-space Delaunaty Tessleation Field Estimator formalism (\cite{Feldbrugge:2024} see appendix~\ref{ap:density} for an outline). 

For each set of $1000$ filament simulation realizations we determine the median density, as a function of the specified smoothing scales and formation times. The median density profile is preferred over the mean density profile, as the latter tends to be dominated by the caustics where the density spikes to infinity. The median density is insensitive to these outliers. In addition, the median also removes effects from the filamentary cluster extensions. Figure~\ref{fig:Median_CoF} presents the median filament mass distribution in a matrix of 12 panels, with formation time running along the horizontal direction -- for $b_c=0.3,0.5,0.8,1.0$ -- and with smoothing scale running along the vertical direction -- for $\sigma=0.1,0.5,1.0$. 

The median filamentary mass distribution has an oval shape extending along the vertical direction. Going from the lefthand panels to the right, \textit{i.e.}, as we proceed from filaments forming early to ones forming later, we see that filaments that form later are thinner and have a concentrated density spike at the center. We may understand this as filaments that form at early times have more time to develop and grow into a denser structure phase-mixed structure. Proceeding from small-scale to large-scale filaments, \textit{i.e.}, from the top to the bottom panels, we find that larger smoothing lengths $\sigma$ entail thicker, more massive filaments. Moreover, on average they define more extended structures along the vertical direction. 

The density profiles along the width of the filaments are presented in figure~\ref{fig:prof}. These underline the impression given by the median filament mass distribution. The filaments that form at earlier times have more time to phase mix, and produce thicker objects that have a density plateau, followed by a rapid decline in density. By contrast, filaments that form at a later time end up with a more peaked density profile.

\subsubsection{Filament Multi-stream Character}
The PS-DTFE formalism also allows us to determine the local streaming structure, \textit{i.e.}, the local number of streams. We find that in the majority of realizations, the central stretching region of the filament is situated in a $3$- or $5$- and even $7$-stream region. For three different values of smoothing length -- $\sigma=0.1,0.5,1.0$ -- four bar histograms depict the distribution of a number of streams for four different filament formation times.

Visual inspection indicates that the configurations with $3$, $5$ or $7$ streams are filaments. For late formation times and small length scales, we find that a significant fraction of realizations entails filaments whose stretching center resides in a $1$-stream region. These configurations tend to be similar to the upper left panel of fig.\ \ref{fig:realizations}, where the filament is about to shell-cross, but has not yet reached that phase. In these instances, the Zel'dovich approximation underpredicts the first shell-crossing time. By contrast, visual inspection reveals that regions with more than $9$ streams tend to correspond to filaments whose stretching center resides in regions that would classify as the extension of a cluster (cf. the lower left panel of fig.~\ref{fig:realizations}).

\section{Alternative filament conditions}\label{sec:alternatives}
In addition to the physical definition for the nature of cosmic filaments forwarded in the present study, there is a range of different definitions in the literature for what constitutes a filament. Here we compare the caustic definition of filaments with that of the two most common definitions. Both entail singularities in the matter distribution, be it singularities in the Eulerian primordial matter distribution.

The alternative filament definitions that we consider here are of an explicit topological nature, involving the connections between singularities in the mass distribution. They are based on the observation that filaments are often the bridges between clusters in the matter and galaxy distribution. Mathematically speaking, the contention is that clusters are maxima in the density field or, following a stronger assumption, that they emanate from maxima in the primordial density field. Within this context, cosmic voids descend from the minima in the primordial density field. Dynamically even more relevant is the identification of the minima in the primordial gravitational potential with the dominant mass concentrations in the Universe, and the maxima of the primordial gravitational potential with the gravitationally dominant voids in the large-scale matter distribution. A major issue within this view remains the determination of the relation between the scale of primordial singularities and that of the resulting structures in the matter distribution, in particular in view of the hierarchical buildup of structure \citep[see \textit{e.g.}][]{Weygaert:2008,Aragon:2024}.

Following the identification of clusters with the maxima in the density field, or with the minima in the gravitational potential field, filaments are seen as the bridges that join the clusters. Mathematically they are then defined to be the integral lines running between the clusters, passing through the corresponding saddle point(s) of the density or potential field. The eigenvectors of the Hessian of the primordial density field, or the Hessian of the gravitational potential, at the saddle point, determine the orientation of the proto-filament. Arguably the most widely known version of this view concerns the density field. The cosmic skeleton model \citep{Pogosyan:2009a,Codis:2018b} analyzes this mathematical model of the cosmic web in considerable detail, while it forms the incentive for the Disperse formalism for cosmic web classification and identification \citep{Sousbie:2011a,Sousbie:2011b}. A similar identification of filaments with the saddles in the gravitational potential field had been forwarded in the earlier studies of the linear constrained gravitational field realizations \citep{Haarlem:1993,Weygaert:1996}. 

These alternative filament definitions share a clear geometric and topological similarity, linking the central regions of the filaments with saddle points. However, they operate on different levels. Whereas the saddle point of the primordial gravitational potential is governed by the first and second derivatives of the primordial gravitational potential (or displacement potential), the saddle points of the density field follow from the third- and fourth-order derivatives of the density field. The caustic skeleton definition proposed in this paper is most closely linked with the former definition, as it is defined in terms of the same eigenvalues and eigenvectors of the Hessian of the gravitational field.

Nonetheless, there are some major conceptual differences between the caustic skeleton description and the specification in terms of saddle points. Most important is that the saddle point definitions do not allow to set the formation time. As such, they do not allow the differentiation of filaments with different formation times and hence lack the ability to describe the hierarchical buildup of the cosmic web. This also implies a problem of a more practical nature, that of the inability to guarantee that the proposed proto-filaments have formed by the current epoch. 

Below we treat the constrained formalism to set up constrained filament realization for the two alternatives of a filament specified on the basis of the saddle point location and orientation in the primordial Gaussian gravitational potential field, and that in the primordial Gaussian density field.

\begin{figure*}
    \centering
    \begin{subfigure}[b]{0.32\textwidth}
        \includegraphics[width=\textwidth]{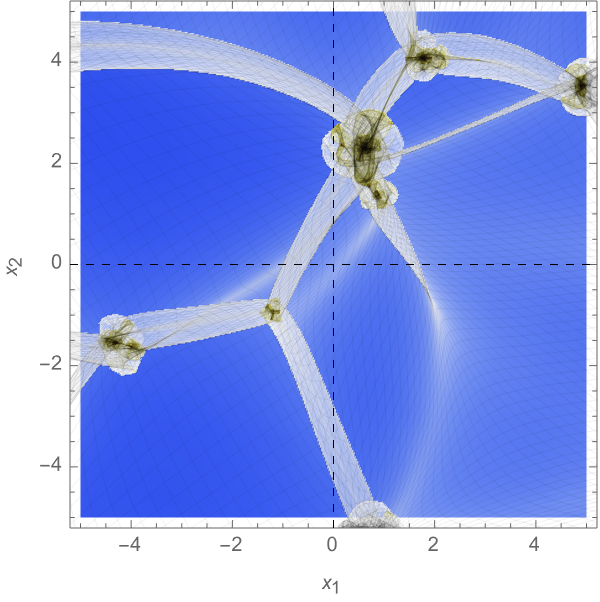}
    \end{subfigure}
    ~ 
    \begin{subfigure}[b]{0.32\textwidth}
        \includegraphics[width=\textwidth]{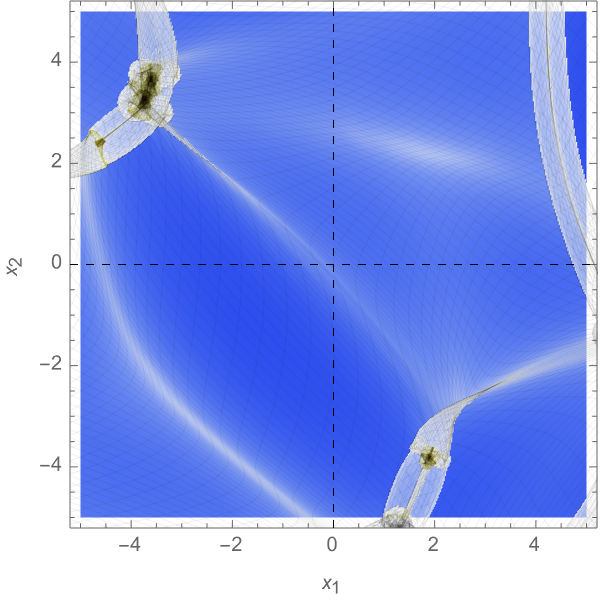}
    \end{subfigure}
    ~
    \begin{subfigure}[b]{0.32\textwidth}
        \includegraphics[width=\textwidth]{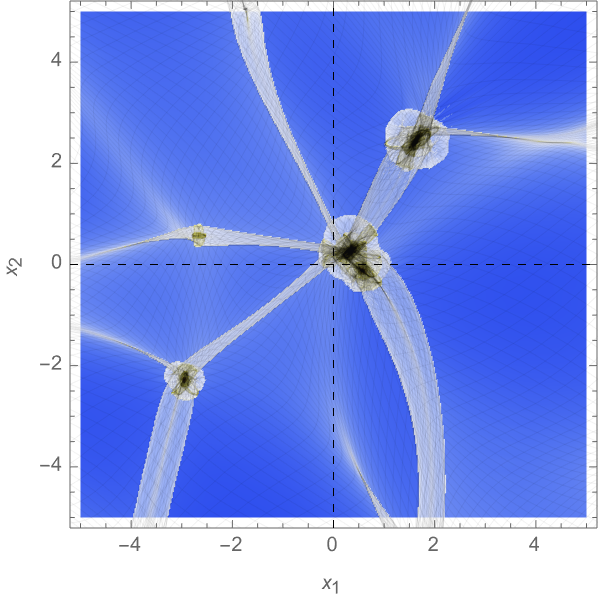}
    \end{subfigure}
    \caption{Realizations of saddle points in the smoothed primordial density perturbation at the scale $\sigma=0.5$. We plot the $N$-body particles and the initial mesh on the corresponding density field $\log(\rho + 1)$.} \label{fig:density_realizations}
    \begin{subfigure}[b]{0.32\textwidth}
        \includegraphics[width=\textwidth]{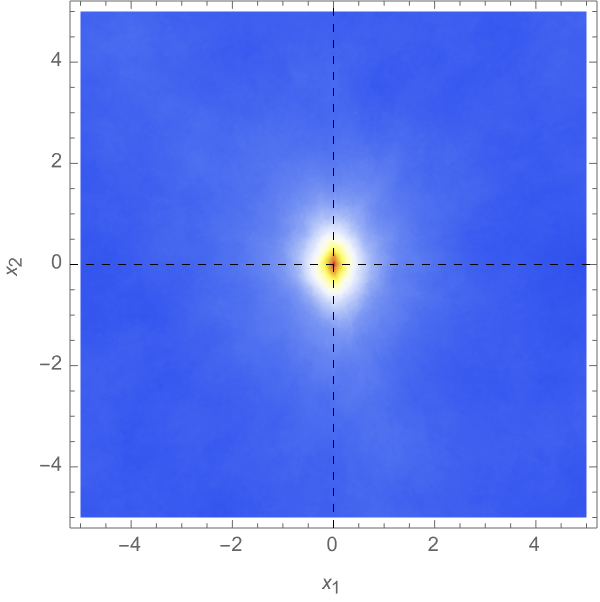}
    \end{subfigure}
    \begin{subfigure}[b]{0.32\textwidth}
        \includegraphics[width=\textwidth]{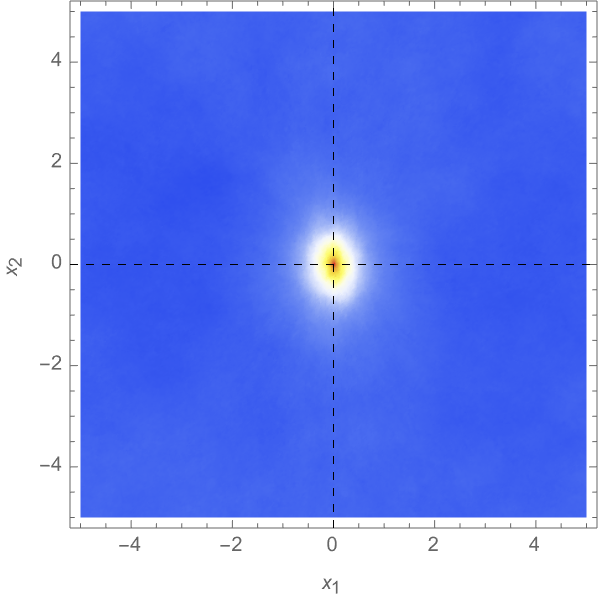}
    \end{subfigure}
    \begin{subfigure}[b]{0.32\textwidth}
        \includegraphics[width=\textwidth]{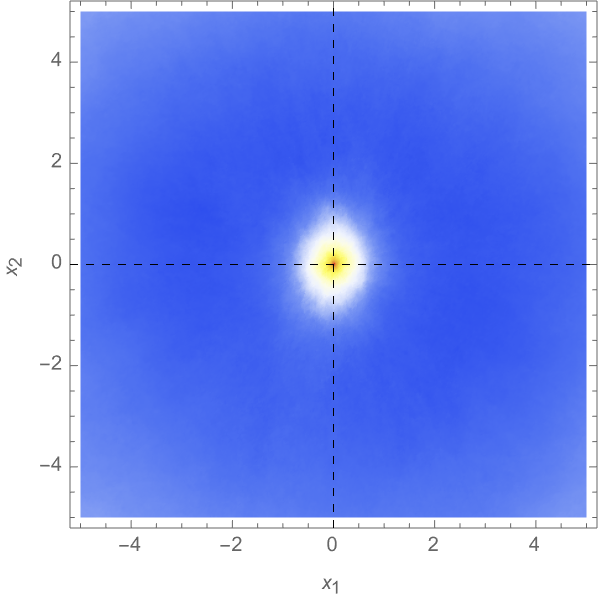}
    \end{subfigure}
    \caption{The median density field of the realizations of the saddle points of the primordial density perturbation as a function of the length scale $\sigma=0.1,0.5,1.0$ from left to right. We plot the median density field on a logarithmic scale $\log(\rho_{median} + 1)$.} \label{fig:Median_density}
\end{figure*}
\begin{figure}
    \centering
    \includegraphics[width=\linewidth]{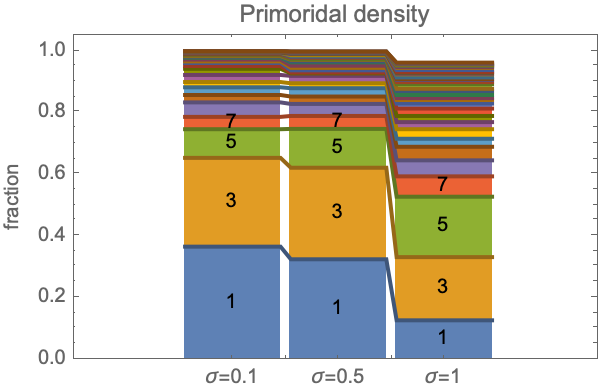}
    \caption{The distribution of the number of streams as a function of the formation time for the length scale $\sigma=0.1, 0.5, 1.0$ (left to right)} \label{fig:streams_density}
\end{figure}

\subsection{Saddle points of the primordial density}
One of the most widely used methods for describing and/or analyzing the cosmic web center around the assumption that filaments are ridges in the cosmic density field, connecting clusters via the saddle points in the density field. A variety of geometric and topological methods are based on this premise for the identification of structure at the current epoch, such as main versions of the MMF/Nexus algorithm \citep{Aragon:2007,Cautun:2013} and the Disperse topological formalism \citep{Sousbie:2011a,Sousbie:2011b}. The related cosmic skeleton model emphasizes the central role of saddles in the primordial density field as sites around which filaments form \citep{Pogosyan:2009a,Codis:2018b}.

While this seems a reasonable assumption for any one moment in time, given the hierarchical buildup of structure and corresponding multiscale nature of the cosmic web it may not have the universal validity it might yield at first sight. Cluster nodes and filamentary arteries may have evolved from proto-structures at a variety of scales, so it may be far from straightforward to identify saddles in the primordial density field with filaments in the current cosmic mass distribution. In other words, while it may be useful for the analysis of the structural pattern at a given cosmic epoch, it remains to be assessed in how far it has any validity for considering the dynamical evolution of the cosmic web.

In this section, we investigate, using constrained realizations, how far saddle points in the primordial density field are related to the emerging filaments in the cosmic web. 

\subsubsection{Saddle Point Definition}
For the primordial density perturbation field smoothed by a Gaussian kernel at length scale $\sigma$, the density perturbation can be written in terms of the displacement potential using the Poisson equation
\begin{equation}
  \delta_\sigma(\bm{q}_c) = \nabla^2\Psi_\sigma(\bm{q}_c) = T_{11} + T_{22}\,.
\end{equation}
Imposing singularity conditions on the density field thus involves constraints on the third-order and fourth-order gradients of the potential field: the point $\bm{q}_c$ is a critical point of the primordial density perturbation $\delta_\sigma$ when its gradient vanishes, \textit{i.e.},
\begin{align}
    \nabla \delta_\sigma(\bm{q}_c) &= (T_{111} + T_{122}, T_{112} + T_{222}) = \bm{0}\,.
\end{align}
It is a saddle point when the determinant of the Hessian
\begin{align}
    \mathcal{H}\delta_\sigma(\bm{q}_c) &= \begin{pmatrix}
        T_{1111} + T_{1122} & T_{1112} + T_{1222}\\
        T_{1112} + T_{1222} & T_{1122} + T_{2222}
    \end{pmatrix}
\end{align}
assumes a negative value (the Hessian has one positive and one negative eigenvalue). For the constrained realizations that we generate for our analysis, we orient the associated filament by requiring that the eigenvector of the Hessian $\mathcal{H}\delta_\sigma(\bm{q}_c)$ associated with the positive eigenvalue is aligned with the vertical direction.

\subsubsection{Implementation}
For the generation of the density field-constrained conditions for a proto-filament, the implementation steps concern the generation of properly sampled third-order and fourth-order gradients of the deformation potential. To this end, we use the fact that the third-order and fourth-order gradients and independently distributed Gaussian variables. Hence, we sample them independently in two subsequent steps. 
\begin{enumerate}
    \item[1.] The third-order derivatives set up the space $\bm{Y}_3=(T_{111}, T_{112}, T_{122}, T_{222})$, and obey the linear constraints
    \begin{align}
    T_{111} + T_{122} &= 0 \,, \nonumber\\
    T_{112} + T_{222} &= 0\,.
    \end{align}

    To exploit the linear constraints on the third-order derivates, we consider the third-order statistic $Y=(T_{111} + T_{122}, T_{112} +T_{222},T_{111} - T_{122}, T_{112} - T_{222})$, which is normally distributed $\mathcal{N}(\bm{0},M)$ with covariance matrix
    \begin{align}
        M = \frac{\sigma_3^2}{4}
        \begin{pmatrix}
            2 &  0 & 1 &  0\\
            0 &  2 & 0 & -1\\
            1 &  0 & 1 &  0\\
            0 & -1 & 0 &  1
        \end{pmatrix}.
    \end{align}
    \item[2.] Given the constraints $T_{111}+T_{122} = T_{112} + T_{222} = 0$, the stochastic variables $T_{111}-T_{122}$ and $T_{112}-T_{222}$ are independent and have an identical normal distribution $\mathcal{N}(0,\sigma_2^2/8)$. Sampling two variables $A$ and $B$ from this normal distribution, a representative sample of the third-order derivatives $T_{ijk}$ is obtained via the relation,  
    \begin{align}
        T_{111} &= A/2\,,\nonumber\\
        T_{122} &= -A/2\,,\nonumber\\
        T_{112} &= B/2 \,,\\
        T_{222} &= -B/2 \,.\nonumber
    \end{align}
    \item[3.] We sample the fourth-order derivatives space $\bm{Y}_4=(T_{1111},\dots,T_{2222})$, which is normally distribution $\mathcal{N}(\bm{0},M_{4,4})$, with zero mean and covariance matrix $M_{4,4}$,
    \begin{align}
        M_{4,4} 
        = \frac{\sigma_4^2}{8}\begin{pmatrix}
            35 & 0 & 5 & 0 & 3 \\
            0 & 5 & 0 & 3 & 0 \\
            5 & 0 & 3 & 0 & 5 \\
            0 & 3 & 0 & 5 & 0 \\
            3 & 0 & 5 & 0 & 35 \\
        \end{pmatrix}.
    \end{align}
    \item[4.] Having sampled the fourth-order derivatives, we evaluate the Hessian $\mathcal{H} \delta_\sigma(\bm{q}_c)$ and reject samples for which its determinant assumes a positive value.
    \item[5.] For the positive eigenvalue of the Hessian $\mathcal{H}\delta_\sigma$, the eigenvector is rotated to the vertical direction, so that the third- and fourth-order derivatives are aligned with the vertical direction. This is accomplished following the transforms derived in appendix E of \cite{Feldbrugge:2023a}.
    \item[6.] Subsequently, we generate a realization of the constrained field with the Hoffman-Ribak algorithm for Gaussian random fields fulfilling a set of linear constraints \citep{Bertschinger:1987,Hoffman:1991,Weygaert:1996}.
    \item[7.] Finally, for each sampled initial condition, two dark-matter-only 2D $N$-body simulations are run \citep{Johan:2020}. The 2D simulations gave $256\times 256$ particles in a $25 R_s \times 25 R_s$ box in an Einstein-de Sitter cosmology. The first simulation evolves the sampled initial conditions. The second simulation evolves a smoothed version of the initial conditions with a Gaussian kernel \eqref{eq:kernel} and the smoothing length $\sigma$ at which the constraints are set. The $N$-body simulation without the smoothing is centered and aligned using the $N$-body simulation with the smoothed initial conditions. 
\end{enumerate}

\begin{figure*}
    \centering
    \begin{subfigure}[b]{0.32\textwidth}
        \includegraphics[width=\textwidth]{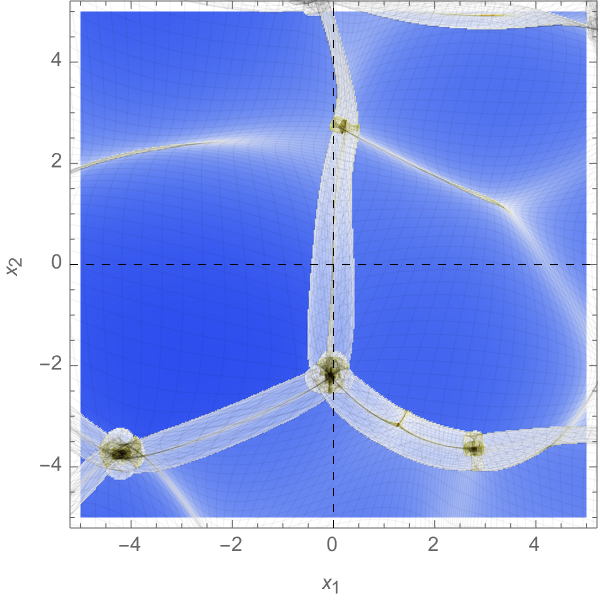}
    \end{subfigure}
    ~ 
    \begin{subfigure}[b]{0.32\textwidth}
        \includegraphics[width=\textwidth]{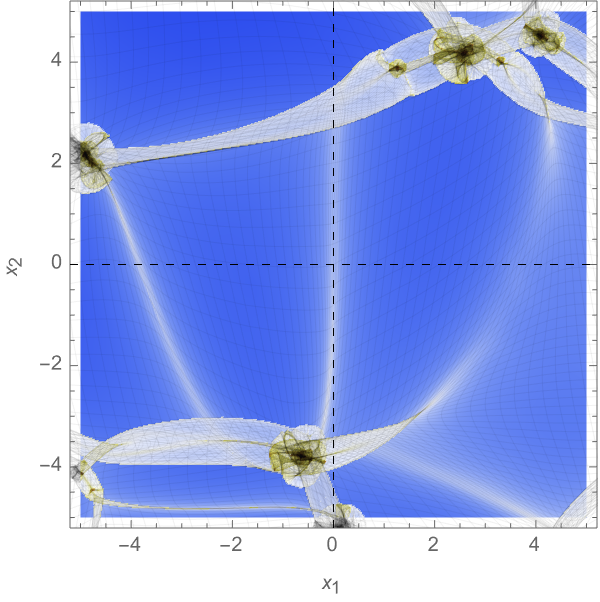}
    \end{subfigure}
    ~
    \begin{subfigure}[b]{0.32\textwidth}
        \includegraphics[width=\textwidth]{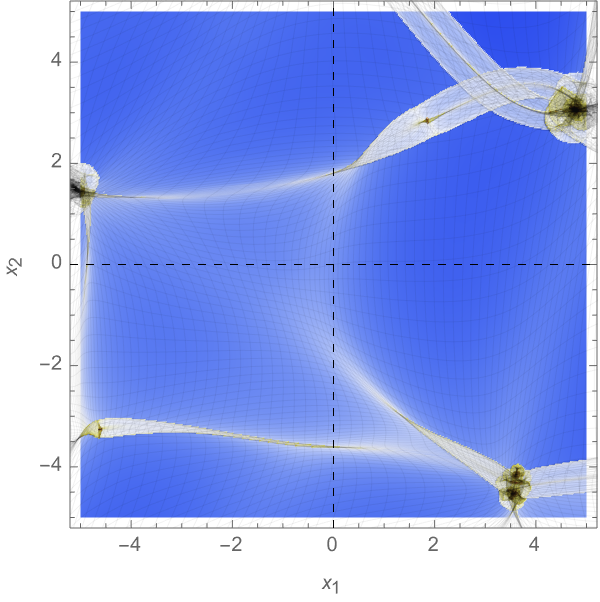}
    \end{subfigure}
    \caption{Realizations of saddle points in the smoothed primordial gravitational potential at the scale $\sigma=0.5$. We plot the $N$-body particles and the initial mesh on the corresponding density field $\log(\rho + 1)$.} \label{fig:deformation_realizations} 
    \begin{subfigure}[b]{0.32\textwidth}
        \includegraphics[width=\textwidth]{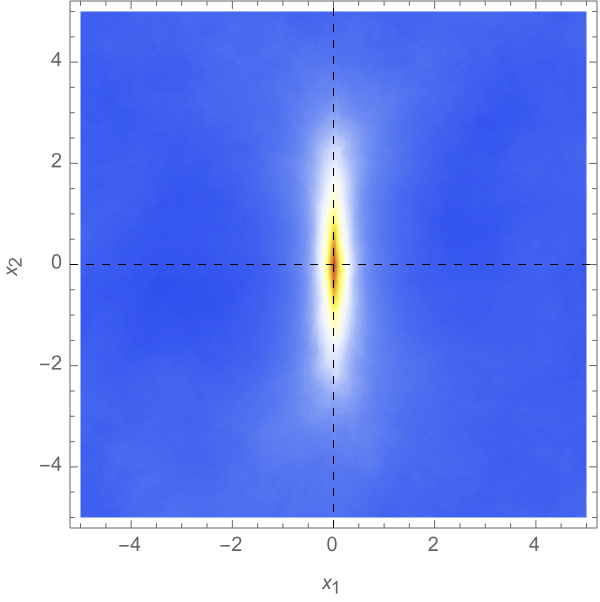}
    \end{subfigure}
    \begin{subfigure}[b]{0.32\textwidth}
        \includegraphics[width=\textwidth]{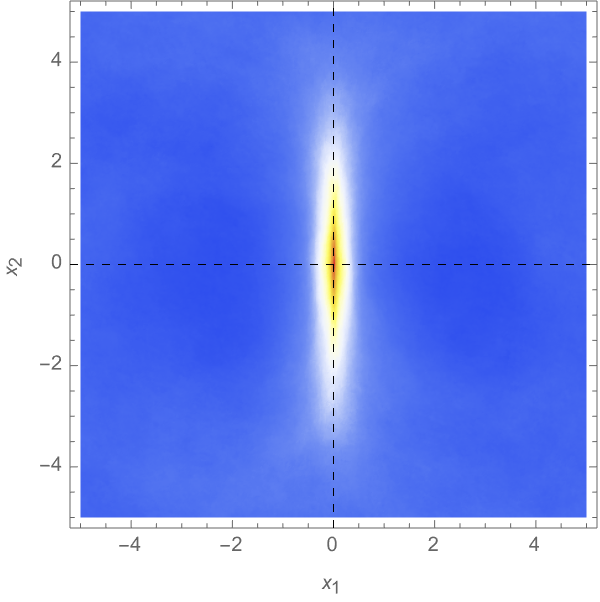}
    \end{subfigure}
    \begin{subfigure}[b]{0.32\textwidth}
        \includegraphics[width=\textwidth]{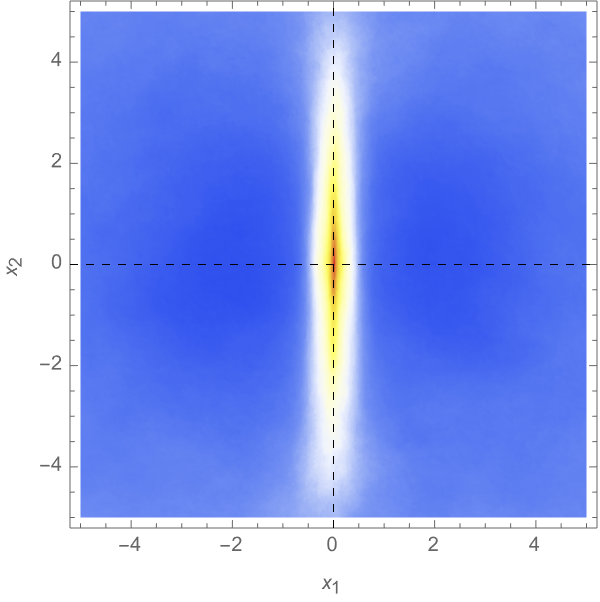}
    \end{subfigure}
    \caption{The median density field of the realizations of the saddle points of the primordial potential field as a function of the length scale $\sigma=0.1,0.5,1.0$ from left to right. We plot the median density field on a logarithmic scale $\log(\rho_{median} + 1)$.} \label{fig:Median_Phi}
\end{figure*}
\begin{figure}
    \centering
    \includegraphics[width=\linewidth]{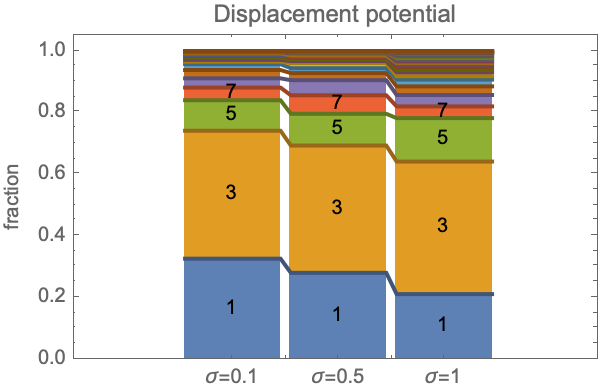}
    \caption{The distribution of the number of streams as a function of the formation time for the length scale $\sigma=0.1, 0.5, 1.0$ (left to right)} \label{fig:streams_displacement}
\end{figure}

\subsubsection{Realizations}
A random set of realizations of filaments implied by the imposed primordial density field saddle constraints is presented in figure~\ref{fig:density_realizations}.

The figure panels clearly reveal the existence of a correlation between the non-linear structure of the cosmic web and the saddle points of the primordial density field. Nonetheless, the correlation is not as strong as in the case of the caustic skeleton filament conditions. In the next subsection, we will see it is also considerably weaker than the filament constraints emanating from saddle conditions in the primordial potential field.

The density field saddle conditions yield configurations that may not even resemble that of filaments. In some cases, the saddle point ends up in a cluster (righthand panels fig.~\ref{fig:density_realizations}). In other cases, it may reside in a region that might develop into a filament in the future, but would at the current epoch most likely be classified as a void (lefthandand central panels). A major factor is that the density field saddle constraints do not specify the formation time so the time at which the filament emerges is left a free parameter. It may also be noticed that the emerging structures do not reveal any of the intended clear alignment with the vertical axis.

To assess the statistical properties of the proto-filaments implied by the primordial density field constraints, we study the corresponding properties of an ensemble consisting of $1000$ $N$-body simulations. This is accomplished for the length scale $\sigma = 0.1, 0.5$ and $1.0$. When assessing the number of streams in the Eulerian region in which the saddle resides, we find a quantitative confirmation of the impressions obtained from figure~\ref{fig:density_realizations}. For a large fraction of the simulations, the saddle point resides in a single stream region. Visual inspection indicates that these are largely void regions. When the saddle point resides in a $3$- or $5$-stream region, we often observe a filamentary structure. Nonetheless, only rarely this structure is aligned with the intended vertical axis. This is emphasized by the median density profiles of the realizations, shown in figure~\ref{fig:Median_density}. Reflecting the fact that the orientation of the proto-filaments is not really constrained, and in fact has a random isotropic distribution, is a median density field that has an approximate rotational symmetry.

\subsection{Saddle points primordial gravitational potential}
From a dynamical perspective, one may assume that the minima of the primordial gravitational potential correspond to regions that are due to gravitational collapse and form clusters, while the maxima will evolve into voidlike regions. From this dynamical perspective, proto-filaments are the ridges in the primordial gravitational potential landscape, centered around the saddle point of the (displacement) potential field $\Psi$ \citep[see \textit{e.g.}][]{Haarlem:1993,Weygaert:1996}. The eigenvectors of its Hessian, $\mathcal{H} \Psi$, determine the orientation of the proto-filament.

Mathematically, this translates into the following conditions with respect to the saddle points in the smoothed primordial gravitational potential. For these, the following conditions hold:
\begin{itemize}
    \item[1.] the gradient of the displacement potential vanishes 
    \begin{align}
        \nabla \Psi_\sigma(\bm{q}_c) = (T_1,T_2) =\bm{0}\,,
    \end{align}
    \item[2.] the determinant of the Hessian
    \begin{align}
        \mathcal{H}\Psi_\sigma(\bm{q}_c) = \begin{pmatrix}
            T_{11} & T_{12} \\
            T_{12} & T_{22}
        \end{pmatrix}
    \end{align}
    assumes a negative value (having one positive and one negative eigenvalue).
\end{itemize}
We orient the proto-filament such that the eigenvector corresponding to the positive eigenvalue of the Hessian $\mathcal{H}\Psi_\sigma(\bm{q}_c)$ is aligned vertically.

\subsubsection{Implementation}
Following the realization that the first-order derivatives $T_i$ and second-order derivates $T_{jk}$ of a Gaussian random field themselves also define a Gaussian random field, and the first-order and second-order derivatives are independently distributed, we set up the following procedure for setting up a proto-filament:
\begin{enumerate}
    \item[1.] The first step consists of sampling the second-order derivatives from the normal distribution $\mathcal{N}(\bm{0},M_{2,2})$. The covariance matrix $M_{2,2}$ is given by equation \eqref{eq:M22}.
    \item[2.] Evaluate the Hessian $\mathcal{H}\Psi_\sigma(\bm{q}_c)$, rejecting sampled Hessians for which the determinant assumes a positive value.
    \item[3.] Evaluate the eigenvector of the Hessian corresponding to the positive eigenvalue. Align the eigenvector with the vertical direction, and rotate the first and second-order derivatives accordingly (following the transforms derived in appendix E of \cite{Feldbrugge:2023a}).
    \item[4.] Generate a realization of the initial Gaussian random field using the Hoffman-Ribak algorithm for Gaussian random field generation with linear constraints \citep{Bertschinger:1987,Hoffman:1991,Weygaert:1996}.
    \item[5.] For each sampled initial field realization, two dark-matter only 2D $N$-body simulations \citep{Johan:2020} are run with $256\times 256$ particles in a $25 R_s \times 25 R_s$ box, within an Einstein-de Sitter cosmology setting. The regular simulation evolves the sampled initial conditions. The second simulation concerns the evolution of a smoothed version of the initial conditions, with a Gaussian kernel \eqref{eq:kernel} of smoothing length $\sigma$, the scale of the imposed constraints. The smoothed $N$-body simulation is used to center and align the regular simulation. 
\end{enumerate}

For this class of potential field saddle point configurations, we generated an ensemble of $1000$ simulations for length scales $\sigma=0.1,0.5,1$. 

\subsubsection{Realizations}
The filament emanating from the primordial gravitational potential criterion appears to represent a more genuine filamentary structure than the one resulting from the primordial density field. This is clearly borne out by the 3 filament realizations shown in figure~\ref{fig:deformation_realizations}. In all cases, we recognize clear filamentary structures aligned along the vertical axis. Nonetheless, there exists a far larger variation than in the case of the caustic skeleton filaments: some of the filaments are fully formed, while other structures still reside in an embryonic stage and will only form later.

The median density field of the filamentary realizations, shown in figure~\ref{fig:Median_Phi}, is that of an oval-shaped configuration aligned with the vertical axis. Not surprisingly, the length -- and width -- of the median filament increases with the length scale $\sigma$, before widening at its tips. 

The visual impression yielded by figure~\ref{fig:deformation_realizations} and the median density fields are supported by the analysis of the multi-stream character of the configurations. In a significant fraction of the realizations, the saddle points lie in a single-stream region. This tends to be either a filament that has not yet formed or one located in a void region. Overall, the $3$-, $5$- and $7$-stream regions appear to represent a wide variety of vertically oriented filaments. Occasionally, they may even end up in a cluster in a cluster in Eulerian space. 

\section{The three-dimensional cosmic web}\label{sec:3d}
The present paper studies the formation of filaments in a two-dimensional model of large structure formation. We propose a non-linear condition on the primordial initial conditions that identifies the proto-filaments forming at a set time and length scale in the cosmic web. In an upcoming paper, we will extend this formalism to the three-dimensional cosmic web consisting of voids, walls, filaments and clusters.

While the details of the three-dimensional caustic skeleton conditions are described in the upcoming paper, here we briefly summarize how the two-dimensional analysis generalizes to the three-dimensional setting. 


In the three-dimensional cosmic web, the cusp caustic is associated with the walls in the matter distribution. It is described by the manifold,
\begin{align}
    A_3 = \{ \bm{q} \in L\,|\, \lambda_1(\bm{q}) \geq 1/b_+(t_0)\,,\  \bm{v}_1(\bm{q}) \cdot \nabla \lambda_1(\bm{q}) = 0\}\,.
\end{align}
Like the filaments of the two-dimensional cosmic web, it is natural to define the center of the wall as a point on the wall that is always being stretched, and -- in the Zel'dovich approximation -- is identified with the point of maximum local stretch.

Formally, in Lagrangian space, the two-dimensional tangent space $T_{\bm{q}_c}A_3$ of the cusp manifold is spanned by the orthogonal basis $\bm{T}_1$ and $\bm{T}_2$. In Eulerian space, the stretching or contraction of the wall is governed by the outer product $\|(\nabla \bm{x}_t \bm{T}_1) \times (\nabla \bm{x}_t \bm{T}_2)\|$.

In the three-dimensional cosmic web, the caustic skeleton entails the existence of two classes of filaments. The first one of these is the \textit{swallowtail filaments}, 
\begin{align}
    A_4 = \{ &\bm{q} \in L\,|\, \lambda_1(\bm{q}) \geq 1/b_+(t_0) \,,\  \bm{v}_1(\bm{q}) \cdot \nabla \lambda_1(\bm{q}) = 0\,\ \nonumber\\
    &\bm{v}_1(\bm{q}) \cdot \nabla(\bm{v}_1(\bm{q}) \cdot \nabla \lambda_1(\bm{q})) = 0\}\,,
\end{align}
the second class those of the \textit{umbilic filaments}, 
\begin{align}
    D_4 = \{ &\bm{q} \in L\,|\, \lambda_1(\bm{q}) = \lambda_2(\bm{q}) \geq 1/b_+(t_0)\}\,.
\end{align}
By determining the typical orientation of the filament $\bm{T}$, the center of the filament is the point that is always stretching -- according to the Zel'dovich approximation -- is the location of maximum stretch. Its location follows from the evaluation of the stretching rate $\|\nabla \bm{x}_t \bm{T}\|$.

\section{Summary and Conclusions}\label{sec:summary}
We propose an unambiguous definition of filaments based on their formation histories rather than their morphology, in the two-dimensional setting. 

The cosmic web can be characterized by its multi-stream nature. Mass elements that have undergone shell-crossing form multi-stream regions where non-linear gravitational collapse takes hold. These multi-stream regions result from the folding of the dark matter sheet in phase-space and can be identified in terms of caustics, famously classified by catastrophe theory. The two-dimensional filament was previously linked to the cusp caustic \citep{Arnold:1982a, Arnold:1982b, Hidding:2014}. For the Zel'dovich approximation, this cusp caustic is expressed in terms of the first eigenvalue field of the primordial tidal tensor and its derivative in the direction of the corresponding eigenvector field. Although the cusp curve neatly follows the filamentary structure of the cosmic web, we show that it by itself does not suffice to identify proto-filaments, as the cusp curve stretches in the filament regions and tangles up in the cluster regions where the Zel'dovich approximation fails to capture structure formation. For this reason, we augment the cusp condition to find the centre of the filament, where the cusp curve is maximally stretched. These additional conditions are expressed in terms of the second eigenvalue field and its derivative in the direction of its corresponding eigenvector field. This yields a family of proto-filaments specified in terms of their orientation, formation time, and length scale.

Using constrained Gaussian random field theory \citep{Feldbrugge:2023b} and dark matter only $N$-body simulations \citep{Johan:2020}, we demonstrate that this filament condition successfully identifies proto-filaments. We compare the structures with previously proposed proto-filament based on the saddle points of the primordial density and gravitational potential. The saddle point of the gravitational potential does a better job than the saddle points of the density perturbation. As it turns out, the saddle point of the primordial density field is correlated to the filaments but the eigenvectors of the Hessian are not a reliable indication of the orientation of the proto-filament. The orientation of the saddle point of the primordial gravitational potential leads to a filamentary structure, however, it does not identify the time at which it is likely to form. The filament definition in terms of the eigenvalue fields performs better than the definition in terms of the gravitational potential. Note that the here proposed filament condition includes the formation time. 

We believe that this unambiguous definition will aid the systematic study of filaments in the cosmic web and their relation to the embedded galaxies and baryonic matter. In the future, we plan to use this definition to study the recently identified filament spin \citep{Wang:2021,Xia:2021}, galaxy alignment \citep{Porciani:2002a,Aragon:2007a,Hahn:2007,Aragon:2007c,Schaefer:2009,Hahn:2009,Tempel:2013,Codis:2012,Codis:2015,Codis:2018a,Ganeshaiah:2018,Ganeshaiah:2019,Lopez:2021,Lopez:2024}, and to the question in how far the mass, stellar content and star formation activity are influenced \citep{Cautun:2014,Kraljic:2018, Kuutma:2017,Laigle:2018,Hellwing:2020,Parente:2024}. 

The present discussion focuses on two-dimensional models of large-scale structure formation, as this enables clear visualization. The work will be extended to three-dimensional structure formation in a sequel paper. Moreover, we focused on the caustic skeleton of the Zel'dovich approximation. In an upcoming paper, we will extend the analysis to higher-order Lagrangian perturbation theory.

\section*{Acknowledgements}
We thank Johan Hidding for providing the two-dimensional $N$-body simulation code used in this project. We thank Benjamin Hertzsch Ma\'e Rodriguez for conversations about the physics underlying cosmic structure formation. We are also grateful to Nynke Niezink for many discussions and for proofreading this manuscript and for an incisive manuscript comment to Lucia van de Weijgaert.  

The work is supported by the STFC Consolidated Grant ‘Particle Physics at the Higgs Centre’ and by a Higgs Fellowship at the University of Edinburgh.

\section*{Data Availability}
No new data were generated or analyzed in support of this research.

\bibliographystyle{mnras}
\bibliography{Library} 

\begin{thebibliography}{}
\makeatletter
\relax
\def\mn@urlcharsother{\let\do\@makeother \do\$\do\&\do\#\do\^\do\_\do\%\do\~}
\def\mn@doi{\begingroup\mn@urlcharsother \@ifnextchar [ {\mn@doi@} {\mn@doi@[]}}
\def\mn@doi@[#1]#2{\def\@tempa{#1}\ifx\@tempa\@empty \href {http://dx.doi.org/#2} {doi:#2}\else \href {http://dx.doi.org/#2} {#1}\fi \endgroup}
\def\mn@eprint#1#2{\mn@eprint@#1:#2::\@nil}
\def\mn@eprint@arXiv#1{\href {http://arxiv.org/abs/#1} {{\tt arXiv:#1}}}
\def\mn@eprint@dblp#1{\href {http://dblp.uni-trier.de/rec/bibtex/#1.xml} {dblp:#1}}
\def\mn@eprint@#1:#2:#3:#4\@nil{\def\@tempa {#1}\def\@tempb {#2}\def\@tempc {#3}\ifx \@tempc \@empty \let \@tempc \@tempb \let \@tempb \@tempa \fi \ifx \@tempb \@empty \def\@tempb {arXiv}\fi \@ifundefined {mn@eprint@\@tempb}{\@tempb:\@tempc}{\expandafter \expandafter \csname mn@eprint@\@tempb\endcsname \expandafter{\@tempc}}}

\bibitem[\protect\citeauthoryear{{Abel}, {Hahn}  \& {Kaehler}}{{Abel} et~al.}{2012}]{Abel:2012}
{Abel} T.,  {Hahn} O.,   {Kaehler} R.,  2012, \mn@doi [\mnras] {10.1111/j.1365-2966.2012.21754.x}, \href {https://ui.adsabs.harvard.edu/abs/2012MNRAS.427...61A} {427, 61}

\bibitem[\protect\citeauthoryear{{Adler}}{{Adler}}{1981}]{Adler:1981}
{Adler} R.~J.,  1981, {The Geometry of Random Fields}.
Society for Industrial and Applied Mathematics

\bibitem[\protect\citeauthoryear{Adler \& Taylor}{Adler \& Taylor}{2009}]{Adler:2009}
Adler R.,  Taylor J.,  2009, Random Fields and Geometry.
Springer Monographs in Mathematics, Springer New York

\bibitem[\protect\citeauthoryear{{Alpaslan} et~al.,}{{Alpaslan} et~al.}{2014}]{Alpaslan:2014}
{Alpaslan} M.,  et~al., 2014, \mn@doi [\mnras] {10.1093/mnrasl/slu019}, \href {https://ui.adsabs.harvard.edu/abs/2014MNRAS.440L.106A} {440, L106}

\bibitem[\protect\citeauthoryear{{Arag{\'o}n Calvo}}{{Arag{\'o}n Calvo}}{2007}]{Aragon:2007c}
{Arag{\'o}n Calvo} M.~A.,  2007, PhD thesis, University of Groningen, Netherlands

\bibitem[\protect\citeauthoryear{{Aragon-Calvo}}{{Aragon-Calvo}}{2016}]{Aragon:2016}
{Aragon-Calvo} M.~A.,  2016, \mn@doi [\mnras] {10.1093/mnras/stv2301}, \href {https://ui.adsabs.harvard.edu/abs/2016MNRAS.455..438A} {455, 438}

\bibitem[\protect\citeauthoryear{{Aragon-Calvo}}{{Aragon-Calvo}}{2024}]{Aragon:2024}
{Aragon-Calvo} M.~A.,  2024, \mn@doi [\mnras] {10.1093/mnras/stae468}, \href {https://ui.adsabs.harvard.edu/abs/2024MNRAS.529...74A} {529, 74}

\bibitem[\protect\citeauthoryear{{Aragon-Calvo} \& {Szalay}}{{Aragon-Calvo} \& {Szalay}}{2013}]{Aragon:2013}
{Aragon-Calvo} M.~A.,  {Szalay} A.~S.,  2013, \mn@doi [\mnras] {10.1093/mnras/sts281}, \href {https://ui.adsabs.harvard.edu/abs/2013MNRAS.428.3409A} {428, 3409}

\bibitem[\protect\citeauthoryear{{Arag{\'o}n-Calvo}, {Jones}, {van de Weygaert}  \& {van der Hulst}}{{Arag{\'o}n-Calvo} et~al.}{2007a}]{Aragon:2007}
{Arag{\'o}n-Calvo} M.~A.,  {Jones} B.~J.~T.,  {van de Weygaert} R.,   {van der Hulst} J.~M.,  2007a, \mn@doi [\aap] {10.1051/0004-6361:20077880}, \href {https://ui.adsabs.harvard.edu/abs/2007A&A...474..315A} {474, 315}

\bibitem[\protect\citeauthoryear{{Arag{\'o}n-Calvo}, {van de Weygaert}, {Jones}  \& {van der Hulst}}{{Arag{\'o}n-Calvo} et~al.}{2007b}]{Aragon:2007a}
{Arag{\'o}n-Calvo} M.~A.,  {van de Weygaert} R.,  {Jones} B. J.~T.,   {van der Hulst} J.~M.,  2007b, \mn@doi [\apjl] {10.1086/511633}, \href {https://ui.adsabs.harvard.edu/abs/2007ApJ...655L...5A} {655, L5}

\bibitem[\protect\citeauthoryear{{Arag{\'o}n-Calvo}, {van de Weygaert}  \& {Jones}}{{Arag{\'o}n-Calvo} et~al.}{2010a}]{Aragon:2010a}
{Arag{\'o}n-Calvo} M.~A.,  {van de Weygaert} R.,   {Jones} B. J.~T.,  2010a, \mn@doi [\mnras] {10.1111/j.1365-2966.2010.17263.x}, \href {https://ui.adsabs.harvard.edu/abs/2010MNRAS.408.2163A} {408, 2163}

\bibitem[\protect\citeauthoryear{{Arag{\'o}n-Calvo}, {Platen}, {van de Weygaert}  \& {Szalay}}{{Arag{\'o}n-Calvo} et~al.}{2010b}]{Aragon:2010b}
{Arag{\'o}n-Calvo} M.~A.,  {Platen} E.,  {van de Weygaert} R.,   {Szalay} A.~S.,  2010b, \mn@doi [\apj] {10.1088/0004-637X/723/1/364}, \href {https://ui.adsabs.harvard.edu/abs/2010ApJ...723..364A} {723, 364}

\bibitem[\protect\citeauthoryear{{Aragon Calvo}, {Neyrinck}  \& {Silk}}{{Aragon Calvo} et~al.}{2019}]{Aragon:2019}
{Aragon Calvo} M.~A.,  {Neyrinck} M.~C.,   {Silk} J.,  2019, \mn@doi [The Open Journal of Astrophysics] {10.21105/astro.1697.07881}, \href {https://ui.adsabs.harvard.edu/abs/2019OJAp....2E...7A} {2, 7}

\bibitem[\protect\citeauthoryear{{Arnol'd}}{{Arnol'd}}{1982}]{Arnold:1982b}
{Arnol'd} V.~I.,  1982, Trudy Seminar imeni G Petrovskogo, \href {http://adsabs.harvard.edu/abs/1982TrPet...8...21A} {8, 21}

\bibitem[\protect\citeauthoryear{{Arnold}}{{Arnold}}{1984}]{Arnold:1984}
{Arnold} V.~I.,  1984, {Catastrophe theory}.
Springer Berlin, Heidelberg

\bibitem[\protect\citeauthoryear{{Arnol'd}, {Shandarin}  \& {Zel'dovich}}{{Arnol'd} et~al.}{1982}]{Arnold:1982a}
{Arnol'd} V.~I.,  {Shandarin} S.~F.,   {Zel'dovich} I.~B.,  1982, \mn@doi [Geophysical and Astrophysical Fluid Dynamics] {10.1080/03091928208209001}, \href {http://adsabs.harvard.edu/abs/1982GApFD..20..111A} {20, 111}

\bibitem[\protect\citeauthoryear{{Bardeen}, {Steinhardt}  \& {Turner}}{{Bardeen} et~al.}{1983}]{Bardeen:1983}
{Bardeen} J.~M.,  {Steinhardt} P.~J.,   {Turner} M.~S.,  1983, \mn@doi [\prd] {10.1103/PhysRevD.28.679}, \href {https://ui.adsabs.harvard.edu/abs/1983PhRvD..28..679B} {28, 679}

\bibitem[\protect\citeauthoryear{{Bardeen}, {Bond}, {Kaiser}  \& {Szalay}}{{Bardeen} et~al.}{1986}]{Bardeen:1986}
{Bardeen} J.~M.,  {Bond} J.~R.,  {Kaiser} N.,   {Szalay} A.~S.,  1986, \mn@doi [The Astrophysical Journal] {10.1086/164143}, \href {http://adsabs.harvard.edu/abs/1986ApJ...304...15B} {304, 15}

\bibitem[\protect\citeauthoryear{{Bermejo}, {Wilding}, {van de Weygaert}, {Jones}, {Vegter}  \& {Efstathiou}}{{Bermejo} et~al.}{2024}]{Bermejo:2024}
{Bermejo} R.,  {Wilding} G.,  {van de Weygaert} R.,  {Jones} B. J.~T.,  {Vegter} G.,   {Efstathiou} K.,  2024, \mn@doi [\mnras] {10.1093/mnras/stae543}, \href {https://ui.adsabs.harvard.edu/abs/2024MNRAS.529.4325B} {529, 4325}

\bibitem[\protect\citeauthoryear{{Bertschinger}}{{Bertschinger}}{1987}]{Bertschinger:1987}
{Bertschinger} E.,  1987, \mn@doi [\apjl] {10.1086/185066}, \href {https://ui.adsabs.harvard.edu/abs/1987ApJ...323L.103B} {323, L103}

\bibitem[\protect\citeauthoryear{{Bi}, {Shlosman}  \& {Romano-D{\'\i}az}}{{Bi} et~al.}{2024}]{Bi:2024}
{Bi} D.,  {Shlosman} I.,   {Romano-D{\'\i}az} E.,  2024, \mn@doi [\mnras] {10.1093/mnras/stad3942}, \href {https://ui.adsabs.harvard.edu/abs/2024MNRAS.52711095B} {527, 11095}

\bibitem[\protect\citeauthoryear{{Bond} \& {Myers}}{{Bond} \& {Myers}}{1996}]{BondMyers:1996}
{Bond} J.~R.,  {Myers} S.~T.,  1996, \mn@doi [\apjs] {10.1086/192267}, \href {https://ui.adsabs.harvard.edu/abs/1996ApJS..103....1B} {103, 1}

\bibitem[\protect\citeauthoryear{{Bond}, {Kofman}  \& {Pogosyan}}{{Bond} et~al.}{1996}]{Bond:1996}
{Bond} J.~R.,  {Kofman} L.,   {Pogosyan} D.,  1996, \mn@doi [\nat] {10.1038/380603a0}, \href {https://ui.adsabs.harvard.edu/abs/1996Natur.380..603B} {380, 603}

\bibitem[\protect\citeauthoryear{{Bonnaire}, {Aghanim}, {Kuruvilla}  \& {Decelle}}{{Bonnaire} et~al.}{2022}]{Bonnaire:2022}
{Bonnaire} T.,  {Aghanim} N.,  {Kuruvilla} J.,   {Decelle} A.,  2022, \mn@doi [\aap] {10.1051/0004-6361/202142852}, \href {https://ui.adsabs.harvard.edu/abs/2022A&A...661A.146B} {661, A146}

\bibitem[\protect\citeauthoryear{{Cadiou}, {Pontzen}, {Peiris}  \& {Lucie-Smith}}{{Cadiou} et~al.}{2021}]{Cadiou:2021}
{Cadiou} C.,  {Pontzen} A.,  {Peiris} H.~V.,   {Lucie-Smith} L.,  2021, \mn@doi [\mnras] {10.1093/mnras/stab2650}, \href {https://ui.adsabs.harvard.edu/abs/2021MNRAS.508.1189C} {508, 1189}

\bibitem[\protect\citeauthoryear{{Canducci} et~al.,}{{Canducci} et~al.}{2022}]{Awad:2022}
{Canducci} M.,  et~al., 2022, \mn@doi [Astronomy and Computing] {10.1016/j.ascom.2022.100658}, \href {https://ui.adsabs.harvard.edu/abs/2022A&C....4100658C} {41, 100658}

\bibitem[\protect\citeauthoryear{{Cautun}, {van de Weygaert}  \& {Jones}}{{Cautun} et~al.}{2013}]{Cautun:2013}
{Cautun} M.,  {van de Weygaert} R.,   {Jones} B. J.~T.,  2013, \mn@doi [\mnras] {10.1093/mnras/sts416}, \href {https://ui.adsabs.harvard.edu/abs/2013MNRAS.429.1286C} {429, 1286}

\bibitem[\protect\citeauthoryear{{Cautun}, {van de Weygaert}, {Jones}  \& {Frenk}}{{Cautun} et~al.}{2014}]{Cautun:2014}
{Cautun} M.,  {van de Weygaert} R.,  {Jones} B. J.~T.,   {Frenk} C.~S.,  2014, \mn@doi [\mnras] {10.1093/mnras/stu768}, \href {https://ui.adsabs.harvard.edu/abs/2014MNRAS.441.2923C} {441, 2923}

\bibitem[\protect\citeauthoryear{{Codis}, {Pichon}, {Devriendt}, {Slyz}, {Pogosyan}, {Dubois}  \& {Sousbie}}{{Codis} et~al.}{2012}]{Codis:2012}
{Codis} S.,  {Pichon} C.,  {Devriendt} J.,  {Slyz} A.,  {Pogosyan} D.,  {Dubois} Y.,   {Sousbie} T.,  2012, \mn@doi [\mnras] {10.1111/j.1365-2966.2012.21636.x}, \href {https://ui.adsabs.harvard.edu/abs/2012MNRAS.427.3320C} {427, 3320}

\bibitem[\protect\citeauthoryear{{Codis}, {Pichon}  \& {Pogosyan}}{{Codis} et~al.}{2015}]{Codis:2015}
{Codis} S.,  {Pichon} C.,   {Pogosyan} D.,  2015, \mn@doi [\mnras] {10.1093/mnras/stv1570}, \href {https://ui.adsabs.harvard.edu/abs/2015MNRAS.452.3369C} {452, 3369}

\bibitem[\protect\citeauthoryear{{Codis}, {Pogosyan}  \& {Pichon}}{{Codis} et~al.}{2018a}]{Codis:2018b}
{Codis} S.,  {Pogosyan} D.,   {Pichon} C.,  2018a, \mn@doi [\mnras] {10.1093/mnras/sty1643}, \href {https://ui.adsabs.harvard.edu/abs/2018MNRAS.479..973C} {479, 973}

\bibitem[\protect\citeauthoryear{{Codis}, {Jindal}, {Chisari}, {Vibert}, {Dubois}, {Pichon}  \& {Devriendt}}{{Codis} et~al.}{2018b}]{Codis:2018a}
{Codis} S.,  {Jindal} A.,  {Chisari} N.~E.,  {Vibert} D.,  {Dubois} Y.,  {Pichon} C.,   {Devriendt} J.,  2018b, \mn@doi [\mnras] {10.1093/mnras/sty2567}, \href {https://ui.adsabs.harvard.edu/abs/2018MNRAS.481.4753C} {481, 4753}

\bibitem[\protect\citeauthoryear{{Colberg}, {Krughoff}  \& {Connolly}}{{Colberg} et~al.}{2005}]{Colberg:2005}
{Colberg} J.~M.,  {Krughoff} K.~S.,   {Connolly} A.~J.,  2005, \mn@doi [\mnras] {10.1111/j.1365-2966.2005.08897.x}, \href {https://ui.adsabs.harvard.edu/abs/2005MNRAS.359..272C} {359, 272}

\bibitem[\protect\citeauthoryear{{Colless} et~al.,}{{Colless} et~al.}{2003}]{Colless:2003}
{Colless} M.,  et~al., 2003, \mn@doi [arXiv e-prints] {10.48550/arXiv.astro-ph/0306581}, \href {https://ui.adsabs.harvard.edu/abs/2003astro.ph..6581C} {pp astro--ph/0306581}

\bibitem[\protect\citeauthoryear{{Colombi}}{{Colombi}}{2015}]{Colombi:2015}
{Colombi} S.,  2015, \mn@doi [\mnras] {10.1093/mnras/stu2308}, \href {https://ui.adsabs.harvard.edu/abs/2015MNRAS.446.2902C} {446, 2902}

\bibitem[\protect\citeauthoryear{{Courtois}, {Pomar{\`e}de}, {Tully}, {Hoffman}  \& {Courtois}}{{Courtois} et~al.}{2013}]{Courtois:2013}
{Courtois} H.~M.,  {Pomar{\`e}de} D.,  {Tully} R.~B.,  {Hoffman} Y.,   {Courtois} D.,  2013, \mn@doi [\aj] {10.1088/0004-6256/146/3/69}, \href {https://ui.adsabs.harvard.edu/abs/2013AJ....146...69C} {146, 69}

\bibitem[\protect\citeauthoryear{{de Lapparent}, {Geller}  \& {Huchra}}{{de Lapparent} et~al.}{1986}]{Lapparent:1986}
{de Lapparent} V.,  {Geller} M.~J.,   {Huchra} J.~P.,  1986, \mn@doi [\apjl] {10.1086/184625}, \href {https://ui.adsabs.harvard.edu/abs/1986ApJ...302L...1D} {302, L1}

\bibitem[\protect\citeauthoryear{{Dhawalikar} \& {Paranjape}}{{Dhawalikar} \& {Paranjape}}{2024}]{Dhawalikar:2024}
{Dhawalikar} S.,  {Paranjape} A.,  2024, \mn@doi [arXiv e-prints] {10.48550/arXiv.2402.18669}, \href {https://ui.adsabs.harvard.edu/abs/2024arXiv240218669D} {p. arXiv:2402.18669}

\bibitem[\protect\citeauthoryear{{Doroshkevich}}{{Doroshkevich}}{1970}]{Doroshkevich:1970}
{Doroshkevich} A.~G.,  1970, \mn@doi [Astrophysics] {10.1007/BF01001625}, \href {http://adsabs.harvard.edu/abs/1970Ap......6..320D} {6, 320}

\bibitem[\protect\citeauthoryear{{Eisenstein}, {Loeb}  \& {Turner}}{{Eisenstein} et~al.}{1997}]{Eisenstein:1997}
{Eisenstein} D.~J.,  {Loeb} A.,   {Turner} E.~L.,  1997, \mn@doi [\apj] {10.1086/303572}, \href {https://ui.adsabs.harvard.edu/abs/1997ApJ...475..421E} {475, 421}

\bibitem[\protect\citeauthoryear{Elek, Burchett, Prochaska  \& Forbes}{Elek et~al.}{2021}]{Elek:2021a}
Elek O.,  Burchett J.~N.,  Prochaska J.~X.,   Forbes A.~G.,  2021, \mn@doi [IEEE Transactions on Visualization and Computer Graphics] {10.1109/TVCG.2020.3030407}, 27, 806

\bibitem[\protect\citeauthoryear{Elek, Burchett, Prochaska  \& Forbes}{Elek et~al.}{2022}]{Elek:2021b}
Elek O.,  Burchett J.~N.,  Prochaska J.~X.,   Forbes A.~G.,  2022, \mn@doi [Artificial Life] {10.1162/artl_a_00351}, 28, 22

\bibitem[\protect\citeauthoryear{{Feldbrugge}}{{Feldbrugge}}{2024}]{Feldbrugge:2024}
{Feldbrugge} J.,  2024, \mn@doi [arXiv e-prints] {10.48550/arXiv.2402.16234}, \href {https://ui.adsabs.harvard.edu/abs/2024arXiv240216234F} {p. arXiv:2402.16234}

\bibitem[\protect\citeauthoryear{{Feldbrugge} \& {van de Weygaert}}{{Feldbrugge} \& {van de Weygaert}}{2023}]{Feldbrugge:2023a}
{Feldbrugge} J.,  {van de Weygaert} R.,  2023, \mn@doi [\jcap] {10.1088/1475-7516/2023/02/058}, \href {https://ui.adsabs.harvard.edu/abs/2023JCAP...02..058F} {2023, 058}

\bibitem[\protect\citeauthoryear{{Feldbrugge}, {Hidding}  \& {van de Weygaert}}{{Feldbrugge} et~al.}{2016}]{Feldbrugge:2016}
{Feldbrugge} J.~L.,  {Hidding} J.,   {van de Weygaert} R.,  2016, in {van de Weygaert} R.,  {Shandarin} S.,  {Saar} E.,   {Einasto} J.,  eds, ~ Vol. 308, The Zel'dovich Universe: Genesis and Growth of the Cosmic Web. pp 107--114, \mn@doi{10.1017/S1743921316009704}

\bibitem[\protect\citeauthoryear{{Feldbrugge}, {van de Weygaert}, {Hidding}  \& {Feldbrugge}}{{Feldbrugge} et~al.}{2018}]{Feldbrugge:2018}
{Feldbrugge} J.,  {van de Weygaert} R.,  {Hidding} J.,   {Feldbrugge} J.,  2018, \mn@doi [\jcap] {10.1088/1475-7516/2018/05/027}, \href {https://ui.adsabs.harvard.edu/abs/2018JCAP...05..027F} {2018, 027}

\bibitem[\protect\citeauthoryear{{Feldbrugge}, {van Engelen}, {van de Weygaert}, {Pranav}  \& {Vegter}}{{Feldbrugge} et~al.}{2019}]{Feldbrugge:2019}
{Feldbrugge} J.,  {van Engelen} M.,  {van de Weygaert} R.,  {Pranav} P.,   {Vegter} G.,  2019, \mn@doi [\jcap] {10.1088/1475-7516/2019/09/052}, \href {https://ui.adsabs.harvard.edu/abs/2019JCAP...09..052F} {2019, 052}

\bibitem[\protect\citeauthoryear{{Feldbrugge}, {Yan}  \& {van de Weygaert}}{{Feldbrugge} et~al.}{2023}]{Feldbrugge:2023b}
{Feldbrugge} J.,  {Yan} Y.,   {van de Weygaert} R.,  2023, \mn@doi [\mnras] {10.1093/mnras/stad2777}, \href {https://ui.adsabs.harvard.edu/abs/2023MNRAS.526.5031F} {526, 5031}

\bibitem[\protect\citeauthoryear{{Forero-Romero}, {Hoffman}, {Gottl{\"o}ber}, {Klypin}  \& {Yepes}}{{Forero-Romero} et~al.}{2009}]{Foreroromero:2009}
{Forero-Romero} J.~E.,  {Hoffman} Y.,  {Gottl{\"o}ber} S.,  {Klypin} A.,   {Yepes} G.,  2009, \mn@doi [\mnras] {10.1111/j.1365-2966.2009.14885.x}, \href {https://ui.adsabs.harvard.edu/abs/2009MNRAS.396.1815F} {396, 1815}

\bibitem[\protect\citeauthoryear{{Gal{\'a}rraga-Espinosa} et~al.,}{{Gal{\'a}rraga-Espinosa} et~al.}{2024}]{Galarraga-Espinosa:2024}
{Gal{\'a}rraga-Espinosa} D.,  et~al., 2024, \mn@doi [\aap] {10.1051/0004-6361/202347982}, \href {https://ui.adsabs.harvard.edu/abs/2024A&A...684A..63G} {684, A63}

\bibitem[\protect\citeauthoryear{{Ganeshaiah Veena}}{{Ganeshaiah Veena}}{2020}]{Ganeshaiah:2020}
{Ganeshaiah Veena} P.,  2020, PhD thesis, University of Groningen, Netherlands

\bibitem[\protect\citeauthoryear{{Ganeshaiah Veena}, {Cautun}, {van de Weygaert}, {Tempel}, {Jones}, {Rieder}  \& {Frenk}}{{Ganeshaiah Veena} et~al.}{2018}]{Ganeshaiah:2018}
{Ganeshaiah Veena} P.,  {Cautun} M.,  {van de Weygaert} R.,  {Tempel} E.,  {Jones} B. J.~T.,  {Rieder} S.,   {Frenk} C.~S.,  2018, \mn@doi [\mnras] {10.1093/mnras/sty2270}, \href {https://ui.adsabs.harvard.edu/abs/2018MNRAS.481..414G} {481, 414}

\bibitem[\protect\citeauthoryear{{Ganeshaiah Veena}, {Cautun}, {Tempel}, {van de Weygaert}  \& {Frenk}}{{Ganeshaiah Veena} et~al.}{2019}]{Ganeshaiah:2019}
{Ganeshaiah Veena} P.,  {Cautun} M.,  {Tempel} E.,  {van de Weygaert} R.,   {Frenk} C.~S.,  2019, \mn@doi [\mnras] {10.1093/mnras/stz1343}, \href {https://ui.adsabs.harvard.edu/abs/2019MNRAS.487.1607G} {487, 1607}

\bibitem[\protect\citeauthoryear{{Gingold} \& {Monaghan}}{{Gingold} \& {Monaghan}}{1977}]{Gingold:1977}
{Gingold} R.~A.,  {Monaghan} J.~J.,  1977, \mn@doi [\mnras] {10.1093/mnras/181.3.375}, \href {https://ui.adsabs.harvard.edu/abs/1977MNRAS.181..375G} {181, 375}

\bibitem[\protect\citeauthoryear{{Giovanelli}, {Haynes}  \& {Chincarini}}{{Giovanelli} et~al.}{1986}]{Haynes:1986}
{Giovanelli} R.,  {Haynes} M.~P.,   {Chincarini} G.~L.,  1986, \mn@doi [\apj] {10.1086/163784}, \href {https://ui.adsabs.harvard.edu/abs/1986ApJ...300...77G} {300, 77}

\bibitem[\protect\citeauthoryear{{Guth} \& {Pi}}{{Guth} \& {Pi}}{1982}]{Guth:1982}
{Guth} A.~H.,  {Pi} S.~Y.,  1982, \mn@doi [\prl] {10.1103/PhysRevLett.49.1110}, \href {https://ui.adsabs.harvard.edu/abs/1982PhRvL..49.1110G} {49, 1110}

\bibitem[\protect\citeauthoryear{{Hahn}, {Porciani}, {Carollo}  \& {Dekel}}{{Hahn} et~al.}{2007}]{Hahn:2007}
{Hahn} O.,  {Porciani} C.,  {Carollo} C.~M.,   {Dekel} A.,  2007, \mn@doi [\mnras] {10.1111/j.1365-2966.2006.11318.x}, \href {https://ui.adsabs.harvard.edu/abs/2007MNRAS.375..489H} {375, 489}

\bibitem[\protect\citeauthoryear{{Hahn}, {Porciani}, {Dekel}  \& {Carollo}}{{Hahn} et~al.}{2009}]{Hahn:2009}
{Hahn} O.,  {Porciani} C.,  {Dekel} A.,   {Carollo} C.~M.,  2009, \mn@doi [\mnras] {10.1111/j.1365-2966.2009.15271.x}, \href {https://ui.adsabs.harvard.edu/abs/2009MNRAS.398.1742H} {398, 1742}

\bibitem[\protect\citeauthoryear{{Hahn}, {Teyssier}  \& {Carollo}}{{Hahn} et~al.}{2010}]{Hahn:2010}
{Hahn} O.,  {Teyssier} R.,   {Carollo} C.~M.,  2010, \mn@doi [\mnras] {10.1111/j.1365-2966.2010.16494.x}, \href {https://ui.adsabs.harvard.edu/abs/2010MNRAS.405..274H} {405, 274}

\bibitem[\protect\citeauthoryear{Harlow, Evans  \& Richtmyer}{Harlow et~al.}{1955}]{Harlow:1955}
Harlow F.,  Evans M.,   Richtmyer R.,  1955, A Machine Calculation Method for Hydrodynamic Problems.
LAMS (Los Alamos Scientific Laboratory), Los Alamos Scientific Laboratory of the University of California

\bibitem[\protect\citeauthoryear{{Harrison}}{{Harrison}}{1970}]{Harrison:1970}
{Harrison} E.~R.,  1970, \mn@doi [\prd] {10.1103/PhysRevD.1.2726}, \href {https://ui.adsabs.harvard.edu/abs/1970PhRvD...1.2726H} {1, 2726}

\bibitem[\protect\citeauthoryear{{Hellwing}, {Cautun}, {van de Weygaert}  \& {Jones}}{{Hellwing} et~al.}{2021}]{Hellwing:2020}
{Hellwing} W.~A.,  {Cautun} M.,  {van de Weygaert} R.,   {Jones} B.~T.,  2021, \mn@doi [\prd] {10.1103/PhysRevD.103.063517}, \href {https://ui.adsabs.harvard.edu/abs/2021PhRvD.103f3517H} {103, 063517}

\bibitem[\protect\citeauthoryear{Hidding}{Hidding}{2020}]{Johan:2020}
Hidding J.,  2020, jhidding/nbody2d: 2d PM n-body code, \mn@doi{10.5281/zenodo.4158731}, \url {https://doi.org/10.5281/zenodo.4158731}

\bibitem[\protect\citeauthoryear{{Hidding}, {Shandarin}  \& {van de Weygaert}}{{Hidding} et~al.}{2014}]{Hidding:2014}
{Hidding} J.,  {Shandarin} S.~F.,   {van de Weygaert} R.,  2014, \mn@doi [\mnras] {10.1093/mnras/stt2142}, \href {https://ui.adsabs.harvard.edu/abs/2014MNRAS.437.3442H} {437, 3442}

\bibitem[\protect\citeauthoryear{{Hoffman} \& {Ribak}}{{Hoffman} \& {Ribak}}{1991}]{Hoffman:1991}
{Hoffman} Y.,  {Ribak} E.,  1991, \mn@doi [\apjl] {10.1086/186160}, \href {https://ui.adsabs.harvard.edu/abs/1991ApJ...380L...5H} {380, L5}

\bibitem[\protect\citeauthoryear{{Hoffman}, {Metuki}, {Yepes}, {Gottl{\"o}ber}, {Forero-Romero}, {Libeskind}  \& {Knebe}}{{Hoffman} et~al.}{2012}]{Hoffman:2012}
{Hoffman} Y.,  {Metuki} O.,  {Yepes} G.,  {Gottl{\"o}ber} S.,  {Forero-Romero} J.~E.,  {Libeskind} N.~I.,   {Knebe} A.,  2012, \mn@doi [\mnras] {10.1111/j.1365-2966.2012.21553.x}, \href {https://ui.adsabs.harvard.edu/abs/2012MNRAS.425.2049H} {425, 2049}

\bibitem[\protect\citeauthoryear{{Huchra} et~al.,}{{Huchra} et~al.}{2012}]{Huchra:2012}
{Huchra} J.~P.,  et~al., 2012, \mn@doi [\apjs] {10.1088/0067-0049/199/2/26}, \href {https://ui.adsabs.harvard.edu/abs/2012ApJS..199...26H} {199, 26}

\bibitem[\protect\citeauthoryear{{Icke}}{{Icke}}{1973}]{Icke:1973}
{Icke} V.,  1973, \aap, \href {https://ui.adsabs.harvard.edu/abs/1973A&A....27....1I} {27, 1}

\bibitem[\protect\citeauthoryear{{Icke} \& {van de Weygaert}}{{Icke} \& {van de Weygaert}}{1987}]{Icke:1987}
{Icke} V.,  {van de Weygaert} R.,  1987, \aap, \href {https://ui.adsabs.harvard.edu/abs/1987A&A...184...16I} {184, 16}

\bibitem[\protect\citeauthoryear{{Jaber}, {Peper}, {Hellwing}, {Arag{\'o}n-Calvo}  \& {Valenzuela}}{{Jaber} et~al.}{2024}]{Jaber:2024}
{Jaber} M.,  {Peper} M.,  {Hellwing} W.~A.,  {Arag{\'o}n-Calvo} M.~A.,   {Valenzuela} O.,  2024, \mn@doi [\mnras] {10.1093/mnras/stad3347}, \href {https://ui.adsabs.harvard.edu/abs/2024MNRAS.527.4087J} {527, 4087}

\bibitem[\protect\citeauthoryear{{Joeveer} \& {Einasto}}{{Joeveer} \& {Einasto}}{1978}]{Joeveer:1978}
{Joeveer} M.,  {Einasto} J.,  1978, in {Longair} M.~S.,  {Einasto} J.,  eds, ~1 Vol. 79, Large Scale Structures in the Universe. p.~241

\bibitem[\protect\citeauthoryear{{Kraljic} et~al.,}{{Kraljic} et~al.}{2018}]{Kraljic:2018}
{Kraljic} K.,  et~al., 2018, \mn@doi [\mnras] {10.1093/mnras/stx2638}, \href {https://ui.adsabs.harvard.edu/abs/2018MNRAS.474..547K} {474, 547}

\bibitem[\protect\citeauthoryear{{Kuchner} et~al.,}{{Kuchner} et~al.}{2020}]{Kuchner:2020}
{Kuchner} U.,  et~al., 2020, \mn@doi [\mnras] {10.1093/mnras/staa1083}, \href {https://ui.adsabs.harvard.edu/abs/2020MNRAS.494.5473K} {494, 5473}

\bibitem[\protect\citeauthoryear{{Kuchner} et~al.,}{{Kuchner} et~al.}{2021}]{Kuchner:2021}
{Kuchner} U.,  et~al., 2021, \mn@doi [\mnras] {10.1093/mnras/stab567}, \href {https://ui.adsabs.harvard.edu/abs/2021MNRAS.503.2065K} {503, 2065}

\bibitem[\protect\citeauthoryear{{Kugul}}{{Kugul}}{2020}]{Kugul:2020}
{Kugul} R.,  2020, Master's thesis, University of Groningen, Netherlands

\bibitem[\protect\citeauthoryear{{Kuutma}, {Tamm}  \& {Tempel}}{{Kuutma} et~al.}{2017}]{Kuutma:2017}
{Kuutma} T.,  {Tamm} A.,   {Tempel} E.,  2017, \mn@doi [\aap] {10.1051/0004-6361/201730526}, \href {https://ui.adsabs.harvard.edu/abs/2017A&A...600L...6K} {600, L6}

\bibitem[\protect\citeauthoryear{{Laigle} et~al.,}{{Laigle} et~al.}{2018}]{Laigle:2018}
{Laigle} C.,  et~al., 2018, \mn@doi [\mnras] {10.1093/mnras/stx3055}, \href {https://ui.adsabs.harvard.edu/abs/2018MNRAS.474.5437L} {474, 5437}

\bibitem[\protect\citeauthoryear{{Libeskind} et~al.,}{{Libeskind} et~al.}{2018}]{Libeskind:2018}
{Libeskind} N.~I.,  et~al., 2018, \mn@doi [\mnras] {10.1093/mnras/stx1976}, \href {https://ui.adsabs.harvard.edu/abs/2018MNRAS.473.1195L} {473, 1195}

\bibitem[\protect\citeauthoryear{{Longuet-Higgins}}{{Longuet-Higgins}}{1957}]{Longuet-Higgins:1957}
{Longuet-Higgins} M.~S.,  1957, \mn@doi [Philosophical Transactions of the Royal Society of London Series A] {10.1098/rsta.1957.0018}, \href {https://ui.adsabs.harvard.edu/abs/1957RSPTA.250..157L} {250, 157}

\bibitem[\protect\citeauthoryear{{L{\'o}pez}}{{L{\'o}pez}}{2024}]{Lopez:2024}
{L{\'o}pez} P.,  2024, \mn@doi [\pasp] {10.1088/1538-3873/ad31c9}, \href {https://ui.adsabs.harvard.edu/abs/2024PASP..136c7001L} {136, 037001}

\bibitem[\protect\citeauthoryear{{L{\'o}pez}, {Cautun}, {Paz}, {Merch{\'a}n}  \& {van de Weygaert}}{{L{\'o}pez} et~al.}{2021}]{Lopez:2021}
{L{\'o}pez} P.,  {Cautun} M.,  {Paz} D.,  {Merch{\'a}n} M.,   {van de Weygaert} R.,  2021, \mn@doi [\mnras] {10.1093/mnras/stab451}, \href {https://ui.adsabs.harvard.edu/abs/2021MNRAS.502.5528L} {502, 5528}

\bibitem[\protect\citeauthoryear{{Lu}, {Mandelker}, {Oh}, {Dekel}, {van den Bosch}, {Springel}, {Nagai}  \& {van de Voort}}{{Lu} et~al.}{2024}]{Lu:2024}
{Lu} Y.~S.,  {Mandelker} N.,  {Oh} S.~P.,  {Dekel} A.,  {van den Bosch} F.~C.,  {Springel} V.,  {Nagai} D.,   {van de Voort} F.,  2024, \mn@doi [\mnras] {10.1093/mnras/stad3779}, \href {https://ui.adsabs.harvard.edu/abs/2024MNRAS.52711256L} {527, 11256}

\bibitem[\protect\citeauthoryear{{Mukhanov} \& {Chibisov}}{{Mukhanov} \& {Chibisov}}{1981}]{Mukhanov:1981}
{Mukhanov} V.~F.,  {Chibisov} G.~V.,  1981, Soviet Journal of Experimental and Theoretical Physics Letters, \href {https://ui.adsabs.harvard.edu/abs/1981JETPL..33..532M} {33, 532}

\bibitem[\protect\citeauthoryear{{Neyrinck}}{{Neyrinck}}{2012}]{Neyrinck:2012}
{Neyrinck} M.~C.,  2012, \mn@doi [\mnras] {10.1111/j.1365-2966.2012.21956.x}, \href {https://ui.adsabs.harvard.edu/abs/2012MNRAS.427..494N} {427, 494}

\bibitem[\protect\citeauthoryear{{Neyrinck}, {Hidding}, {Konstantatou}  \& {van de Weygaert}}{{Neyrinck} et~al.}{2018}]{Neyrinck:2018}
{Neyrinck} M.~C.,  {Hidding} J.,  {Konstantatou} M.,   {van de Weygaert} R.,  2018, \mn@doi [Royal Society Open Science] {10.1098/rsos.171582}, \href {https://ui.adsabs.harvard.edu/abs/2018RSOS....571582N} {5, 171582}

\bibitem[\protect\citeauthoryear{Okabe, Boots, Sugihara  \& Chiu}{Okabe et~al.}{2000}]{Okabe:2000}
Okabe A.,  Boots B.,  Sugihara K.,   Chiu S.~N.,  2000, Spatial Tessellations: Concepts and Applications of {V}oronoi Diagrams, 2nd ed. edn.
Series in Probability and Statistics, John Wiley and Sons, Inc.

\bibitem[\protect\citeauthoryear{{Paranjape}}{{Paranjape}}{2021}]{Paranjape:2021}
{Paranjape} A.,  2021, \mn@doi [\mnras] {10.1093/mnras/stab359}, \href {https://ui.adsabs.harvard.edu/abs/2021MNRAS.502.5210P} {502, 5210}

\bibitem[\protect\citeauthoryear{{Paranjape}, {Hahn}  \& {Sheth}}{{Paranjape} et~al.}{2018}]{Paranjape:2018}
{Paranjape} A.,  {Hahn} O.,   {Sheth} R.~K.,  2018, \mn@doi [\mnras] {10.1093/mnras/sty633}, \href {https://ui.adsabs.harvard.edu/abs/2018MNRAS.476.5442P} {476, 5442}

\bibitem[\protect\citeauthoryear{{Parente} et~al.,}{{Parente} et~al.}{2023}]{Parente:2024}
{Parente} M.,  et~al., 2023, \mn@doi [arXiv e-prints] {10.48550/arXiv.2312.10146}, \href {https://ui.adsabs.harvard.edu/abs/2023arXiv231210146P} {p. arXiv:2312.10146}

\bibitem[\protect\citeauthoryear{{Park} \& {Lee}}{{Park} \& {Lee}}{2009}]{Park:2009}
{Park} D.,  {Lee} J.,  2009, \mn@doi [\mnras] {10.1111/j.1365-2966.2009.15524.x}, \href {https://ui.adsabs.harvard.edu/abs/2009MNRAS.400.1105P} {400, 1105}

\bibitem[\protect\citeauthoryear{{Pelupessy}, {Schaap}  \& {van de Weygaert}}{{Pelupessy} et~al.}{2003}]{Pelupessy:2003}
{Pelupessy} F.~I.,  {Schaap} W.~E.,   {van de Weygaert} R.,  2003, \mn@doi [\aap] {10.1051/0004-6361:20030314}, \href {https://ui.adsabs.harvard.edu/abs/2003A&A...403..389P} {403, 389}

\bibitem[\protect\citeauthoryear{{Pogosyan}, {Gay}  \& {Pichon}}{{Pogosyan} et~al.}{2009a}]{Pogosyan:2009b}
{Pogosyan} D.,  {Gay} C.,   {Pichon} C.,  2009a, \mn@doi [\prd] {10.1103/PhysRevD.80.081301}, \href {https://ui.adsabs.harvard.edu/abs/2009PhRvD..80h1301P} {80, 081301}

\bibitem[\protect\citeauthoryear{{Pogosyan}, {Pichon}, {Gay}, {Prunet}, {Cardoso}, {Sousbie}  \& {Colombi}}{{Pogosyan} et~al.}{2009b}]{Pogosyan:2009a}
{Pogosyan} D.,  {Pichon} C.,  {Gay} C.,  {Prunet} S.,  {Cardoso} J.~F.,  {Sousbie} T.,   {Colombi} S.,  2009b, \mn@doi [\mnras] {10.1111/j.1365-2966.2009.14753.x}, \href {https://ui.adsabs.harvard.edu/abs/2009MNRAS.396..635P} {396, 635}

\bibitem[\protect\citeauthoryear{{Pomar{\`e}de}, {Hoffman}, {Courtois}  \& {Tully}}{{Pomar{\`e}de} et~al.}{2017}]{Pomarede:2017}
{Pomar{\`e}de} D.,  {Hoffman} Y.,  {Courtois} H.~M.,   {Tully} R.~B.,  2017, \mn@doi [\apj] {10.3847/1538-4357/aa7f78}, \href {https://ui.adsabs.harvard.edu/abs/2017ApJ...845...55P} {845, 55}

\bibitem[\protect\citeauthoryear{{Pontzen}, {Slosar}, {Roth}  \& {Peiris}}{{Pontzen} et~al.}{2016}]{Pontzen:2016}
{Pontzen} A.,  {Slosar} A.,  {Roth} N.,   {Peiris} H.~V.,  2016, \mn@doi [\prd] {10.1103/PhysRevD.93.103519}, \href {https://ui.adsabs.harvard.edu/abs/2016PhRvD..93j3519P} {93, 103519}

\bibitem[\protect\citeauthoryear{{Porciani}, {Dekel}  \& {Hoffman}}{{Porciani} et~al.}{2002}]{Porciani:2002a}
{Porciani} C.,  {Dekel} A.,   {Hoffman} Y.,  2002, \mn@doi [\mnras] {10.1046/j.1365-8711.2002.05305.x}, \href {https://ui.adsabs.harvard.edu/abs/2002MNRAS.332..325P} {332, 325}

\bibitem[\protect\citeauthoryear{Poston \& Stewart}{Poston \& Stewart}{1996}]{Poston:1978}
Poston T.,  Stewart I.,  1996, Catastrophe Theory and Its Applications.
Dover books on mathematics, Dover Publications, \url {https://books.google.nl/books?id=7Zm5zTh8rLAC}

\bibitem[\protect\citeauthoryear{{Rice}}{{Rice}}{1944}]{Rice:1944}
{Rice} S.~O.,  1944, \mn@doi [Bell System Technical Journal] {10.1002/j.1538-7305.1944.tb00874.x}, \href {https://ui.adsabs.harvard.edu/abs/1944BSTJ...23..282R} {23, 282}

\bibitem[\protect\citeauthoryear{{Rice}}{{Rice}}{1945}]{Rice:1945}
{Rice} S.~O.,  1945, Bell System Technical Journal, \href {https://ui.adsabs.harvard.edu/abs/1945BSTJ...24...46R} {24, 46}

\bibitem[\protect\citeauthoryear{{Rieder}, {van de Weygaert}, {Cautun}, {Beygu}  \& {Portegies Zwart}}{{Rieder} et~al.}{2013}]{Rieder:2013}
{Rieder} S.,  {van de Weygaert} R.,  {Cautun} M.,  {Beygu} B.,   {Portegies Zwart} S.,  2013, \mn@doi [\mnras] {10.1093/mnras/stt1288}, \href {https://ui.adsabs.harvard.edu/abs/2013MNRAS.435..222R} {435, 222}

\bibitem[\protect\citeauthoryear{{Schaap}}{{Schaap}}{2007}]{Schaap:2007}
{Schaap} W.~E.,  2007, PhD thesis, University of Groningen, Netherlands

\bibitem[\protect\citeauthoryear{{Sch{\"a}fer}}{{Sch{\"a}fer}}{2009}]{Schaefer:2009}
{Sch{\"a}fer} B.~M.,  2009, \mn@doi [International Journal of Modern Physics D] {10.1142/S0218271809014388}, \href {https://ui.adsabs.harvard.edu/abs/2009IJMPD..18..173S} {18, 173}

\bibitem[\protect\citeauthoryear{{Shandarin}}{{Shandarin}}{2011}]{Shandarin:2011}
{Shandarin} S.~F.,  2011, \mn@doi [\jcap] {10.1088/1475-7516/2011/05/015}, \href {https://ui.adsabs.harvard.edu/abs/2011JCAP...05..015S} {2011, 015}

\bibitem[\protect\citeauthoryear{{Shandarin} \& {Zel'dovich}}{{Shandarin} \& {Zel'dovich}}{1989}]{Shandarin:1989}
{Shandarin} S.~F.,  {Zel'dovich} Y.~B.,  1989, \mn@doi [Reviews of Modern Physics] {10.1103/RevModPhys.61.185}, \href {https://ui.adsabs.harvard.edu/abs/1989RvMP...61..185S} {61, 185}

\bibitem[\protect\citeauthoryear{{Shandarin}, {Habib}  \& {Heitmann}}{{Shandarin} et~al.}{2012}]{Shandarin:2012}
{Shandarin} S.,  {Habib} S.,   {Heitmann} K.,  2012, \mn@doi [\prd] {10.1103/PhysRevD.85.083005}, \href {https://ui.adsabs.harvard.edu/abs/2012PhRvD..85h3005S} {85, 083005}

\bibitem[\protect\citeauthoryear{{Shectman}, {Landy}, {Oemler}, {Tucker}, {Lin}, {Kirshner}  \& {Schechter}}{{Shectman} et~al.}{1996}]{Shectman:1996}
{Shectman} S.~A.,  {Landy} S.~D.,  {Oemler} A.,  {Tucker} D.~L.,  {Lin} H.,  {Kirshner} R.~P.,   {Schechter} P.~L.,  1996, \mn@doi [\apj] {10.1086/177858}, \href {https://ui.adsabs.harvard.edu/abs/1996ApJ...470..172S} {470, 172}

\bibitem[\protect\citeauthoryear{{Shim}, {Park}, {Kim}  \& {Hwang}}{{Shim} et~al.}{2021}]{Shim:2021}
{Shim} J.,  {Park} C.,  {Kim} J.,   {Hwang} H.~S.,  2021, \mn@doi [\apj] {10.3847/1538-4357/abd0f6}, \href {https://ui.adsabs.harvard.edu/abs/2021ApJ...908..211S} {908, 211}

\bibitem[\protect\citeauthoryear{{Shim}, {Park}, {Kim}  \& {Hong}}{{Shim} et~al.}{2023}]{Shim:2023}
{Shim} J.,  {Park} C.,  {Kim} J.,   {Hong} S.~E.,  2023, \mn@doi [\apj] {10.3847/1538-4357/acd852}, \href {https://ui.adsabs.harvard.edu/abs/2023ApJ...952...59S} {952, 59}

\bibitem[\protect\citeauthoryear{{Shivashankar}, {Pranav}, {Natarajan}, {van de Weygaert}, {Bos}  \& {Rieder}}{{Shivashankar} et~al.}{2016}]{Shivashankar:2016}
{Shivashankar} N.,  {Pranav} P.,  {Natarajan} V.,  {van de Weygaert} R.,  {Bos} E.~G.~P.,   {Rieder} S.,  2016, \mn@doi [IEEE Transactions on Visualizations and Computer Graphics. 2016. Vol. 22(6] {10.1109/TVCG.2015.2452919}, \href {https://ui.adsabs.harvard.edu/abs/2016ITVCG..22.1745S} {22, 1745}

\bibitem[\protect\citeauthoryear{{Sousbie}}{{Sousbie}}{2011}]{Sousbie:2011a}
{Sousbie} T.,  2011, \mn@doi [\mnras] {10.1111/j.1365-2966.2011.18394.x}, \href {https://ui.adsabs.harvard.edu/abs/2011MNRAS.414..350S} {414, 350}

\bibitem[\protect\citeauthoryear{{Sousbie}, {Pichon}  \& {Kawahara}}{{Sousbie} et~al.}{2011}]{Sousbie:2011b}
{Sousbie} T.,  {Pichon} C.,   {Kawahara} H.,  2011, \mn@doi [\mnras] {10.1111/j.1365-2966.2011.18395.x}, \href {https://ui.adsabs.harvard.edu/abs/2011MNRAS.414..384S} {414, 384}

\bibitem[\protect\citeauthoryear{{Springel} et~al.,}{{Springel} et~al.}{2008}]{Aquarius:2008}
{Springel} V.,  et~al., 2008, \mn@doi [\mnras] {10.1111/j.1365-2966.2008.14066.x}, \href {https://ui.adsabs.harvard.edu/abs/2008MNRAS.391.1685S} {391, 1685}

\bibitem[\protect\citeauthoryear{{Starobinsky}}{{Starobinsky}}{1982}]{Starobinsky:1982}
{Starobinsky} A.~A.,  1982, \mn@doi [Physics Letters B] {10.1016/0370-2693(82)90541-X}, \href {https://ui.adsabs.harvard.edu/abs/1982PhLB..117..175S} {117, 175}

\bibitem[\protect\citeauthoryear{{Stopyra}, {Peiris}  \& {Pontzen}}{{Stopyra} et~al.}{2021}]{Stopyra:2021}
{Stopyra} S.,  {Peiris} H.~V.,   {Pontzen} A.,  2021, \mn@doi [\mnras] {10.1093/mnras/staa3587}, \href {https://ui.adsabs.harvard.edu/abs/2021MNRAS.500.4173S} {500, 4173}

\bibitem[\protect\citeauthoryear{{Stopyra}, {Peiris}, {Pontzen}, {Jasche}  \& {Lavaux}}{{Stopyra} et~al.}{2024}]{Stopyra:2024}
{Stopyra} S.,  {Peiris} H.~V.,  {Pontzen} A.,  {Jasche} J.,   {Lavaux} G.,  2024, \mn@doi [\mnras] {10.1093/mnras/stae1251}, \href {https://ui.adsabs.harvard.edu/abs/2024MNRAS.tmp.1259S} {}

\bibitem[\protect\citeauthoryear{{Tegmark} et~al.,}{{Tegmark} et~al.}{2004}]{Tegmark:2004}
{Tegmark} M.,  et~al., 2004, \mn@doi [\prd] {10.1103/PhysRevD.69.103501}, \href {https://ui.adsabs.harvard.edu/abs/2004PhRvD..69j3501T} {69, 103501}

\bibitem[\protect\citeauthoryear{{Tempel} \& {Libeskind}}{{Tempel} \& {Libeskind}}{2013}]{Tempel:2013}
{Tempel} E.,  {Libeskind} N.~I.,  2013, \mn@doi [\apjl] {10.1088/2041-8205/775/2/L42}, \href {https://ui.adsabs.harvard.edu/abs/2013ApJ...775L..42T} {775, L42}

\bibitem[\protect\citeauthoryear{{Tempel}, {Stoica}, {Mart{\'\i}nez}, {Liivam{\"a}gi}, {Castellan}  \& {Saar}}{{Tempel} et~al.}{2014}]{Tempel:2014}
{Tempel} E.,  {Stoica} R.~S.,  {Mart{\'\i}nez} V.~J.,  {Liivam{\"a}gi} L.~J.,  {Castellan} G.,   {Saar} E.,  2014, \mn@doi [\mnras] {10.1093/mnras/stt2454}, \href {https://ui.adsabs.harvard.edu/abs/2014MNRAS.438.3465T} {438, 3465}

\bibitem[\protect\citeauthoryear{Thom}{Thom}{1975}]{Thom:1972}
Thom R.,  1975, Structural Stability and Morphogenesis: An Outline of a General Theory of Models.
Advanced book program, W. A. Benjamin, \url {https://books.google.nl/books?id=3jtRAAAAMAAJ}

\bibitem[\protect\citeauthoryear{{Tully} et~al.,}{{Tully} et~al.}{2023}]{Tully:2023}
{Tully} R.~B.,  et~al., 2023, \mn@doi [\apj] {10.3847/1538-4357/ac94d8}, \href {https://ui.adsabs.harvard.edu/abs/2023ApJ...944...94T} {944, 94}

\bibitem[\protect\citeauthoryear{{van de Weygaert}}{{van de Weygaert}}{1994}]{Weygaert:1994}
{van de Weygaert} R.,  1994, \aap, \href {https://ui.adsabs.harvard.edu/abs/1994A&A...283..361V} {283, 361}

\bibitem[\protect\citeauthoryear{{van de Weygaert} \& {Bertschinger}}{{van de Weygaert} \& {Bertschinger}}{1996}]{Weygaert:1996}
{van de Weygaert} R.,  {Bertschinger} E.,  1996, \mn@doi [\mnras] {10.1093/mnras/281.1.84}, \href {https://ui.adsabs.harvard.edu/abs/1996MNRAS.281...84V} {281, 84}

\bibitem[\protect\citeauthoryear{{van de Weygaert} \& {Bond}}{{van de Weygaert} \& {Bond}}{2008}]{Weygaert:2008}
{van de Weygaert} R.,  {Bond} J.~R.,  2008, in {Plionis} M.,  {L{\'o}pez-Cruz} O.,   {Hughes} D.,  eds, , Vol.~740, A Pan-Chromatic View of Clusters of Galaxies and the Large-Scale Structure.
p.~335, \mn@doi{10.1007/978-1-4020-6941-3_10}

\bibitem[\protect\citeauthoryear{{van de Weygaert} \& {Schaap}}{{van de Weygaert} \& {Schaap}}{2009}]{Weygaert:2009}
{van de Weygaert} R.,  {Schaap} W.,  2009, in {Mart{\'\i}nez} V.~J.,  {Saar} E.,  {Mart{\'\i}nez-Gonz{\'a}lez} E.,   {Pons-Border{\'\i}a} M.~J.,  eds, , Vol.~665, Data Analysis in Cosmology.
pp 291--413, \mn@doi{10.1007/978-3-540-44767-2_11}

\bibitem[\protect\citeauthoryear{{van Haarlem} \& {van de Weygaert}}{{van Haarlem} \& {van de Weygaert}}{1993}]{Haarlem:1993}
{van Haarlem} M.,  {van de Weygaert} R.,  1993, \mn@doi [\apj] {10.1086/173416}, \href {https://ui.adsabs.harvard.edu/abs/1993ApJ...418..544V} {418, 544}

\bibitem[\protect\citeauthoryear{{Vazza} \& {Feletti}}{{Vazza} \& {Feletti}}{2020}]{Vazza:2020}
{Vazza} F.,  {Feletti} A.,  2020, \mn@doi [Frontiers in Physics] {10.3389/fphy.2020.525731}, \href {https://ui.adsabs.harvard.edu/abs/2020FrP.....8..491V} {8, 491}

\bibitem[\protect\citeauthoryear{{Wang}, {Libeskind}, {Tempel}, {Kang}  \& {Guo}}{{Wang} et~al.}{2021}]{Wang:2021}
{Wang} P.,  {Libeskind} N.~I.,  {Tempel} E.,  {Kang} X.,   {Guo} Q.,  2021, \mn@doi [Nature Astronomy] {10.1038/s41550-021-01380-6}, \href {https://ui.adsabs.harvard.edu/abs/2021NatAs...5..839W} {5, 839}

\bibitem[\protect\citeauthoryear{{Wang} et~al.,}{{Wang} et~al.}{2024}]{Wang:2024}
{Wang} W.,  et~al., 2024, \mn@doi [arXiv e-prints] {10.48550/arXiv.2402.11678}, \href {https://ui.adsabs.harvard.edu/abs/2024arXiv240211678W} {p. arXiv:2402.11678}

\bibitem[\protect\citeauthoryear{{Wilde}, {Elek}, {Burchett}, {Nagai}, {Prochaska}, {Werk}, {Tuttle}  \& {Forbes}}{{Wilde} et~al.}{2023}]{Wilde:2023}
{Wilde} M.~C.,  {Elek} O.,  {Burchett} J.~N.,  {Nagai} D.,  {Prochaska} J.~X.,  {Werk} J.,  {Tuttle} S.,   {Forbes} A.~G.,  2023, \mn@doi [arXiv e-prints] {10.48550/arXiv.2301.02719}, \href {https://ui.adsabs.harvard.edu/abs/2023arXiv230102719W} {p. arXiv:2301.02719}

\bibitem[\protect\citeauthoryear{Wilding}{Wilding}{2022}]{Wilding:2022}
Wilding G.,  2022, PhD thesis, University of Groningen, \mn@doi{10.33612/diss.250010290}

\bibitem[\protect\citeauthoryear{{Xia}, {Neyrinck}, {Cai}  \& {Arag{\'o}n-Calvo}}{{Xia} et~al.}{2021}]{Xia:2021}
{Xia} Q.,  {Neyrinck} M.~C.,  {Cai} Y.-C.,   {Arag{\'o}n-Calvo} M.~A.,  2021, \mn@doi [\mnras] {10.1093/mnras/stab1713}, \href {https://ui.adsabs.harvard.edu/abs/2021MNRAS.506.1059X} {506, 1059}

\bibitem[\protect\citeauthoryear{{Zakharova}, {Vulcani}, {De Lucia}, {Xie}, {Hirschmann}  \& {Fontanot}}{{Zakharova} et~al.}{2023}]{Zakharova:2023}
{Zakharova} D.,  {Vulcani} B.,  {De Lucia} G.,  {Xie} L.,  {Hirschmann} M.,   {Fontanot} F.,  2023, \mn@doi [\mnras] {10.1093/mnras/stad2562}, \href {https://ui.adsabs.harvard.edu/abs/2023MNRAS.525.4079Z} {525, 4079}

\bibitem[\protect\citeauthoryear{Zeeman}{Zeeman}{1977}]{Zeeman:1977}
Zeeman E.,  1977, Catastrophe Theory: Selected Papers, 1972-1977.
Advanced book program, Addison-Wesley Publishing Company, Advanced Book Program, \url {https://books.google.nl/books?id=iVPvAAAAMAAJ}

\bibitem[\protect\citeauthoryear{{Zel'dovich}}{{Zel'dovich}}{1970}]{Zeldovich:1970}
{Zel'dovich} Y.~B.,  1970, \aap, \href {https://ui.adsabs.harvard.edu/abs/1970A&A.....5...84Z} {5, 84}

\bibitem[\protect\citeauthoryear{{Zel'dovich}}{{Zel'dovich}}{1972}]{Zeldovich:1972}
{Zel'dovich} Y.~B.,  1972, \mn@doi [\mnras] {10.1093/mnras/160.1.1P}, \href {https://ui.adsabs.harvard.edu/abs/1972MNRAS.160P...1Z} {160, 1P}

\makeatother
\end{thebibliography}

\begin{figure*}
    \centering
    \begin{subfigure}[b]{0.49\textwidth}
        \includegraphics[width=\textwidth]{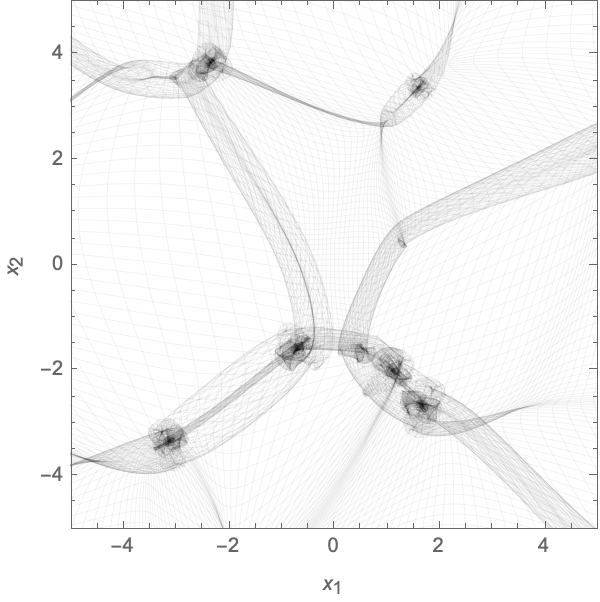}
        \caption{$N$-body simulation}
    \end{subfigure}
    \hfill
    \begin{subfigure}[b]{0.49\textwidth}
    \end{subfigure}\\
    \vspace{0.5cm}
    \begin{subfigure}[b]{0.49\textwidth}
        \includegraphics[width=\textwidth]{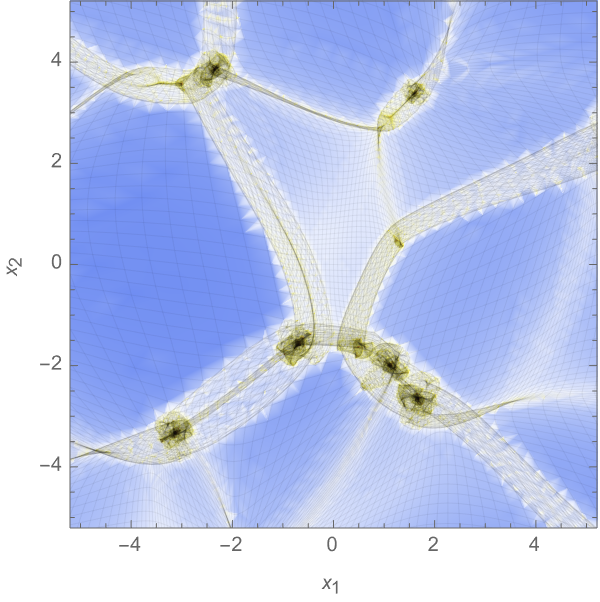}
        \caption{DTFE density field}
    \end{subfigure}
    \hfill
    \begin{subfigure}[b]{0.49\textwidth}
        \includegraphics[width=\textwidth]{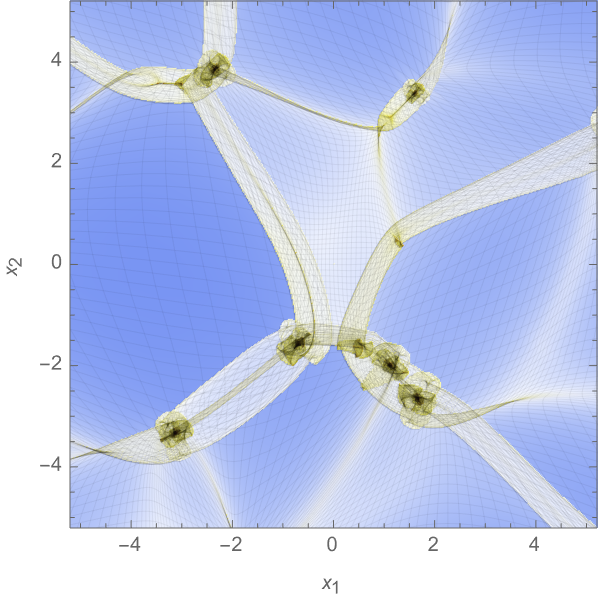}
        \caption{Phase-space DTFE density field}
    \end{subfigure}
    \caption{The phase-space sheet of a two-dimensional $N=256^2$ dark matter only $N$-body simulation in a $25 \times 25$ box (upper left) with the corresponding DTFE (lower left) and phase-space DTFE logarithmic density field $\log(\rho + 1)$ (lower right).} \label{fig:DTFE-vs-PS-DTFE}
\end{figure*}

\appendix
\section{Phase-Space Delaunay Tessellation Field Estimation}\label{ap:density}
Lagrangian fluid dynamics describes the evolution of the dark matter fluid in terms of the Lagrangian map $\bm{x}_t(\bm{q}):\bm{q} + \bm{s}_t(\bm{q})$ describing the flow of a mass element starting at $\bm{q}$ to the position $\bm{x}_t$ in time $t$. The density field in a point $\bm{x}$ is a derived quantity given by a sum over the initial positions that map to $\bm{x}$ in the given time
\begin{align}
    \rho(\bm{x}) = \sum_{\bm{q} \in \bm{x}_t^{-1}(\bm{q})} \frac{\bar{\rho}}{|\det \nabla \bm{x}_t(\bm{q})|}\,.\label{eq:densityEstimator_Lagrangian}
\end{align}

An $N$-body simulation can be interpreted as a discretization of the Lagrangian map $\bm{x}_t$, following the evolution of a finite set of $N$-body particles starting at the locations $\{\bm{q}_i\}$ leading to the configuration $\{\bm{x}_i\}$. There exist various methods to estimate the density field $\rho$ from the final configuration $\{\bm{x}_i\}$. See for example gird-based particle-in-cell \cite{Harlow:1955}, smoothed particle hydrodynamics \cite{Gingold:1977}, and the now commonly used Delaunay Tessellation Field Estimator (DTFE) method  \cite{Pelupessy:2003, Schaap:2007, Weygaert:2009}. 

More recently, \cite{Shandarin:2011, Abel:2012} proposed a density estimator based on the discrete phase-space structure of the $N$-body particles, taking both the initial and final positions of the particles into account. Inspired by this novel approach, one of the authors has developed a phase-space Delaunay Tessellation Field Estimator that is used in the present paper. For completeness, we summarise the procedure here. For a full exposition of the method and a comparison to other density estimators, we refer to \cite{Feldbrugge:2024}.
\bigskip

The traditional DTFE density field is based on a Delaunay tessellation of the particles in Eulerian space $\{\bm{x}_i\}$. For the Phase-Space DTFE estimator, we evaluate a Delaunay tessellation of an early time of structure formation, before the first shell-crossing events occurred. At this time, the Lagrangian fluid consists of a single-stream region, and the DTFE density field is close to the continuum density field \eqref{eq:densityEstimator_Lagrangian}. Keeping the tessellation fixed, we evolve the $N$-body particles to the final time, potentially including multi-stream regions. The tessellation is a discrete version of the dark matter sheet in phase-space. The subsequent steps are identical for the DTFE and phase-space DTFE estimators.

Next, for each $N$-body particle $\bm{x}_i$, we evaluate the volume of the simplices adjoining it $V(W_i)$ and compute the density estimate at the vertex 
\begin{align}
    \rho_i = \frac{(d+1)m_i}{V(W_i)}\,,
\end{align}
with $d$ the dimension and $m_i$ the mass of the particle.

Finally, when estimating the density in a general point $\bm{x}$, we compute the simplices $\{D_j\}$ in which it lies (exactly one for the DTFE and an odd number for the phase-space DTFE scheme). For each of these simplices, we linearly interpolate the density 
\begin{align}
    \rho_{l_0} + [\nabla \rho](\bm{x}-\bm{x}_{l_0})
\end{align}
with $l_0$ a reference vertex and the gradient $[\nabla \rho]$ of the simplex $D_j$. The density field is defined as the sum of these interpolated densities. This yields a smooth density field that performs well in both the void regions -- with a low density of particles -- and multi-stream regions -- where Eulerian density estimators often yield artifacts. The resulting density field spikes at the caustics, where the volume $V(W_i)$ becomes very small.

We illustrate the DTFE and the phase-space DTFE density field in fig.\ \ref{fig:DTFE-vs-PS-DTFE}. In the single stream regions of the $N$-body simulation, the two density fields coincide. In the multi-stream regions, the DTFE density field shows artifacts that can be traced to the fact that points on different sheets are tessellated as if they are direct neighbors. The phase-space DTFE density field removes and fixes these artifacts.


\bsp	
\label{lastpage}
\end{document}